\documentclass{aa}  
\usepackage{txfonts}
\usepackage{hyperref} 

\usepackage{array} 
\usepackage{tabularx} 

\newcommand{\dwone}{DE0615$-$01}
\newcommand{\dwtwo}{DE0630$-$18}
\newcommand{\dwthree}{DE0644$-$28}
\newcommand{\dwfour}{DE0652$-$25}
\newcommand{\dwfive}{DE0716$-$06}
\newcommand{\dwsix}{DE0751$-$25}
\newcommand{\dwseven}{DE0805$-$31}
\newcommand{\dweight}{DE0812$-$24}
\newcommand{\dwnine}{DE0823$-$49}
\newcommand{\dwten}{DE0828$-$13}
\newcommand{\dweleven}{DE1048$-$52}
\newcommand{\dwtwelve}{DE1157$-$48}
\newcommand{\dwthirt}{DE1159$-$52}
\newcommand{\dwfourt}{DE1253$-$57}
\newcommand{\dwfift}{DE1520$-$44}
\newcommand{\dwsixt}{DE1705$-$54}
\newcommand{\dwsevent}{DE1733$-$16}
\newcommand{\dweightt}{DE1745$-$16}
\newcommand{\dwninet}{DE1756$-$45}
\newcommand{\dwtwenty}{DE1756$-$48}

\defcitealias{Sahlmann:2013kk}{JS13}
\defcitealias{Lazorenko:2013kk}{paper\,II} 	

\begin{document}
   \title{Astrometric planet search around southern ultracool dwarfs}
   \subtitle{I. First results, including parallaxes of 20 M8--L2 dwarfs\thanks{\textit{Based on observations made with ESO telescopes at the La Silla Paranal Observatory under programme IDs 086.C-0680, 087.C-0567, 088.C-0679, and 089.C-0397.}}}
   
\author{J. Sahlmann\inst{1,2}
		\and P. F. Lazorenko\inst{3}
		\and D. S\'egransan\inst{2}
		\and E. L. Mart\'in\inst{4} 
	        \and M. Mayor\inst{2} 
		\and D. Queloz\inst{2,5} 
		\and S. Udry\inst{2}}	

\institute{European Space Agency, European Space Astronomy Centre, P.O. Box 78, Villanueva de la Ca\~nada, 28691 Madrid, Spain\\
		\email{johannes.sahlmann@sciops.esa.int}				
		\and
		Observatoire de Gen\`eve, Universit\'e de Gen\`eve, 51 Chemin Des Maillettes, 1290 Versoix, Switzerland
		\and
		Main Astronomical Observatory, National Academy of Sciences of the Ukraine, Zabolotnogo 27, 03680 Kyiv, Ukraine
		\and  
		INTA-CSIC Centro de Astrobiolog\'ia, 28850 Torrej\'on de Ardoz, Madrid, Spain
		\and
		University of Cambridge, Cavendish Laboratory, J J Thomson Avenue, Cambridge, CB3 0HE, UK
}

\date{Received 6 December 2013 / Accepted 10 March 2014} 

\abstract
{Extrasolar-planet searches that target very low-mass stars and brown dwarfs are hampered by intrinsic or instrumental limitations. Time series of astrometric measurements with precisions better than one milli-arcsecond can yield new evidence on the planet occurrence around these objects.}
{We present first results of an astrometric search for planets around 20 nearby dwarf stars with spectral types M8--L2.}
{Over a time-span of two years, we obtained $I$-band images of the target fields with the FORS2 camera at the Very Large Telescope. Using background stars as references, we monitored the targets' astrometric trajectories, which allowed us to measure parallax and proper motions, set limits on the presence of planets, and to discover the orbital motions of two binary systems.}
{We determined trigonometric parallaxes with an average accuracy of 0.09 mas ($\simeq$\,0.2\,\%), which resulted in a reference sample for the study of ultracool dwarfs at the M/L transition, whose members are located at distances of 9.5--40 pc. This sample contains two newly discovered tight binaries (\dwtwo\ and \dwnine) and one previously known wide binary (\dwfift). Only one target shows $I$-band variability $>$5 mmag r.m.s.  We derived planet exclusion limits that set an upper limit of 9 \% on the occurrence of giant planets with masses $\gtrsim$\,5\,$M_J$ in intermediate-separation (0.01--0.8 AU) orbits around M8--L2 dwarfs.}
{We demonstrate that astrometric observations with an accuracy of 120 $\mu$as over two years are feasible from the ground and can be used for a planet-search survey. The detection of two tight very low-mass binaries shows that our search strategy is efficient and may lead to the detection of planetary-mass companions through follow-up observations.}

\maketitle

\section{Introduction}
Extrasolar planets are common around stars in the solar neighbourhood \citep{Mayor:1995lr, Mayor:2011fj}, but little is known about their existence around very low-mass stars and brown dwarfs, also known as ultracool dwarfs (UCDs) with spectral types M7 and later \citep{Martin:1999yf}, because of their low luminosities and the associated observational limitations. The presence of planets is expected because UCDs provide the necessary ingredients for planet formation and are commonly surrounded by disks in which grain growth and dust settling has been observed \citep{Apai:2005kx, Riaz:2012ys, Ricci:2012fk, Luhman:2012vn}. The potential planet mass depends on the amount of material available in the disk, which is generally lower than for main-sequence stars. Extended disks with masses higher than Jupiter-mass are observed, but not common \citep{Scholz:2006vl, Harvey:2012vn}, and smaller disk masses are found frequently, which provides the material for the formation of sub-Jupiter-mass planets \citep{Payne:2007ad}. 

The discovery of giant planets around UCDs can on one hand be used to probe the predictions of planet formation theories. {According to the core-accretion theory, giant-planet occurrence scales with central star mass and is expected be low around M dwarfs \citep{Laughlin:2004uq}, hence especially low around UCDs. Disk instability may be able to form giant planets around UCDs if their disks are \emph{suitably unstable} \citep{Boss:2006kx}.} On the other hand, the search for planets with Neptune-mass and lighter is a first step towards characterising the population of small and terrestrial planets around UCDs, some of which may reside in the habitable zones and therefore become prime targets for future attempts to detect the constituents of their atmospheres \citep{Belu:2013aa,Bolmont:2011lr}. 

Radial-velocity measurements of UCDs were used to exclude a large population of giant planets $\gtrsim$2\,{Jupiter-mass} ($M_J$) on very tight orbits $<0.05$ AU \citep{Blake:2010lr, Rodler:2012uq}. At wider separations $\gtrsim$2\,AU, direct-imaging searches equally excluded a large population of giant planets \citep{Stumpf:2010lr}. Two very low-mass stars were found to host Earth-mass (\citealt{Kubas:2010fk}, using microlensing) and Mars-sized (\citealt{Muirhead:2012fk}, using \textit{Kepler}) planets. Recently, a $\sim$$2\,M_J$ planetary mass object\footnote{The assignment of planet status to individual objects in the literature is debated because of their unknown formation paths and the observed overlapping mass range of planets and brown dwarfs. For UCDs, we propose to use mass ratio {and separation thresholds} of 0.1 and 10 AU, respectively, below which companions may be called planets.} was discovered at 0.87 AU around a 0.022 $M_\sun$ brown dwarf using gravitational microlensing \citep{Han:2013aa}.

\subsection{Very low-mass binaries}
Ultracool dwarfs are thought to form like stars, but this view is challenged by the apparent properties of UCD binaries that show significant differences to stellar binaries \citep{Bouy:2006aa, Burgasser:2007ix, Duchene:2013aa}. In particular, the secondary-to-primary mass ratio ($q=M_2/M_1$) distribution is strongly skewed towards unity in contrast to Sun-like stars and M dwarfs that show a nearly uniform $q$-distribution, which may suggest different formation mechanisms \citep{Goodwin:2013fk}. By mapping for example the $q$- and orbital-eccentricity distribution, the discovery and characterisation of UCD binaries {yields new observational results that can help to examine very low-mass binary formation.} 

{In this work, we consider binaries to be \emph{tight} if their relative semi-major axis is $\lesssim 1$\,AU.}

\subsection{Astrometric planet search}
Astrometry consists of measuring the apparent sky-position of stars and is a powerful method for the discovery and characterisation of extrasolar planets, provided that the achieved accuracy is better than 1 milli-arcsecond (mas), {a threshold that} corresponds to the reflex motion amplitude induced on a Sun-like star at 10 pc by a 5 $M_J$ giant planet on a three-year orbit \citep{Sozzetti:2005qy, Sahlmann:2012fk2}. Ultracool dwarfs have been targeted by several astrometric planet searches \citep{Pravdo:1996fk, Boss:2009ff, Forbrich:2013aa}, but have not yet yielded new exoplanet discoveries. So far, the importance of astrometry for UCD research stemmed therefore primarily from its ability to yield precise trigonometric distances, which are central to determine the luminosity, mass, and age relationships for UCDs and required to understand the physics of these objects \citep{Dahn:2002zr, Andrei:2011lr, Dupuy:2012fk, Dupuy:2013aa, Smart:2013aa}.  

The currently most precise ground-based instrument for astrometry of faint ($\gtrsim$10th mag) optical sources is {\small FORS2} at the Very Large Telescope, achieving accuracies of 50--100 micro-arcsecond ($\mu$as) \citep{Lazorenko:2009ph}. At this level, astrometry opens a new observational window to low-mass companions of UCDs at small-to-intermediate separations ($\sim$\,0.05\,--\,2\,AU). {For instance, the reflex motion amplitude induced on a 0.08 $M_\sun$ object at 10 pc by a Neptune-mass planet on a three-year orbit is 60 $\mu$as.} We therefore began an astrometric survey of UCDs using {\small FORS2} in 2010 ({also known as the PALTA project: planets around L-dwarfs with astrometry.}) and announced its first discovery, a low-mass companion to an L\,dwarf, in \cite{Sahlmann:2013kk}, hereafter \citetalias{Sahlmann:2013kk}. Here, we report first results of the survey covering a time-span of two years, which allowed us to screen the target sample, measure parallaxes, and discover new binaries. The paper is structured as follows: Sections \ref{sec:targetsel} and \ref{sec:obs} describe the target selection and the observations. The astrometric data analysis is detailed in Sect.~\ref{sec:analysis} and the results are presented in Sect.~\ref{sec:results}. We conclude in Sect.~\ref{sec:concl}. In an accompanying paper, we describe the data reduction procedures in detail and present a new deep astrometric catalogue of reference stars in the target fields.

\section{Target selection and observation strategy}\label{sec:targetsel}
A large sample of targets is desirable for a planet-search survey to increase the detection probability and to eventually draw conclusions on planet occurrence. However, high-precision astrometric monitoring is expensive in telescope time. We found a reasonable compromise in a sample size of 20 targets and searched the literature for UCDs that fulfil the following criteria: a) sufficiently bright to be observable at high signal-to-noise ratio (S/N) in the optical; b) nearby with a distance estimate lower than 30 pc, because the astrometric orbital signature decreases with distance; and c) availability of a large number of reference stars of similar magnitude within the field of view to obtain the best achievable astrometric precision.

\begin{table} 
\caption{Survey sample. The first column lists the target number and the third column gives the identifier used throughout this paper. The spectral types and $J$-band magnitudes are taken from \cite{Phan-Bao:2008fr}, while the $I$-band magnitudes are reproduced from Table~\ref{tab:targsobj}.}
\label{tab:paltatargets} 
\begin{tabular}{r@{\;\;} c@{\;\;\;} c@{\;\;\;} c@{\;\;\;} c r} 
\hline\hline  
Nr & DENIS-P & ID & $m_I$ & $m_J$ & Sp. \\ 
     &             &     &   (mag)          &  (mag)          & Type \\ 
     \hline
 1 & \href{http://simbad.u-strasbg.fr/simbad/sim-basic?Ident=DENIS-P+J061549.3-010041&submit=SIMBAD+search}{J0615493-010041} & \dwone & 17.0 & 12.7 & L2.5 \\ 
 2 & \href{http://simbad.u-strasbg.fr/simbad/sim-basic?Ident=DENIS-P+J063001.4-184014&submit=SIMBAD+search}{J0630014-184014} & \dwtwo & 15.7 & 11.3 & M8.5 \\ 
 3 & \href{http://simbad.u-strasbg.fr/simbad/sim-basic?Ident=DENIS-P+J064414.3-284141&submit=SIMBAD+search}{J0644143-284141} & \dwthree & 16.9 & 11.7 & M9.5 \\ 
 4 & \href{http://simbad.u-strasbg.fr/simbad/sim-basic?Ident=DENIS-P+J065219.7-253450&submit=SIMBAD+search}{J0652197-253450} & \dwfour & 16.0 & 11.8 & L0.0 \\ 
 5 & \href{http://simbad.u-strasbg.fr/simbad/sim-basic?Ident=DENIS-P+J071647.8-063037&submit=SIMBAD+search}{J0716478-063037} & \dwfive & 17.5 & 12.2 & L1.0 \\ 
 6 & \href{http://simbad.u-strasbg.fr/simbad/sim-basic?Ident=DENIS-P+J075116.4-253043&submit=SIMBAD+search}{J0751164-253043} & \dwsix & 16.5 & 12.4 & L1.5 \\ 
 7 & \href{http://simbad.u-strasbg.fr/simbad/sim-basic?Ident=DENIS-P+J080511.0-315811&submit=SIMBAD+search}{J0805110-315811} & \dwseven & 16.0 & 11.1 & M8.0 \\ 
 8 & \href{http://simbad.u-strasbg.fr/simbad/sim-basic?Ident=DENIS-P+J081231.6-244442&submit=SIMBAD+search}{J0812316-244442} & \dweight & 17.2 & 12.4 & L1.5 \\ 
 9 & \href{http://simbad.u-strasbg.fr/simbad/sim-basic?Ident=DENIS-P+J082303.1-491201&submit=SIMBAD+search}{J0823031-491201} & \dwnine & 17.1 & 12.4 & L1.5 \\ 
10 & \href{http://simbad.u-strasbg.fr/simbad/sim-basic?Ident=DENIS-P+J082834.3-130919&submit=SIMBAD+search}{J0828343-130919} & \dwten & 16.1 & 12.2 & L1.0 \\ 
11 & \href{http://simbad.u-strasbg.fr/simbad/sim-basic?Ident=DENIS-P+J104827.8-525418&submit=SIMBAD+search}{J1048278-525418} & \dweleven & 17.5 & 12.4 & L1.5 \\ 
12 & \href{http://simbad.u-strasbg.fr/simbad/sim-basic?Ident=DENIS-P+J115748.0-484442&submit=SIMBAD+search}{J1157480-484442} & \dwtwelve & 17.3 & 12.2 & L1.0 \\ 
13 & \href{http://simbad.u-strasbg.fr/simbad/sim-basic?Ident=DENIS-P+J115927.4-524718&submit=SIMBAD+search}{J1159274-524718} & \dwthirt & 14.6 & 11.5 & M9.0 \\ 
14 & \href{http://simbad.u-strasbg.fr/simbad/sim-basic?Ident=DENIS-P+J125310.8-570924&submit=SIMBAD+search}{J1253108-570924} & \dwfourt & 16.7 & 12.0 & L0.5 \\ 
15 & \href{http://simbad.u-strasbg.fr/simbad/sim-basic?Ident=DENIS-P+J152002.2-442242&submit=SIMBAD+search}{J1520022-442242} & \dwfift & 16.8 & 12.2 & L1.0 \\ 
16 & \href{http://simbad.u-strasbg.fr/simbad/sim-basic?Ident=DENIS-P+J170547.4-544151&submit=SIMBAD+search}{J1705474-544151} & \dwsixt & 16.5 & 11.3 & M8.5 \\ 
17 & \href{http://simbad.u-strasbg.fr/simbad/sim-basic?Ident=DENIS-P+J173342.3-165449&submit=SIMBAD+search}{J1733423-165449} & \dwsevent & 16.9 & 12.2 & L1.0 \\ 
18 & \href{http://simbad.u-strasbg.fr/simbad/sim-basic?Ident=DENIS-P+J174534.6-164053&submit=SIMBAD+search}{J1745346-164053} & \dweightt & 17.0 & 12.4 & L1.5 \\ 
19 & \href{http://simbad.u-strasbg.fr/simbad/sim-basic?Ident=DENIS-P+J175629.6-451822&submit=SIMBAD+search}{J1756296-451822} & \dwninet & 15.5 & 11.5 & M9.0 \\ 
20 & \href{http://simbad.u-strasbg.fr/simbad/sim-basic?Ident=DENIS-P+J175656.1-480509&submit=SIMBAD+search}{J1756561-480509} & \dwtwenty & 16.7 & 11.8 & L0.0 \\ 
\hline\hline
\end{tabular} 
\end{table}

\begin{figure}
\center
\includegraphics[width= \linewidth]{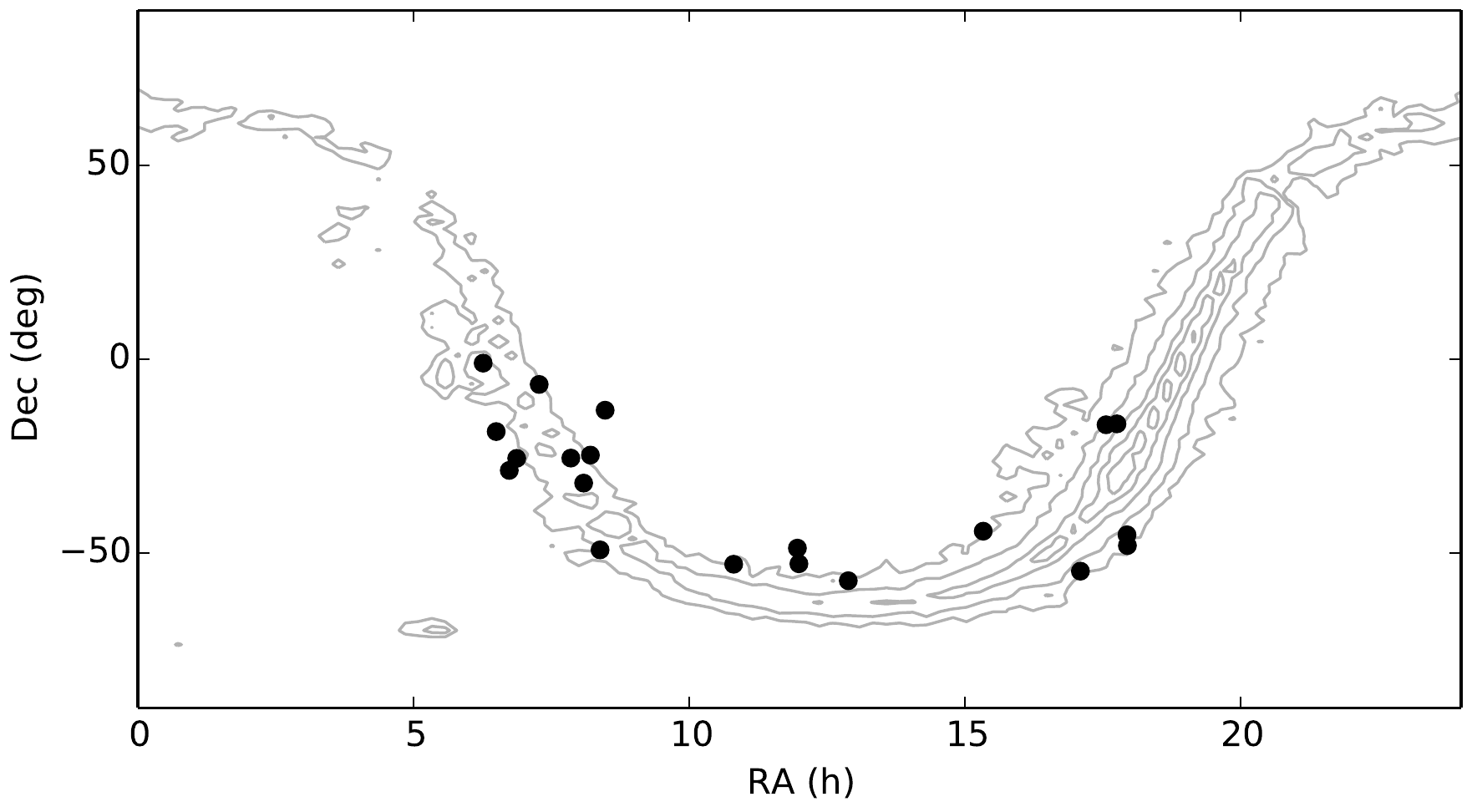}
\caption{Location of the observed objects in equatorial coordinates (filled circles). The contour lines correspond to IRAS \citep{Joint-Iras-Science:1994aa} source density and trace the Galactic plane.}
\label{fig:sky}
\end{figure}

The nearby M and L dwarfs that were identified in the DENIS database and spectroscopically confirmed by \cite{Phan-Bao:2008fr} satisfy our criteria, because they are located close to the Galactic plane, where the density of potential reference stars is high. Their corresponding VizieR catalogue (J/MNRAS/383/831) has 26 entries and we excluded four entries located too far south for observation at small airmass from Paranal by imposing declination (Dec) $> -60\degr$. Of the remaining 22 objects, we removed the two faintest ones. The 20 selected targets, fourteen early-type L dwarfs and six late-type M dwarfs, with spectral types of M8.0 - L2.5 are presented in Table~\ref{tab:paltatargets}. Finding charts are shown in Figs. \ref{fig:fcs1} and \ref{fig:fcs2}.

Because they are nearby, the astrometric trajectories of our targets are dominated by proper and parallactic motion. A possible orbital signature becomes apparent as periodic offsets to these regular displacements and, to detect these deviations, the proper motion and parallax parameters have to be measured first. At least five one-dimensional measurements, ideally evenly spaced in time, over one year are required to separate proper motion and parallactic motion. To obtain accurate astrometric parameters and detect indications of the superimposed orbital motion, we observed each target approximately ten times, effectively yielding 20 one-dimensional measurements, over a time-span of 18 months, that is fives times in each of the two seasonal observation windows. Because our targets span a broad range in right ascension (RA), see Fig. \ref{fig:sky}, these observations took place over two years.

The only previously known multiple system in the sample is \dwfift, a visual binary with projected separation of $\sim$1.1\arcsec~\citep{Burgasser:2007gd}. Its components are resolved in the {\small FORS2} images (Sect.~\ref{sec:dw15}) and we searched for planetary companions of the primary.

\section{Observations and data reduction}\label{sec:obs}
\begin{figure}
\center
\includegraphics[width= \linewidth]{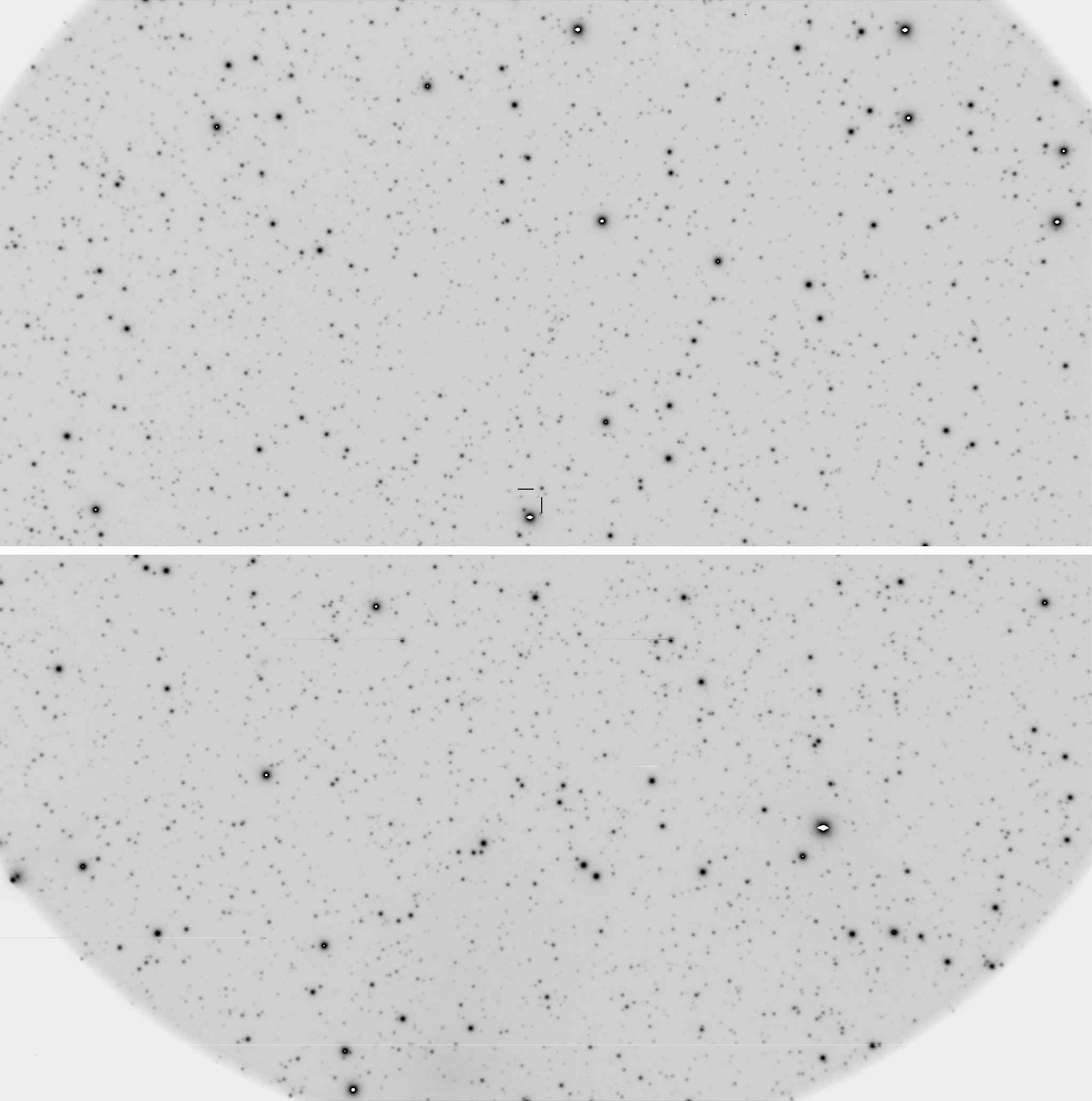}
\caption{Raw FORS2 image of \dwsevent, whose location is marked close to the centre of the $4.2\arcmin \times 4.2\arcmin$ field of view. North is up, east is left. The two CCD chips are separated by a gap that is effectively $\sim$2.0\arcsec\ wide.}
\label{fig:sampleImage}
\end{figure}

The observations were made with the FORS2 instrument \citep{Appenzeller:1998lr} in imaging mode, which is attached to the UT1 telescope of the ESO Very Large Telescope. The $4.2\arcmin \times 4.2\arcmin$ field of view is imaged on two CCD chips measuring $2048 \times 1024$ pixels (px) each. We used the red-optimised MIT CCDs and configured the camera in high-resolution mode with 2$\times$2 binning, resulting in an on-sky pixel scale of $\sim$0.126~$\arcsec$px$^{-1}$. The selected $I$-band filter (I\_BESS+77) has a central wavelength of 768 nm and a FWHM of 138 nm\footnote{\url{http://www.eso.org/sci/facilities/paranal/instruments/fors}}. Figure \ref{fig:sampleImage} shows an example {\small FORS2} image. 

All observations were carried out in service mode by ESO personnel. Atmospheric seeing is a critical parameter for the achievable astrometric precision and we requested it to be $<0.9\arcsec$. To minimise the effects of differential chromatic refraction (DCR), we also limited the acceptable airmass to $\lesssim 1.2$ and requested observations taken near meridian ($|$hour angle$|<1$h). The background level was minimised by avoiding observations with thick clouds and close to the moon in bright time. At every observation epoch, several (20-50) individual exposures were obtained.

Table \ref{tab:paltaobs} summarises the observations taken between October 2010 and September 2012 (PI: Sahlmann; title: \emph{Exploring planetary and substellar companions of L dwarfs}; ESO periods P86--P89) and lists their time span $\Delta T$, the number of epochs $N_e$ and frames $N_f$ and their ratio $N_{f/e}$, the number of used reference stars $N_\star$ in the images, and the range of individual exposure times (DIT). Our targets have magnitudes of $m_I=14.5-17.5$ (median of 16.7) and require exposure times of single frames of $3-70$ s to obtain well-exposed images. We limited the telescope time spent for one epoch to 0.7 h including overheads, regardless of the target magnitude, therefore the number of frames varied for different targets. On average, every target was observed at 11 epochs with 31 individual frames and an effective exposure time of 0.24 h  each. The total telescope time included overheads used over two years was $\sim$15 nights. 

\begin{table} 
\caption{Summary of FORS2 observations in P86--P89.}
\label{tab:paltaobs} 
\centering
\begin{tabular}{r@{\;\;} c c r c c  r r r r} 
\hline\hline  
Nr&ID & $\Delta T$ & $N_e$&$N_f$&$N_{f/e}$ & $N_{\star}$ & DIT\\ 
  &   &   (d)     &           &    &        &                     & (s)  \\ 
     \hline
1 & \dwone & 448 & 11 & 260 & 24 & 365 & 44,60 \\
2 & \dwtwo & 473 & 10 & 376 & 38 & 246 & 5--21 \\
3 & \dwthree & 471 & 11 & 280 & 25 & 259 & 35,58 \\
4 & \dwfour & 475 & 11 & 396 & 36 & 130 & 14,23 \\
5 & \dwfive & 476 & 10 & 214 & 21 & 676 & 47,92 \\
6 & \dwsix & 470 & 12 & 371 & 31 & 371 & 22,39 \\
7 & \dwseven & 471 & 11 & 384 & 35 & 441 & 17--38 \\
8 & \dweight & 473 & 11 & 213 & 19 & 461 & 50--77 \\
9 & \dwnine & 471 & 12 & 240 & 20 & 850 & 50--68 \\
10 & \dwten & 472 & 11 & 381 & 35 & 193 & 20,26 \\
11 & \dweleven & 469 & 12 & 223 & 19 & 508 & 20--60 \\
12 & \dwtwelve & 517 & 10 & 290 & 29 & 465 & 10--60 \\
13 & \dwthirt & 471 & 9 & 491 & 55 & 376 & 3,4 \\
14 & \dwfourt & 502 & 10 & 286 & 29 & 438 & 28,33 \\
15 & \dwfift & 459 & 10 & 281 & 28 & 635 & 26--30 \\
16 & \dwsixt & 499 & 11 & 338 & 31 & 1387 & 26 \\
17 & \dwsevent & 481 & 10 & 307 & 31 & 1218 & 30 \\
18 & \dweightt & 483 & 10 & 259 & 26 & 1056 & 35,39 \\
19 & \dwninet & 483 & 10 & 561 & 56 & 366 & 8,9 \\
20 & \dwtwenty & 483 & 10 & 346 & 35 & 1111 & 28 \\
\hline
\end{tabular} 
\end{table}

The data reduction methods used to extract the target's astrometry from the raw images are described in \cite{Lazorenko:2006qf} and \cite{Lazorenko:2011lr,Lazorenko:2009ph,Lazorenko:2007ul} (see also \citetalias{Sahlmann:2013kk} and \citealt{Sahlmann:2012fk2}). However, because of the large amount of collected data in the present programme, it was possible to significantly improve the reduction, in particular the mitigation of systematic errors. A detailed description of these improvements and a global analysis of the achieved astrometric performance is therefore presented in an accompanying paper (Lazorenko et al., \emph{in prep.}, hereafter \citetalias{Lazorenko:2013kk}). Here, we present a short summary and a few specific steps.

The {\small FORS2} raw frames were flat-fielded and bias-corrected following standard procedures. The photocentres of the hundreds of stars detectable in an image were determined by modelling the stellar images with an analytical model. The large number of reference stars, whose  proper and parallax motions are taken into account, makes it possible to model and mitigate atmospheric image motion, field distortions introduced by the optical system, and systematic image displacements due to a variety of effects, for example small relative motions of the two CCD chips and seeing-dependent errors. The pixel scale was determined to {a relative precision of better than $10^{-3}$} for every target using a large number of reference stars included in the USNO-B astrometric catalogue, which also allowed us to obtain absolute ICRF astrometry at the 0.05--0.10\arcsec\ level. The final output of the reduction procedure is the position of the nearby target relative to the field of reference stars in each frame. 

The seeing as measured in the accepted images ranged between 0.3\arcsec\ and 0.9\arcsec. Frames taken in poorer conditions were not considered in the reduction, because they introduce large systematic errors. About 14 \% of our observations were taken in very good seeing conditions $<$0.5\arcsec, which in some cases led to saturation of a few central pixels of the target's image. Although this occurred in only 4 \% of all frames and concerned 11 targets, these images were reduced with a dedicated procedure. A characteristic of the reduction procedure is that astrometric measurements taken within one epoch are correlated and the corresponding covariance matrix has non-zero off-diagonal entries. This has to be accounted for in the data analysis \citepalias{Sahlmann:2013kk,Lazorenko:2013kk}. 

\section{Data analysis}\label{sec:analysis} 
For every target, the data reduction yields a two-dimensional position measurement in RA and Dec in each frame, the associated uncertainties, the epoch, the observing conditions, and additional information that serves to characterise the individual observation. We illustrate the chain of analysis steps in detail with the help of \dwfour~and summarise the results for the other targets. The first step is to determine the relative parallax and proper motions of a target.

\subsection{Fit for parallax and proper motion}
The astrometric measurements $\alpha^{\star}_m$ and $\delta_m$ in RA and Dec, respectively, in frame $m$ at time $t_m$ relative to the reference frame of background stars are modelled with seven parameters (\citealt{Lazorenko:2011lr}, \citetalias{Sahlmann:2013kk})
\begin{equation}\label{eq:axmodel}
\begin{array}{ll@{\hspace{2mm}}l}
\!\alpha^{\star}_{m} =& \Delta \alpha^{\star}_0 + \mu_{\alpha^\star} \, t_m + \varpi \, \Pi_{\alpha,m} &-\;\; \rho\, f_{1,x,m}-  d \,f_{2,x,m} \\
\delta_{m} = & \underbrace{\Delta \delta_0 + \,\mu_\delta      \,  \;                      t_m \;+ \varpi \, \Pi_{\delta,m}}_{\mbox{Standard model} }  &\underbrace{+\;\; \rho \,f_{1,y,m}+  d \,f_{2,y,m}}_{\mbox{Refraction}},
\end{array}
\end{equation}
where $\Delta\alpha^{\star}_0, \Delta\delta_0$ are coordinate offsets, $\mu_{\alpha^\star}, \mu_\delta$ are proper motions, and the parallactic motion is expressed as the product of relative parallax $\varpi$ and the parallax factors $\Pi_\alpha, \Pi_\delta$. The parallax factors are computed as in \cite{Woolard:1966fk} on the basis of rectangular geocentric coordinates of the solar system barycentre obtained from the JPL Horizons system \citep{Giorgini:1996qy}. The parallax $\varpi$ is determined relative to the reference stars that are not located at infinite distances, therefore has to be corrected to obtain the absolute parallax (Sect.~\ref{sec:parcor}). DCR is modelled with the two parameters $\rho$ and $d$, and the parameters $f_{(1,2)}$ depend on zenith angle, temperature, and pressure as described in \citetalias{Sahlmann:2013kk}. The model Eq.~(\ref{eq:axmodel}) therefore has seven free parameters, five related to positions, proper motions, and parallax, and two to model the DCR.

Equation (\ref{eq:axmodel}) defines a system of $2\times N_{f}$ linear equations, whose least-squares solution is determined using matrix inversion \citep{Press:1986eu}, taking into account the individual data weights and covariances via the covariance matrix. Note that we set the reference epoch to the arithmetic mean $<\!t_m\!>$ that is tabulated in Table \ref{tab:targsobj} to minimise parameter correlations and that the solution is found by considering both coordinates simultaneously.

\begin{table*} 
\caption{Relative parallaxes and proper motions. The standard uncertainties were computed from the parameter variances that correspond to the diagonal of the problem's inverse matrix.}
\small
\label{tab:ppmres}  \centering  
\begin{tabular}{r@{\;\;} c@{\;\;} r@{\,$\pm$\,}l@{\;\;} r@{\,$\pm$\,}l@{\;\;} r@{\,$\pm$\,}l@{\;\;} r@{\,$\pm$\,}l@{\;\;} r@{\,$\pm$\,}l r@{\,$\pm$\,}l@{\;\;} r@{\,$\pm$\,}l c c}
\hline\hline %
Nr & ID & \multicolumn{2}{c}{$\Delta \alpha^\star_0$} &\multicolumn{2}{c}{$\Delta \delta_0$} &\multicolumn{2}{c}{$\varpi$} &\multicolumn{2}{c}{$\mu_{\alpha^\star}$} &\multicolumn{2}{c}{$\mu_{\delta}$} &\multicolumn{2}{c}{$\rho$} &\multicolumn{2}{c}{$d$}  & $\sigma$ & $\Sigma$\\ 
     &    & \multicolumn{2}{c}{(mas)}& \multicolumn{2}{c}{(mas)}& \multicolumn{2}{c}{(mas)} &\multicolumn{2}{c}{(mas yr$^{-1}$)}&\multicolumn{2}{c}{(mas yr$^{-1}$)} & \multicolumn{2}{c}{(mas)}& \multicolumn{2}{c}{(mas)}& (mas)& (mas) \\
\hline
1 & \dwone & 108.74 & 0.05 & -13.39 & 0.15 & 45.26 & 0.09 & 198.72 & 0.11 & -56.21 & 0.11 & 16.3 & 1.2 & -22.6 & 1.1 & 0.65 & 0.20 \\ 
2\tablefootmark{a} & \dwtwo & 375.32 & 0.25 &  -600.09 & 0.12 & 51.29 & 0.09 & 322.69 & 0.48 & -501.46 & 0.16 & 17.6 & 0.9 & -24.6 & 0.8 &1.00 & 0.24 \\ 
3 & \dwthree & 141.73 & 0.07 & -32.18 & 0.07 & 24.76 & 0.07 & 222.73 & 0.12 & -80.49 & 0.11 & 17.9 & 1.0 & -23.9 & 0.8 & 0.57 & 0.12 \\ 
4 & \dwfour & -139.06 & 0.07 & 113.84 & 0.06 & 61.50 & 0.06 & -235.32 & 0.11 & 87.40 & 0.09 & 18.7 & 0.8 & -24.8 & 0.6 & 0.69 & 0.14 \\ 
5 & \dwfive & -69.05 & 0.07 & 57.55 & 0.22 & 40.53 & 0.12 & -11.03 & 0.14 & 148.50 & 0.13 & 15.4 & 2.8 & -20.5 & 2.4 & 0.87 & 0.15 \\ 
6 & \dwsix & -612.51 & 0.06 & 121.02 & 0.07 & 55.98 & 0.08 & -874.35 & 0.11 & 143.79 & 0.10 & 16.3 & 0.8 & -22.5 & 0.7 & 0.80 & 0.17 \\ 
7 & \dwseven & -10.91 & 0.06 & 107.69 & 0.06 & 42.09 & 0.08 & -237.18 & 0.09 & 87.44 & 0.08 & 22.6 & 0.9 & -27.9 & 0.7 & 0.64 & 0.12 \\ 
8 & \dweight & 69.02 & 0.07 & -68.53 & 0.07 & 46.96 & 0.08 & 136.49 & 0.12 & -142.72 & 0.10 & 17.5 & 0.8 & -23.1 & 0.7 & 0.56 & 0.20 \\ 
9\tablefootmark{b}& \dwnine & -137.65 & 0.19 & -30.72 & 0.43 & 48.09 & 0.18 & -154.30 & 0.12 & 7.46 & 0.09 & 20.5 & 2.0 & -24.5 & 1.7 & 0.94 & 0.33 \\ 
10 & \dwten & -418.03 & 0.06 & 83.41 & 0.15 & 85.26 & 0.13 & -576.16 & 0.12 & 26.39 & 0.10 & 12.7 & 2.1 & -20.1 & 1.6 & 0.84 & 0.16 \\ 
11 & \dweleven & -113.49 & 0.06 & 74.42 & 0.13 & 35.94 & 0.07 & -233.01 & 0.11 & 45.55 & 0.11 & 22.9 & 2.8 & -27.4 & 2.2 & 0.71 & 0.20 \\ 
12 & \dwtwelve & -46.13 & 0.06 & 33.15 & 0.10 & 34.39 & 0.07 & -59.10 & 0.11 & -18.11 & 0.12 & 13.7 & 1.8 & -19.6 & 1.4 & 1.06 & 0.24 \\ 
13 & \dwthirt & -701.58 & 0.06 & -0.54 & 0.16 & 105.21 & 0.12 & -1056.23 & 0.13 & -129.20 & 0.15 & 21.2 & 3.8 & -27.7 & 3.1 & 1.39 & 0.31 \\ 
14 & \dwfourt & -672.51 & 0.05 & -158.94 & 0.20 & 59.87 & 0.05 & -1549.76 & 0.09 & -429.94 & 0.09 & 15.3 & 3.9 & -21.0 & 3.1 & 0.57 & 0.21 \\ 
15 & \dwfift & -424.99 & 0.06 & -200.13 & 0.10 & 53.84 & 0.10 & -620.41 & 0.10 & -378.67 & 0.10 & 21.8 & 1.8 & -27.8 & 1.5 & 0.70 & 0.15 \\ 
16 & \dwsixt & -89.70 & 0.05 & 65.32 & 0.16 & 37.51 & 0.08 & -82.17 & 0.10 & 28.41 & 0.09 & 8.1 & 3.7 & -14.8 & 3.0 & 0.64 & 0.21 \\ 
17 & \dwsevent & 21.39 & 0.06 & -34.81 & 0.07 & 55.11 & 0.07 & 73.82 & 0.12 & -35.12 & 0.11 & 16.9 & 0.8 & -24.9 & 0.6 & 0.61 & 0.21 \\ 
18 & \dweightt & 84.83 & 0.08 & -40.82 & 0.09 & 50.84 & 0.09 & 110.14 & 0.15 & -97.16 & 0.15 & 14.6 & 1.3 & -21.3 & 1.0 & 0.64 & 0.17 \\ 
19 & \dwninet & 20.30 & 0.05 & -10.96 & 0.09 & 43.38 & 0.06 & 46.22 & 0.12 & -180.21 & 0.10 & 10.4 & 1.5 & -17.9 & 1.2 & 0.76 & 0.20 \\ 
20 & \dwtwenty & 65.98 & 0.04 & -28.70 & 0.07 & 46.98 & 0.05 & 79.09 & 0.07 & 35.04 & 0.08 & 19.3 & 1.6 & -25.7 & 1.3 & 0.67 & 0.19 \\ 
\hline
\end{tabular}
\tablefoot{
\tablefoottext{a}{Astrometric binary \citep{Sahlmann:2014}. The parameters are preliminary.}
\tablefoottext{b}{Astrometric binary \citepalias{Sahlmann:2013kk}. The parameters are from the discovery paper.}
}

\end{table*}

\begin{figure}
\center
\includegraphics[width= \linewidth]{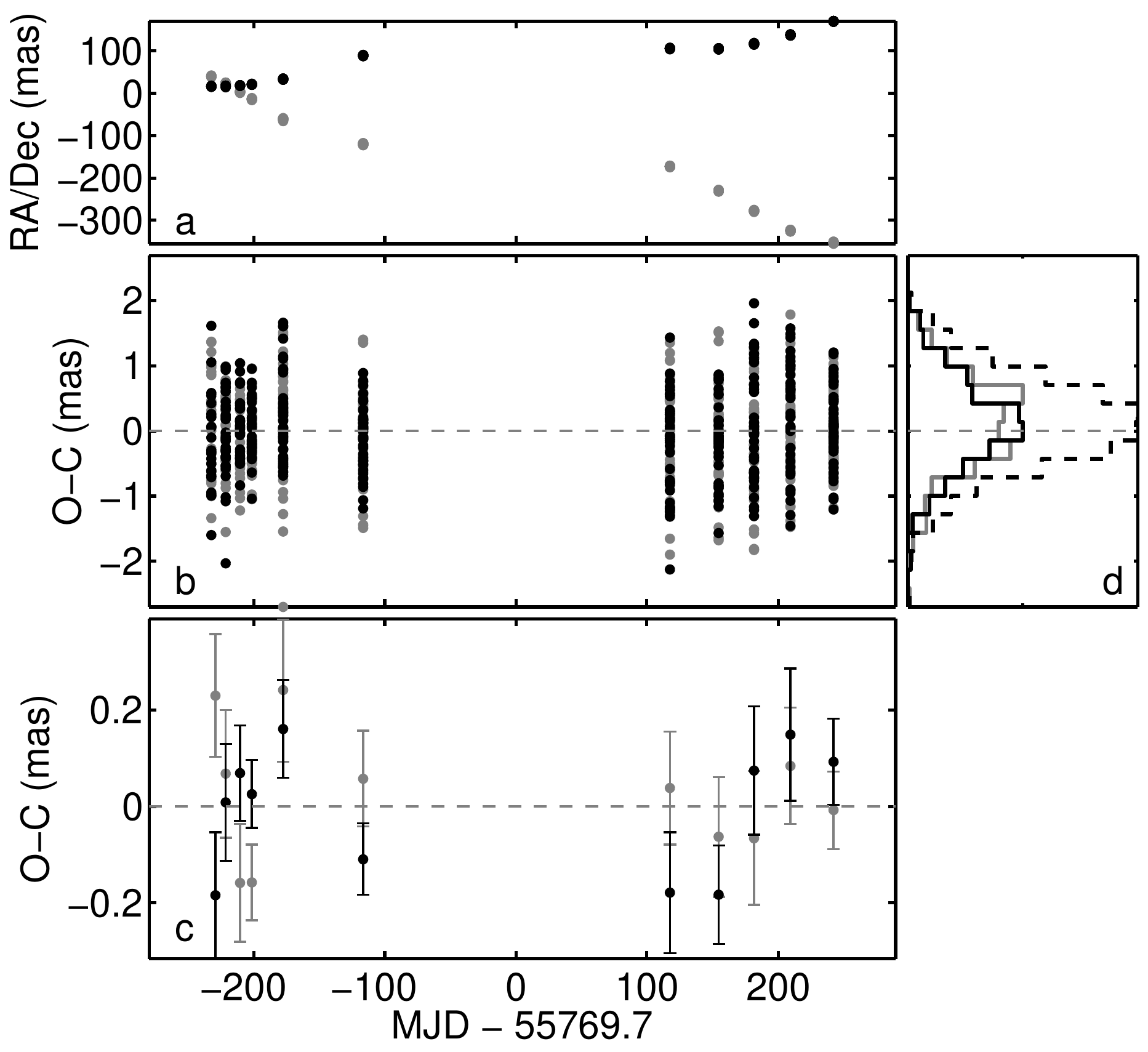}
\caption{Residuals of the astrometry solution for \dwfour. Panel a) shows the reduced data in RA (grey symbols) and Dec (black symbols) and panel b) shows the frame residuals after solving Eq. (\ref{eq:axmodel}). Panel d) displays the residual histogram in RA and Dec separately and combined (dashed curve), which appear to be normally distributed. Panel c) shows the epoch-averaged residuals with their mean uncertainties.}
\label{fig:dw04_residuals}
\end{figure}

Table \ref{tab:ppmres} shows the numerical results for all targets and Fig. \ref{fig:dw04_residuals} illustrates the case of \dwfour\ graphically. The last two columns of Table \ref{tab:ppmres} give the residual r.m.s. of individual frames ($\sigma$) and of epoch-averages\footnote{Epoch residuals were computed as weighted averages \citepalias{Lazorenko:2013kk}.} ($\Sigma$). A detailed discussion of the measurement uncertainties and their relation to the residual amplitude is given in \citetalias{Lazorenko:2013kk}, where we conclude that the $\chi^2$ statistics correspond to the theoretical expectations. Table \ref{tab:dw4corr} displays the parameter correlation matrix \citep{Press:1986eu} for \dwfour, which is a case with a high average correlation amplitude. There is nearly total anticorrelation between $\rho$ and $d$, which is expected, because both parameters model essentially the same effect of DCR. Moderate levels of correlation are observed between  a few parameters, but the main information is that the correlation between parallax and proper motions is low. 

\begin{figure}[h!] 
\center
\includegraphics[width= \linewidth]{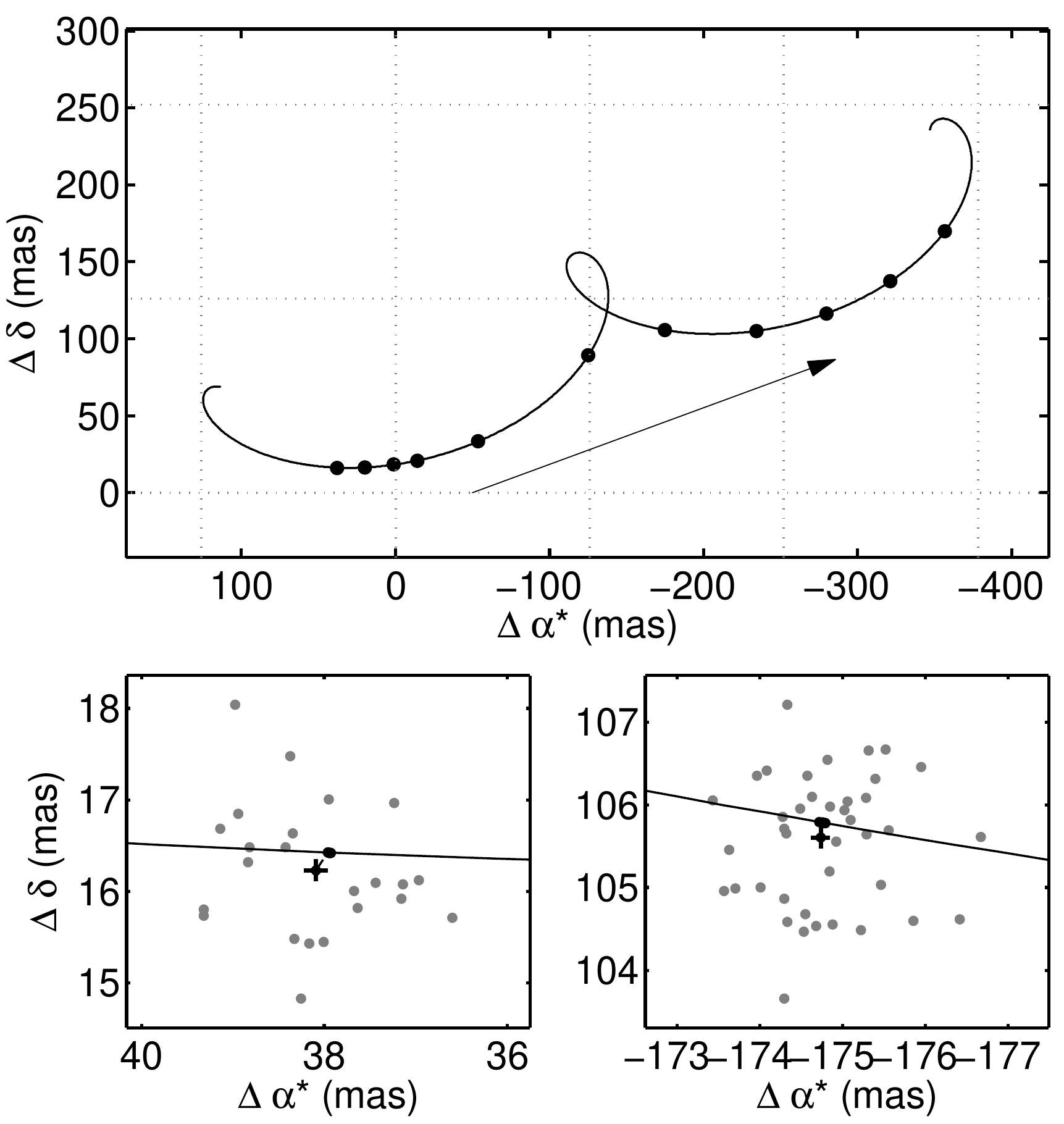}
\caption{\emph{Top:} motion of \dwfour\ in the sky. North is up, east is left. Circles show the epoch-averaged measurements and the curve corresponds to the best-fit model of parallax and proper motion. The arrow represents the proper-motion vector and the dotted grid indicates the sky-projected pixels of FORS2 measuring $126\times126$ mas$^2$. \emph{Bottom:} the two panels show close-up views of epochs Nr. 2 and 7, where individual frame measurements are shown in grey and the epoch averages are shown in black with their mean uncertainties. A dashed line connects the epoch average to the model position.}
\label{fig:dw01_2D}
\end{figure}

\begin{figure*}[h!]
\center
\includegraphics[width= 0.4\linewidth]{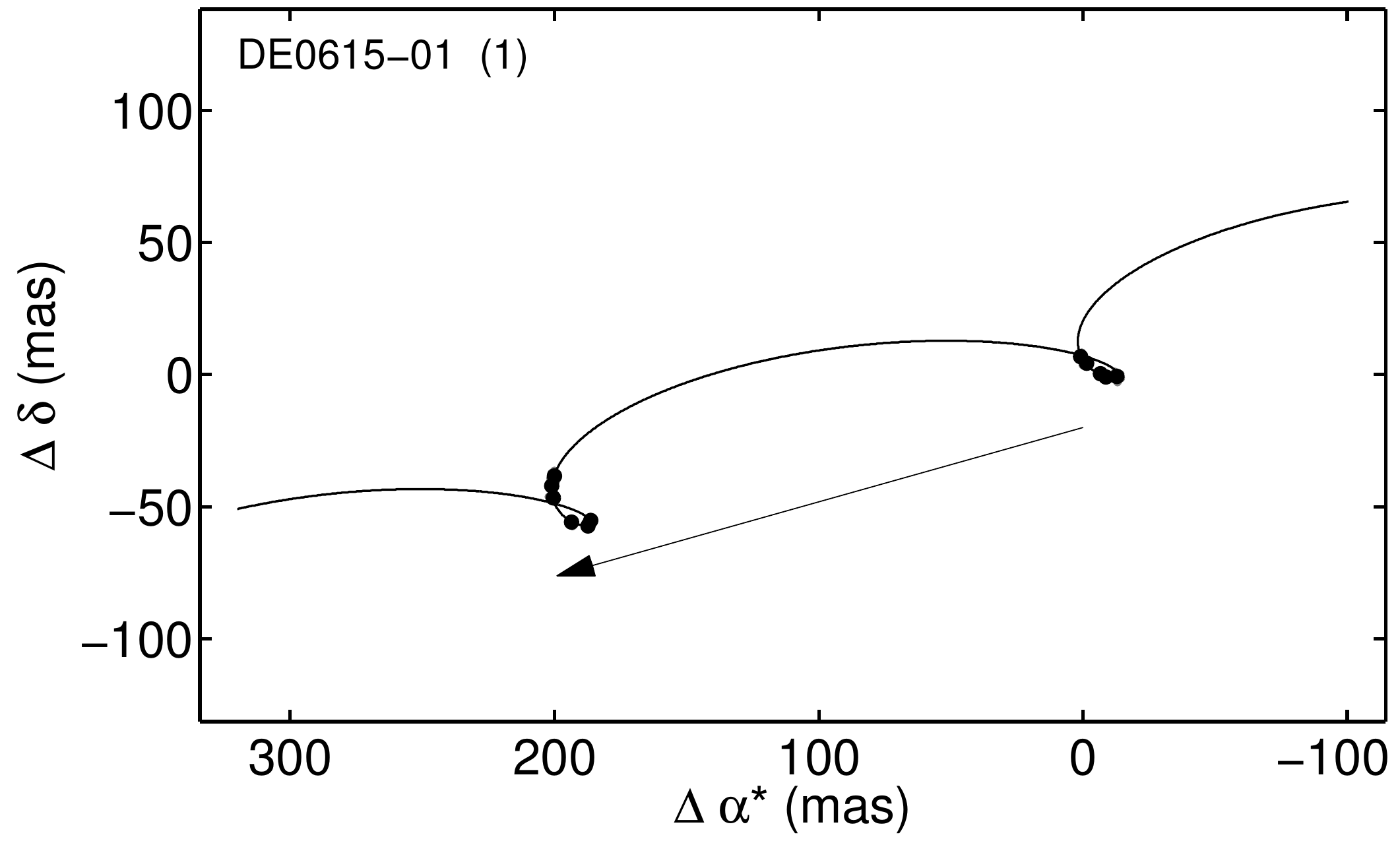}
\includegraphics[width= 0.4\linewidth]{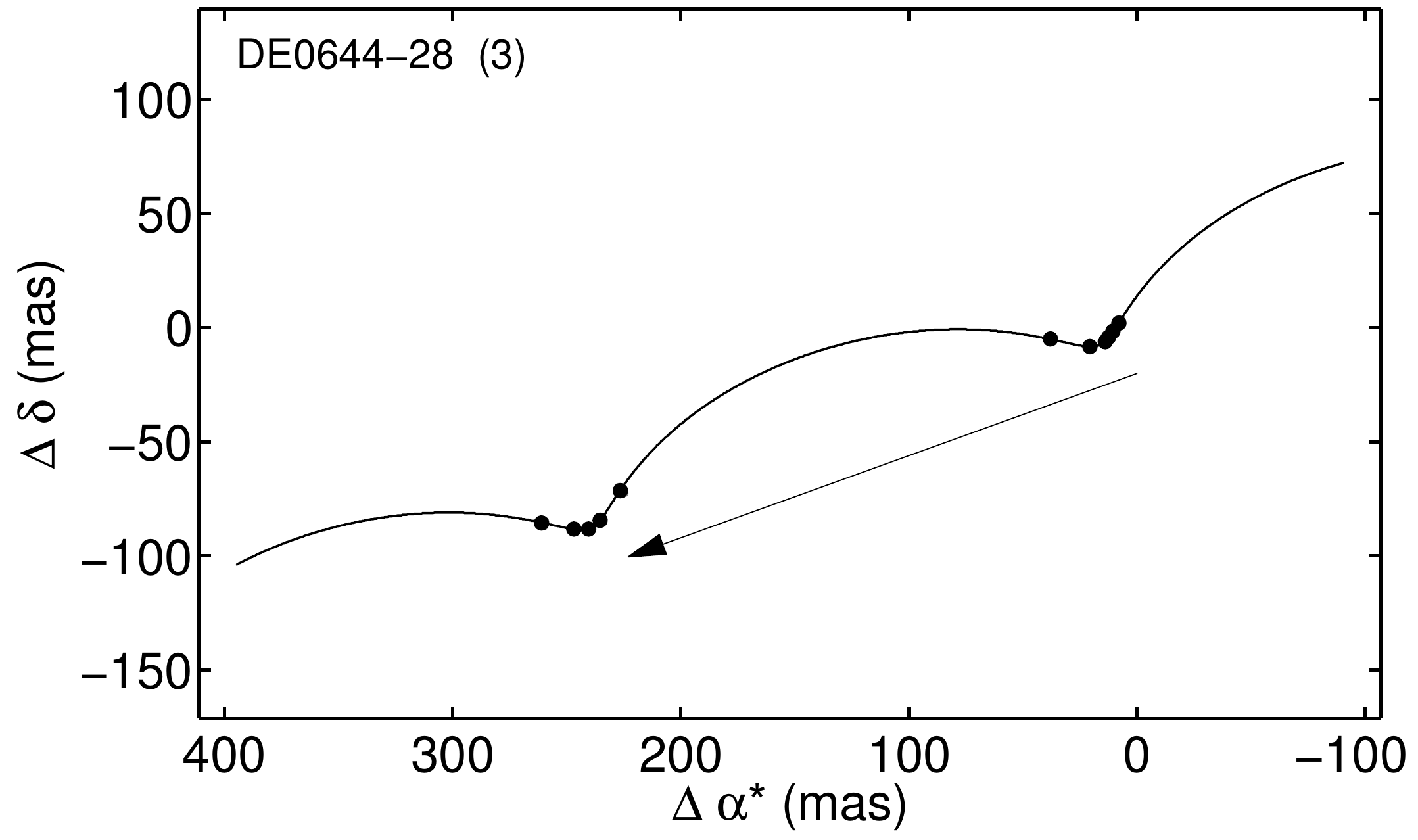}
\includegraphics[width= 0.4\linewidth]{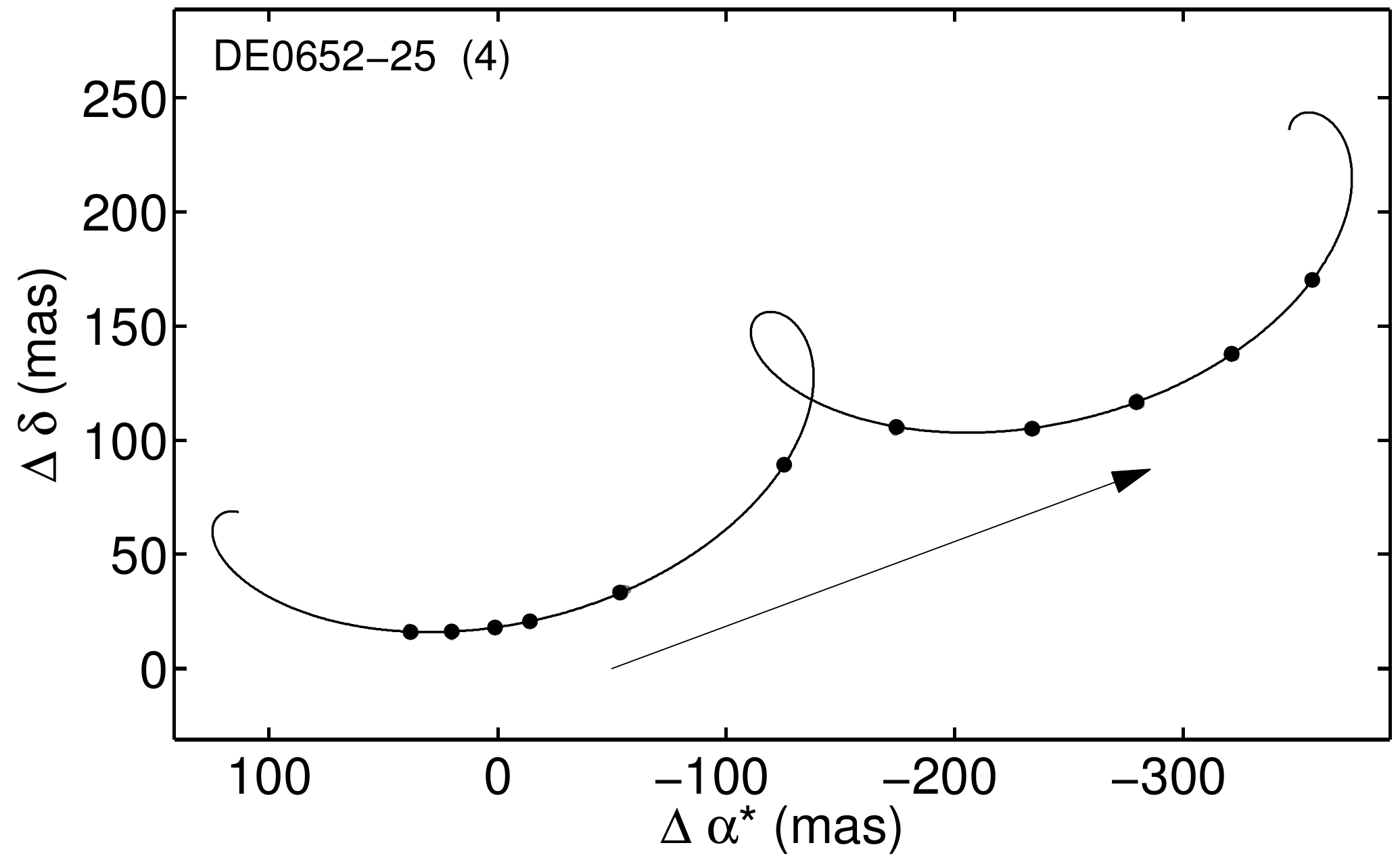}
\includegraphics[width= 0.4\linewidth]{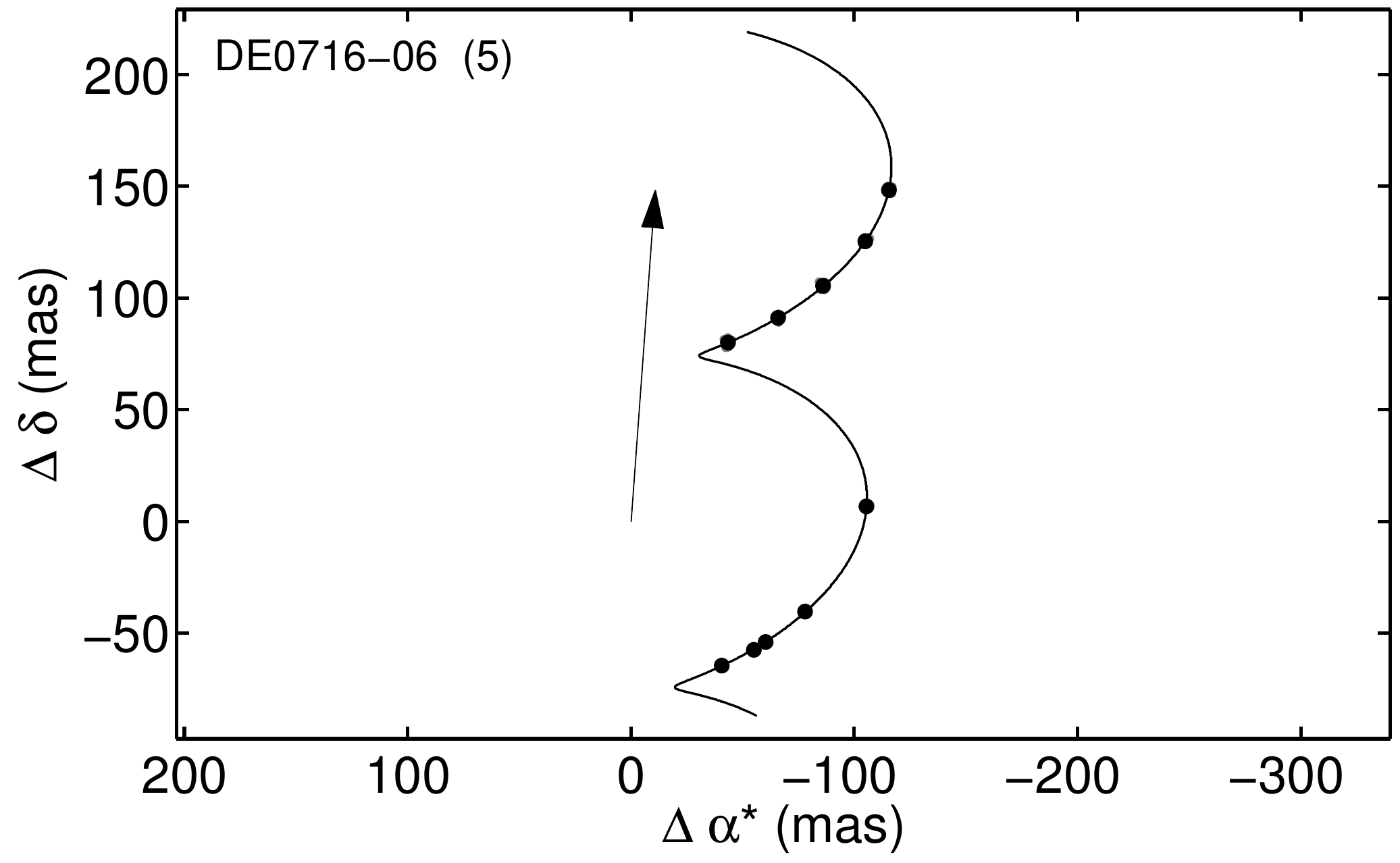}
\includegraphics[width= 0.4\linewidth]{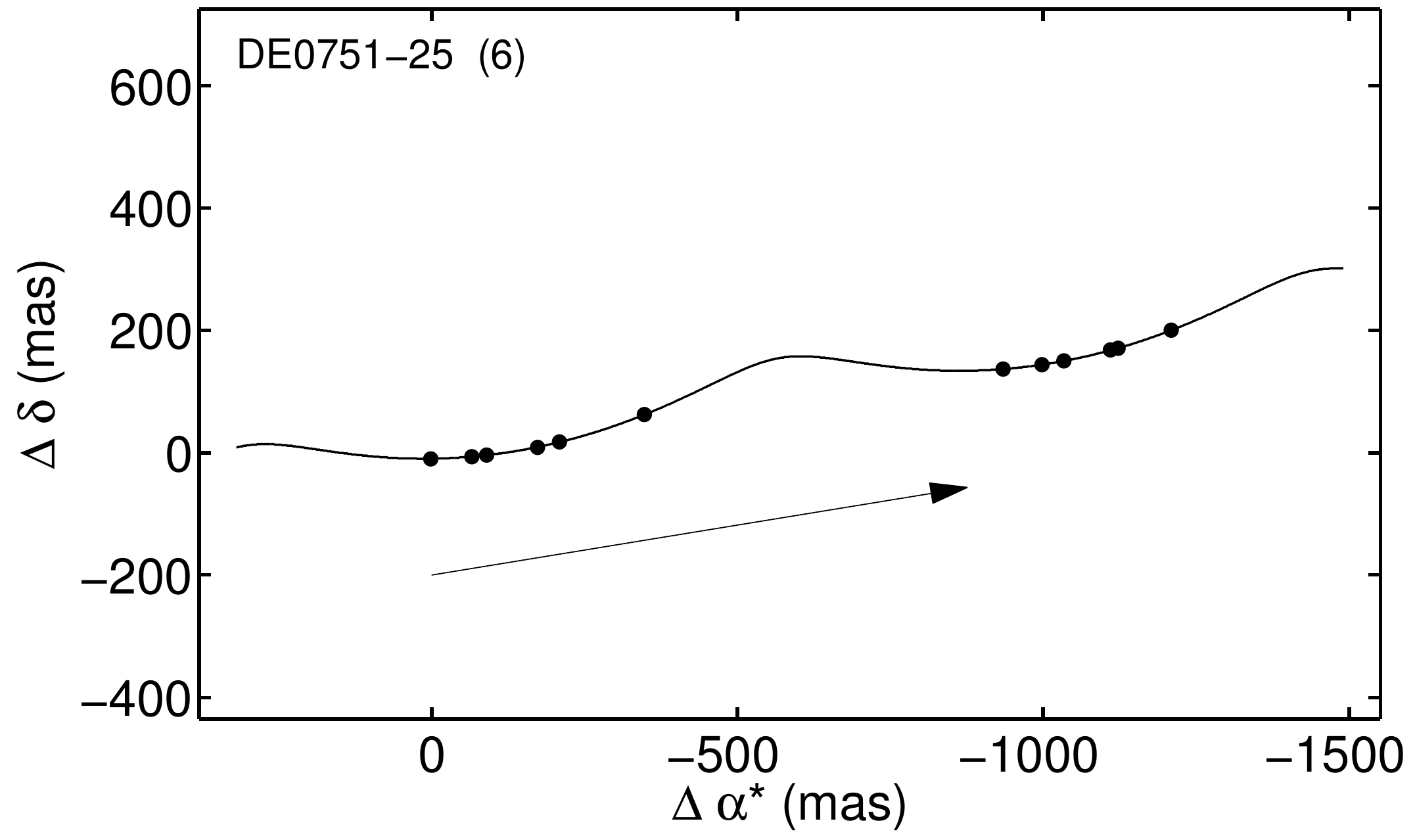}
\includegraphics[width= 0.4\linewidth]{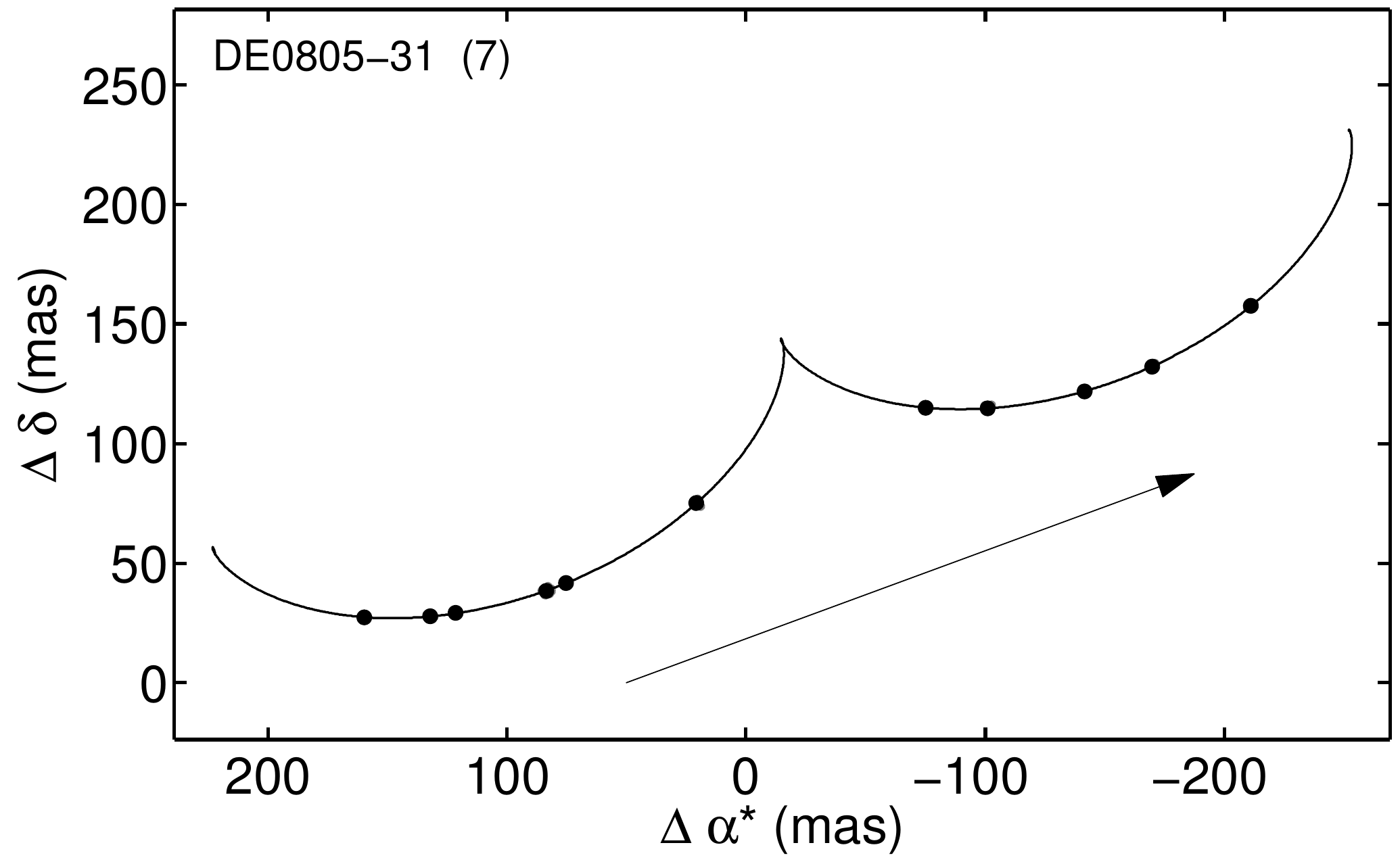}
\includegraphics[width= 0.4\linewidth]{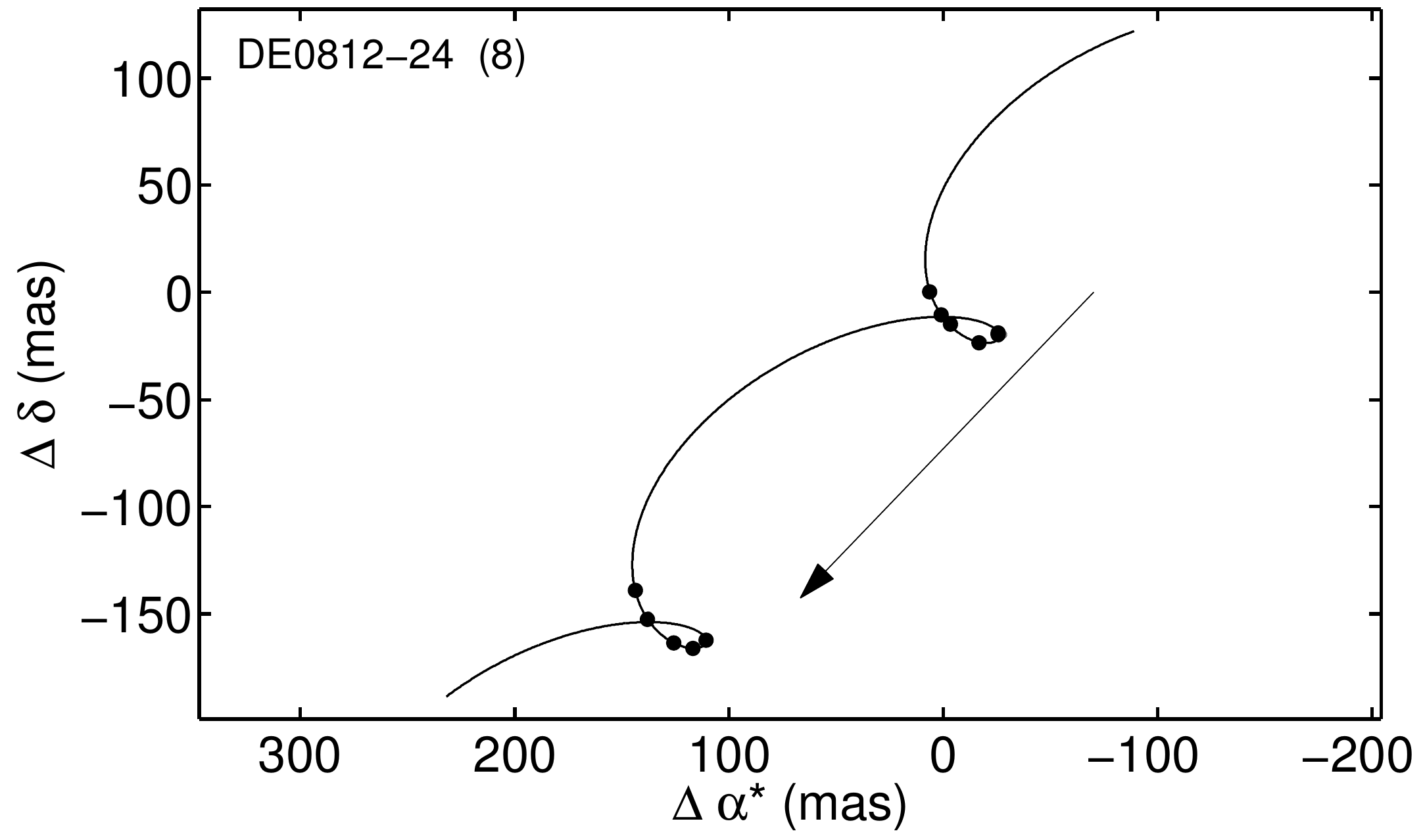}
\includegraphics[width= 0.4\linewidth]{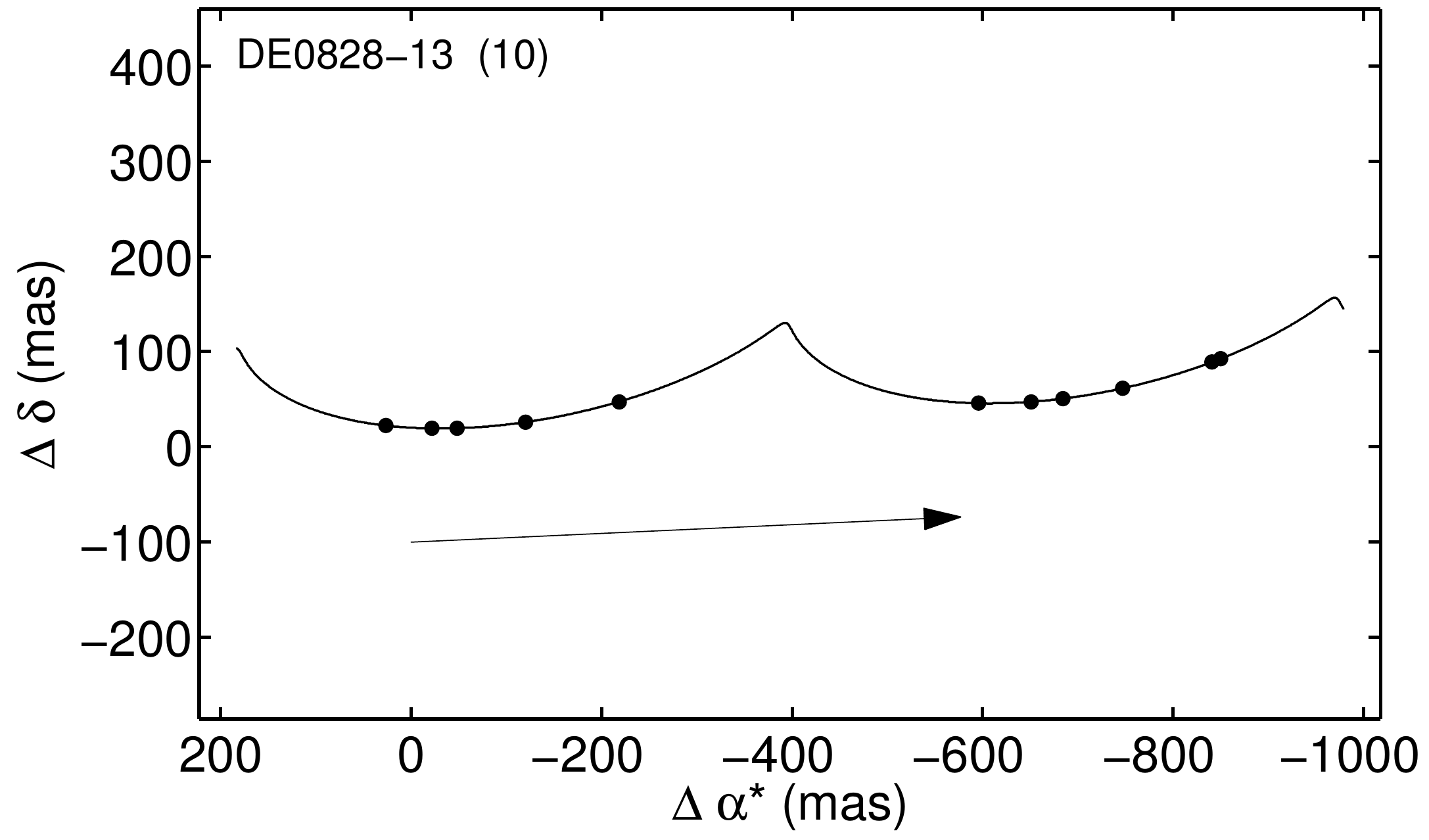}
\caption{Astrometric motions of targets number 1, 3--8, and 10 in the sky showing the data used in this paper. The target ID is indicated in the top-left corner of every panel and the target number is shown between parentheses. The displays are equivalent to Fig. \ref{fig:dw01_2D}. The solid curve indicates the best-fit model of proper motion and parallax. North is up, east is left.}
\label{fig:dwarfs1_2D}
\end{figure*}

\begin{figure*}[h!]
\center
\includegraphics[width= 0.4\linewidth]{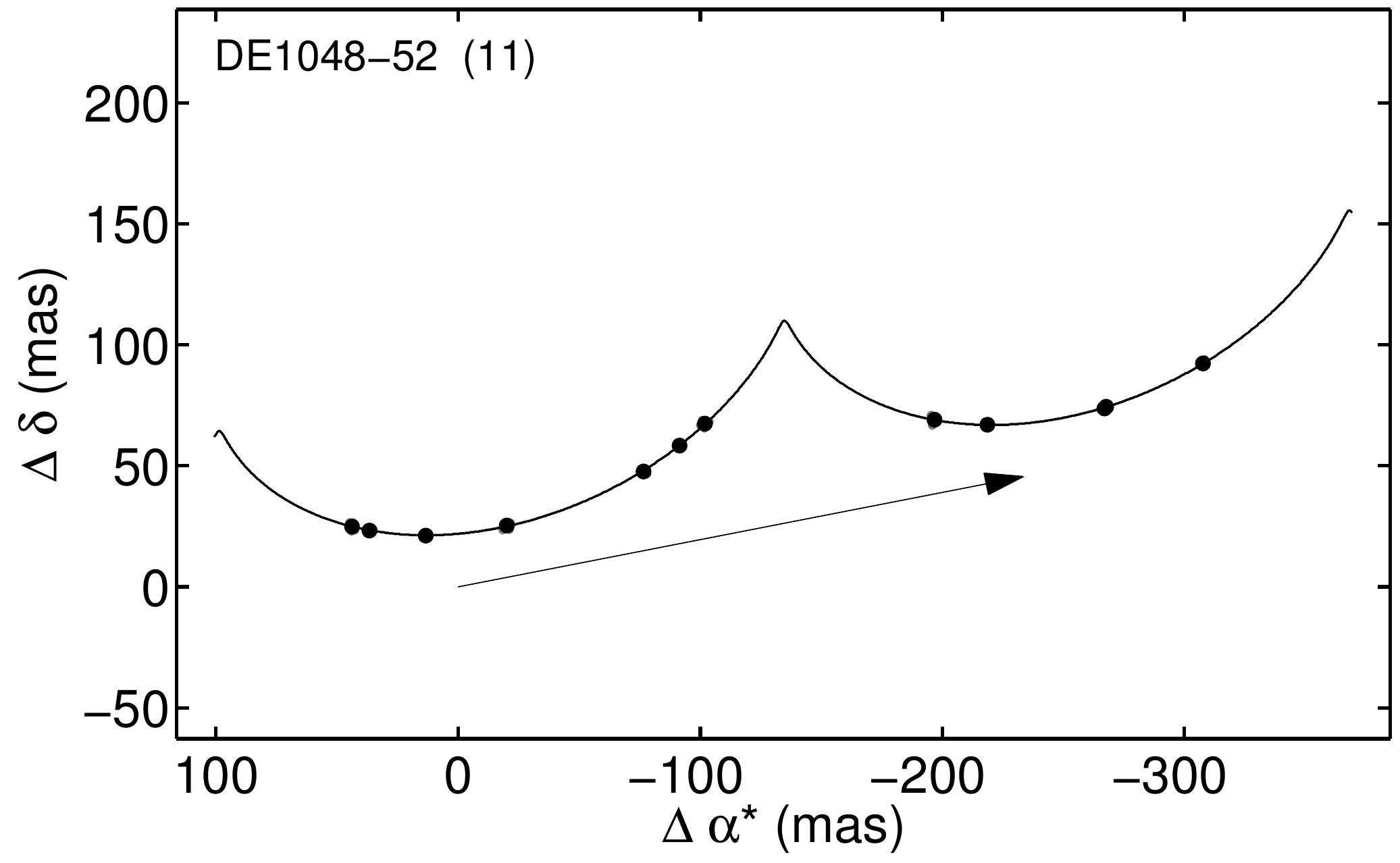}
\includegraphics[width= 0.4\linewidth]{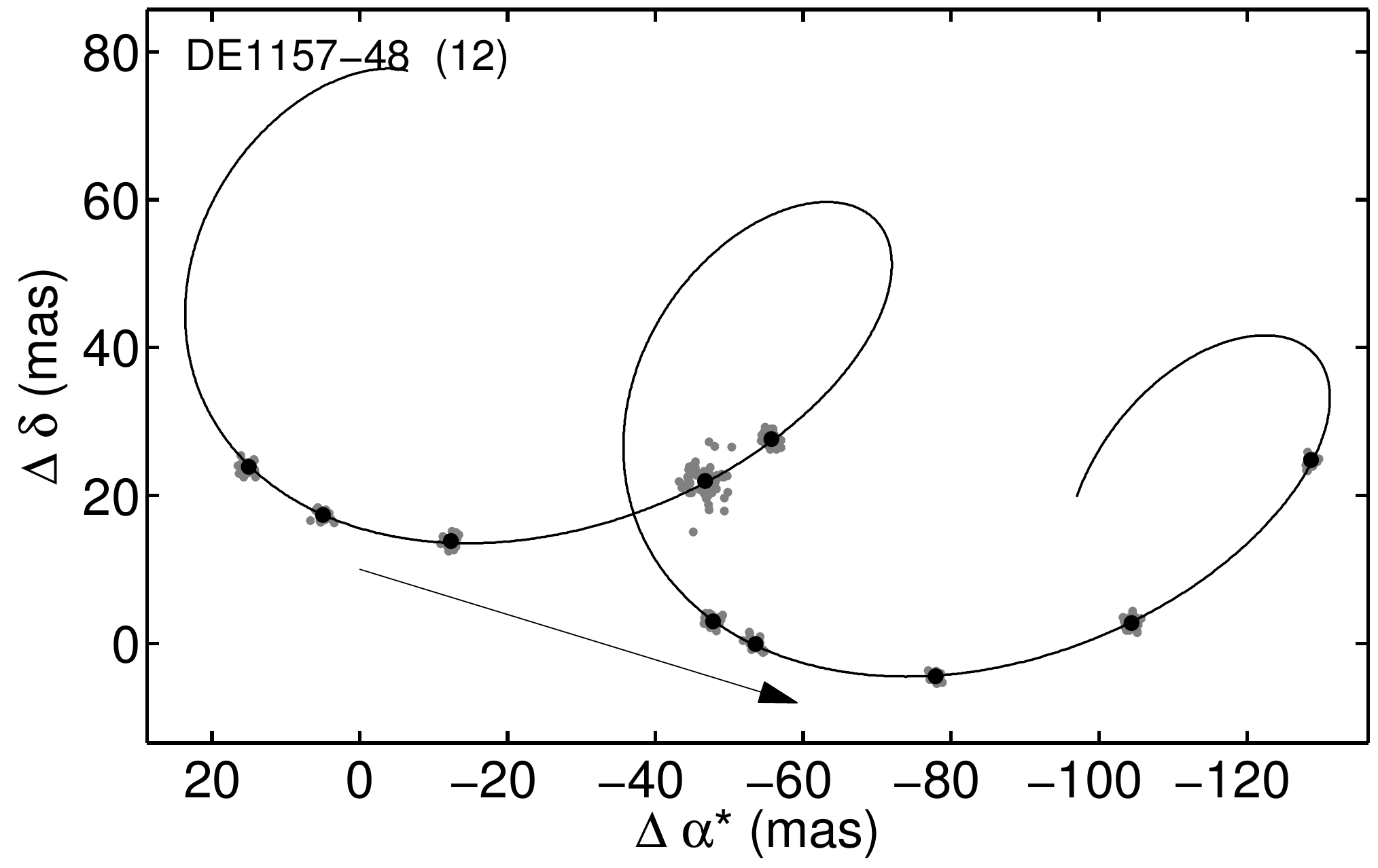}
\includegraphics[width= 0.4\linewidth]{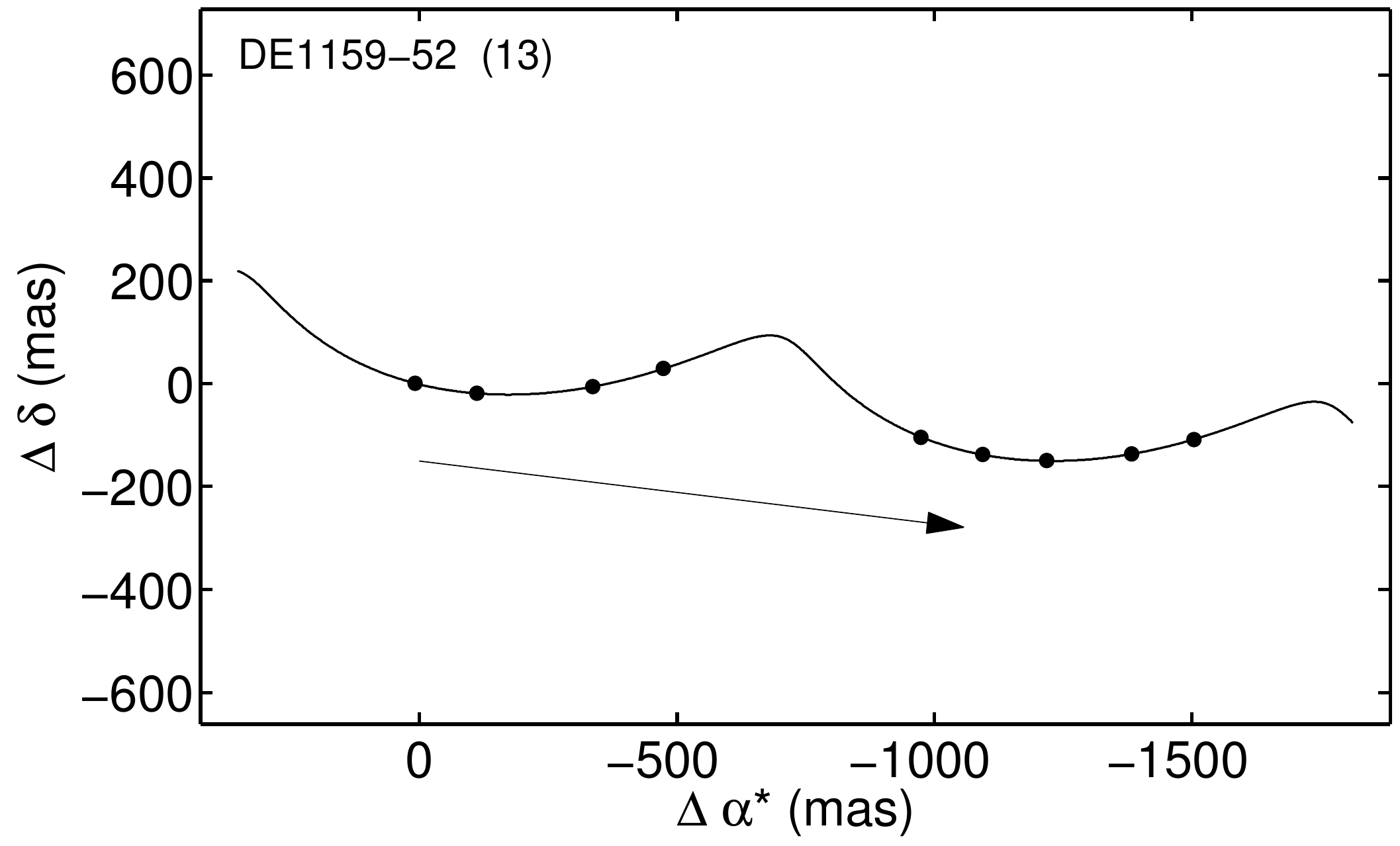}
\includegraphics[width= 0.4\linewidth]{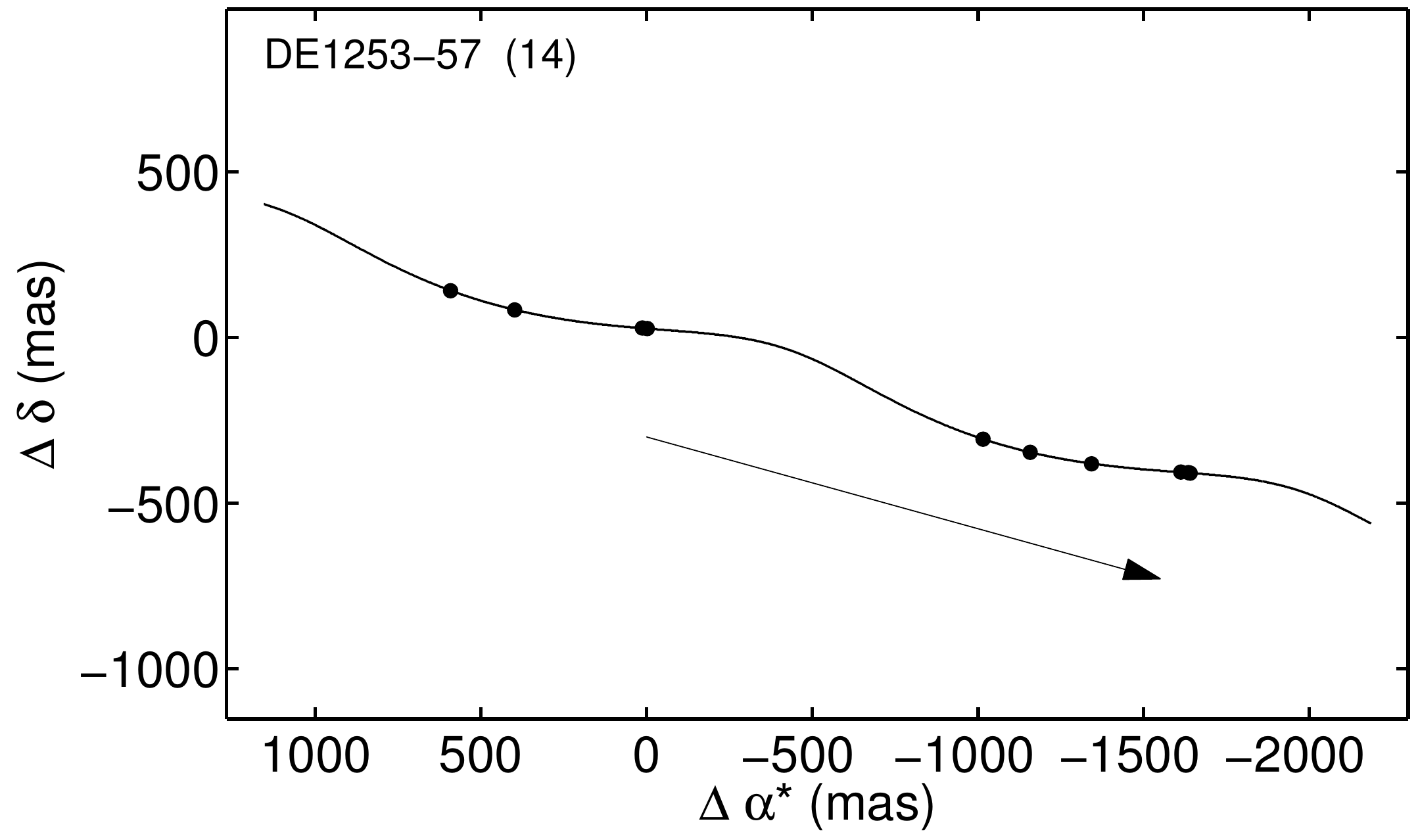}
\includegraphics[width= 0.4\linewidth]{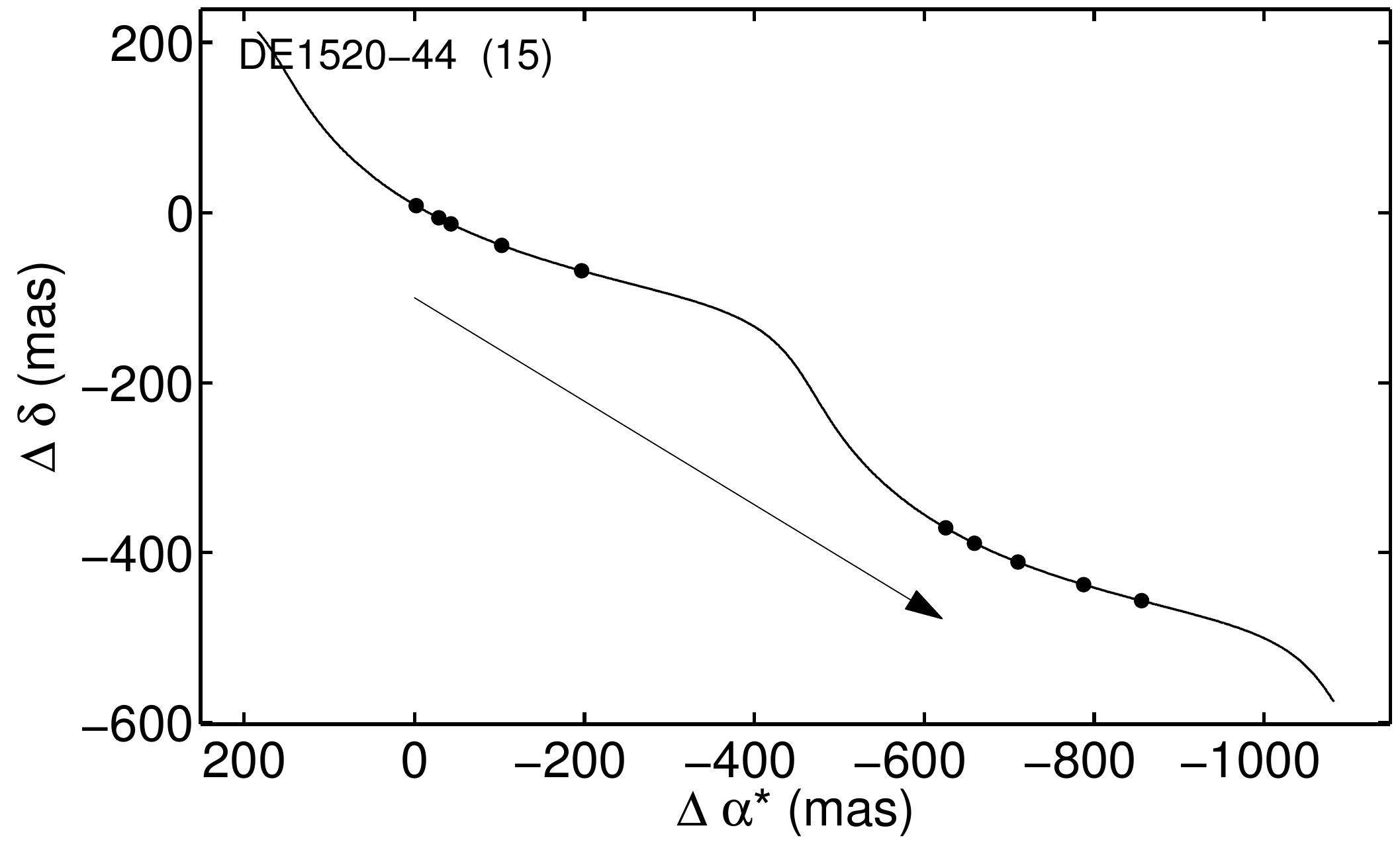}
\includegraphics[width= 0.4\linewidth]{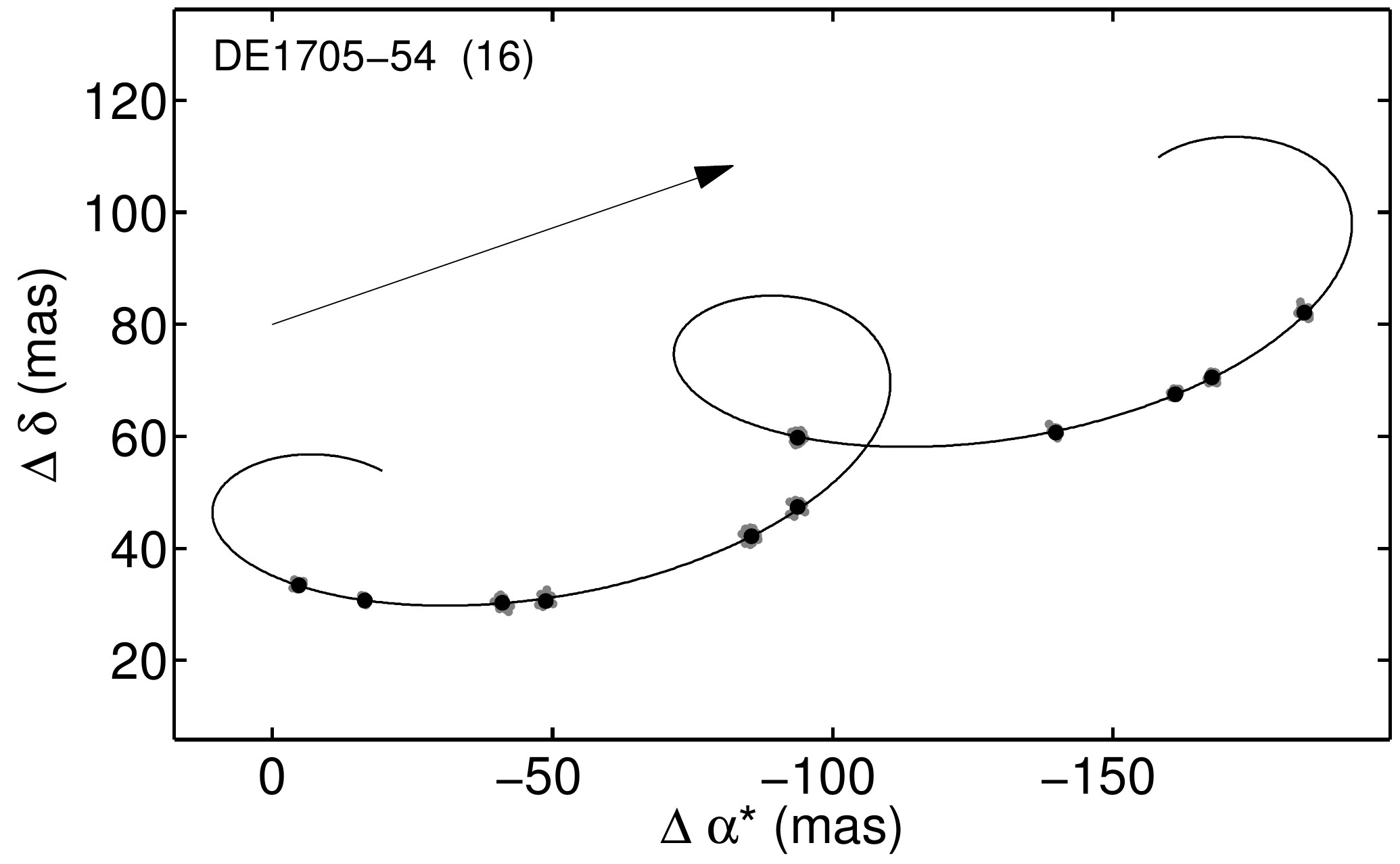}
\includegraphics[width= 0.4\linewidth]{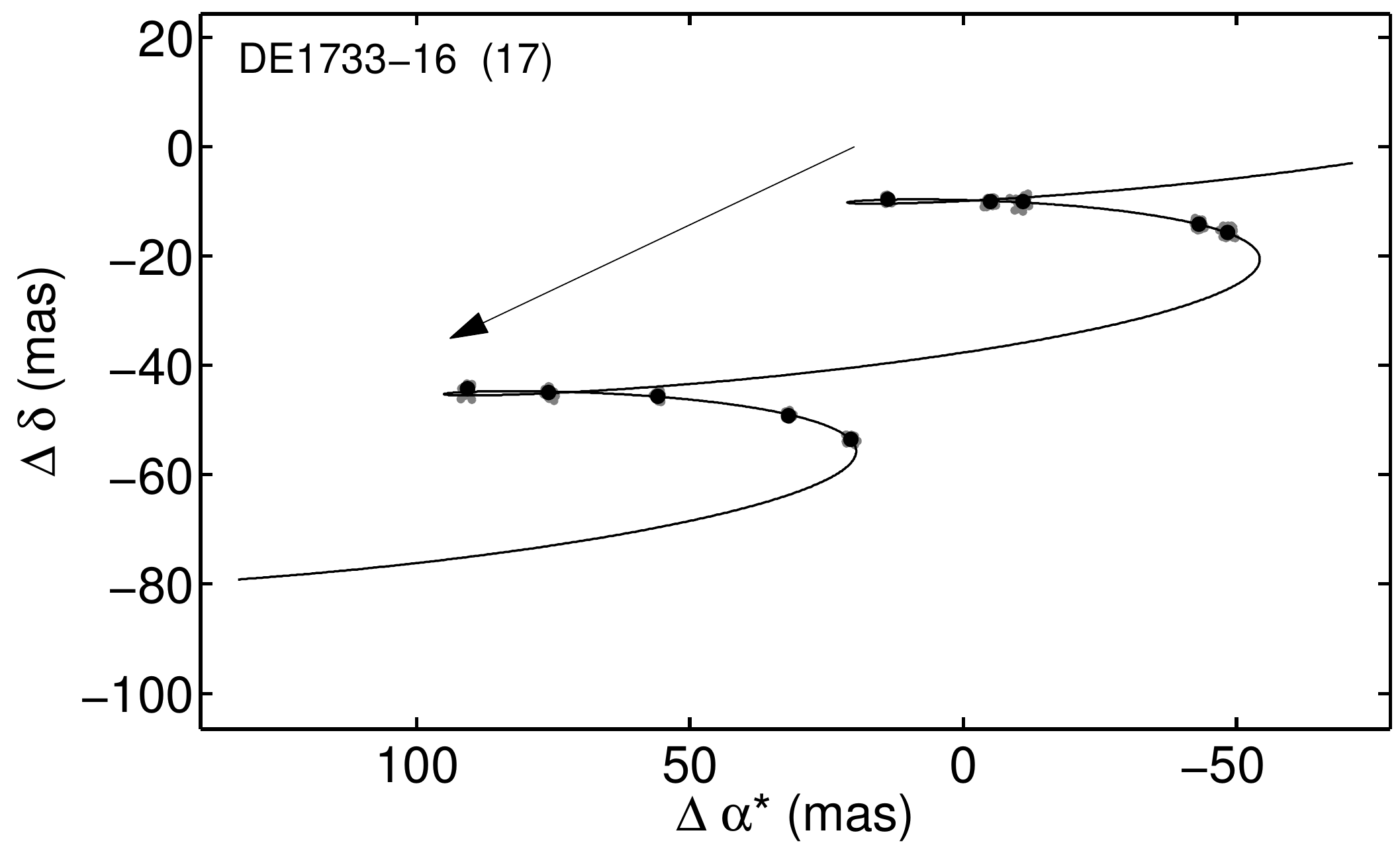}
\includegraphics[width= 0.4\linewidth]{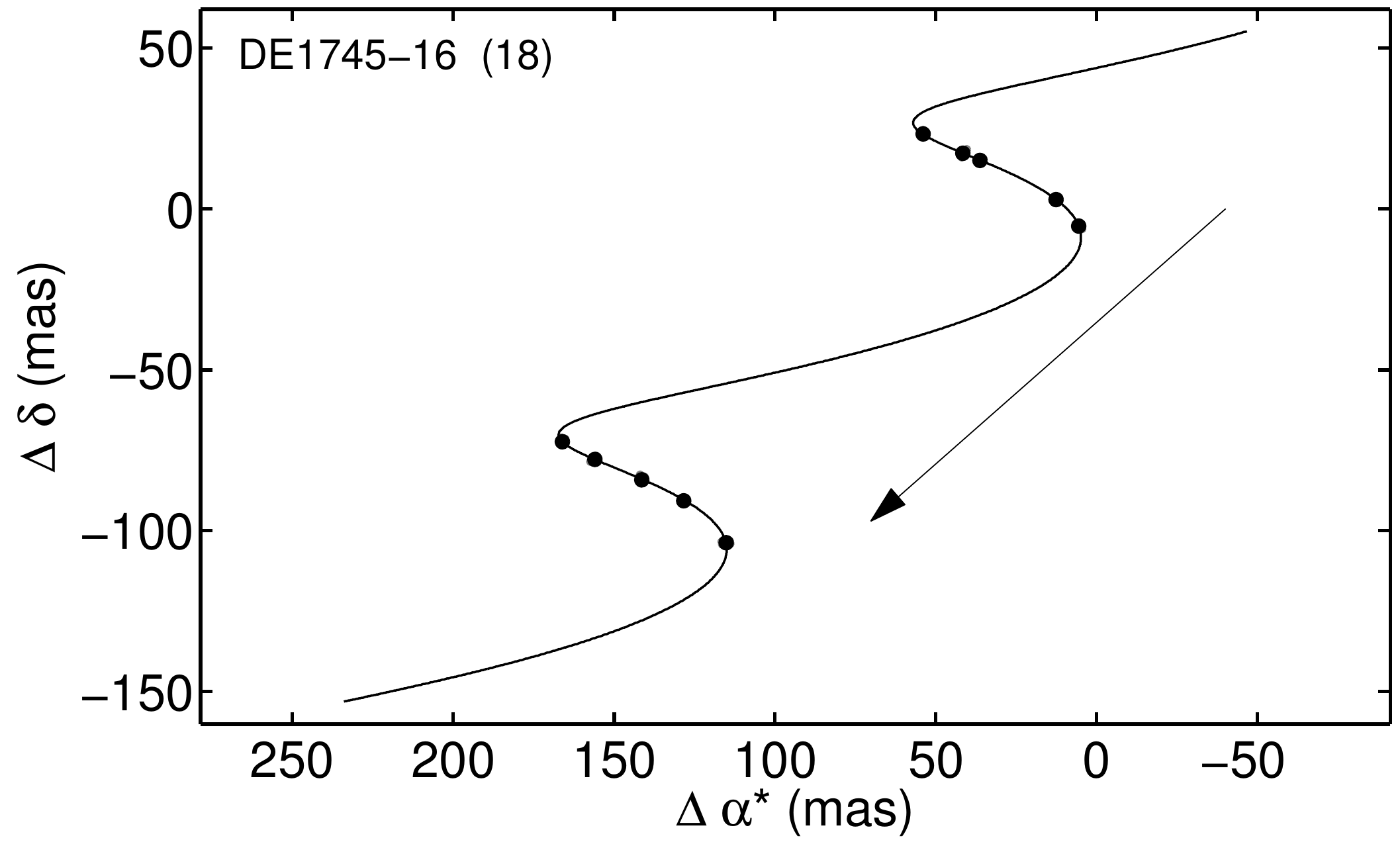}
\includegraphics[width= 0.4\linewidth]{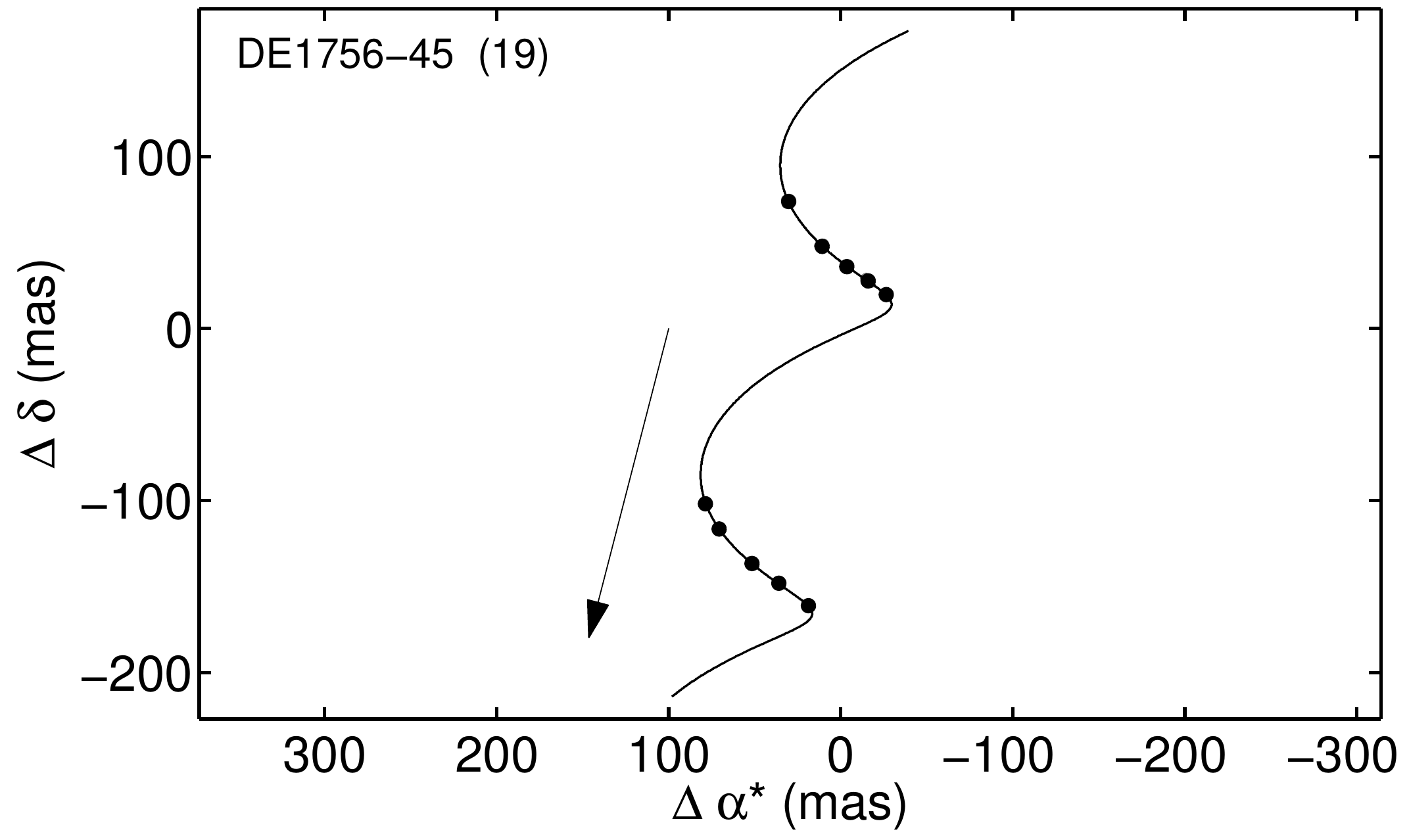}
\includegraphics[width= 0.4\linewidth]{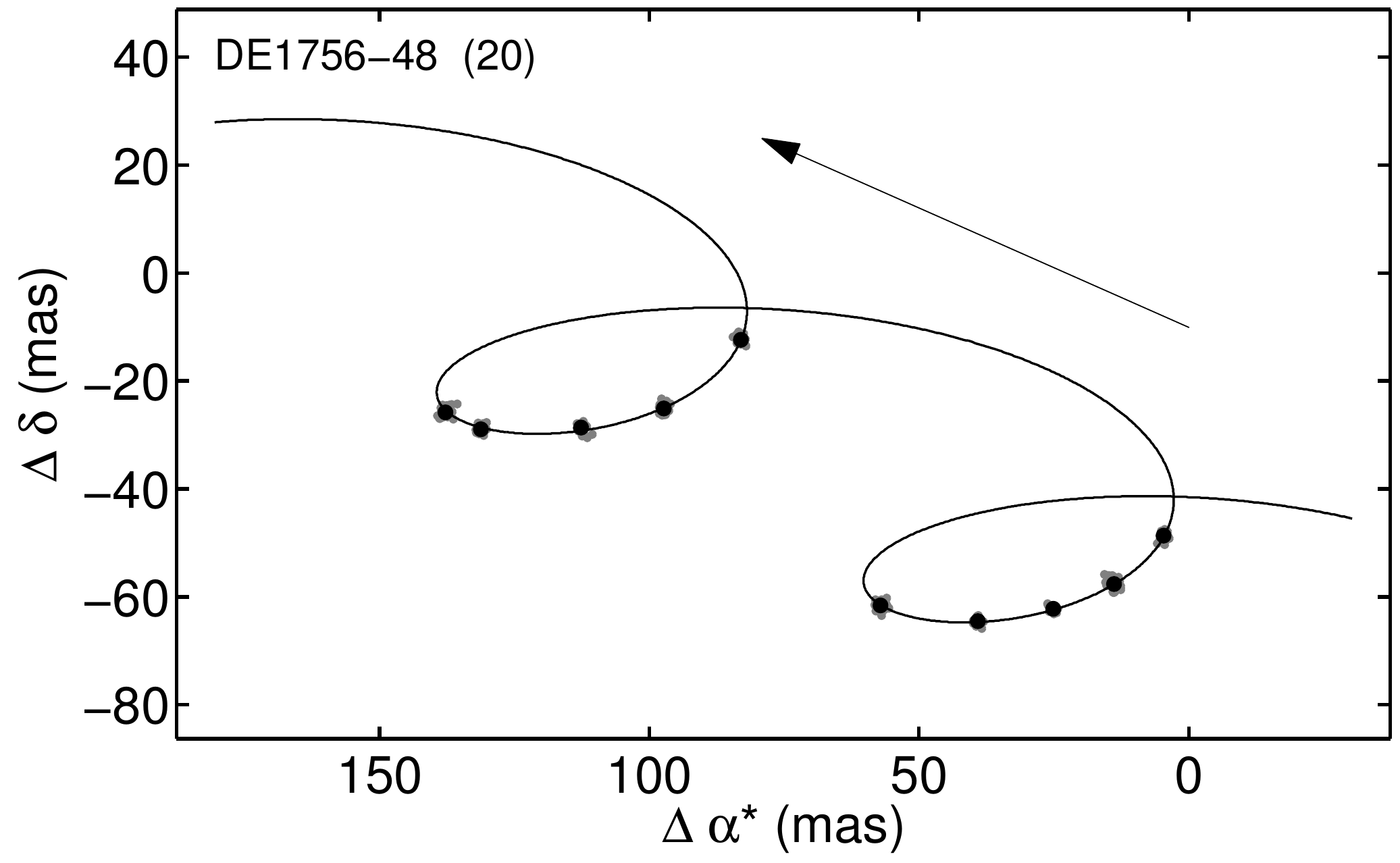}
\caption{Astrometric motions of targets number 11--20 in the sky. Display equivalent to Fig. \ref{fig:dwarfs1_2D}.}
\label{fig:dwarfs2_2D_2}
\end{figure*}

\begin{figure*}[h!]
\center
\includegraphics[width= 0.4\linewidth]{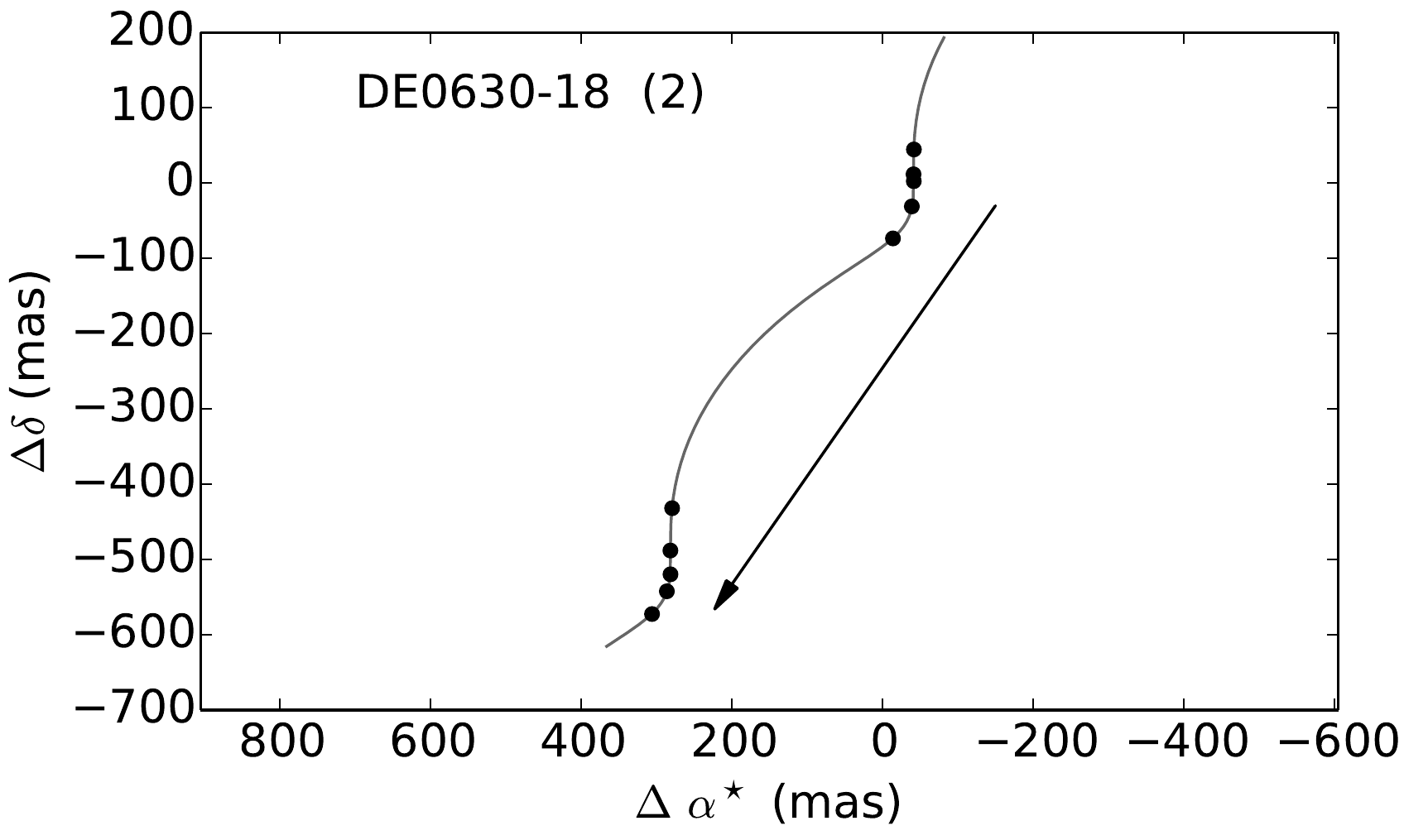}
\includegraphics[width= 0.4\linewidth]{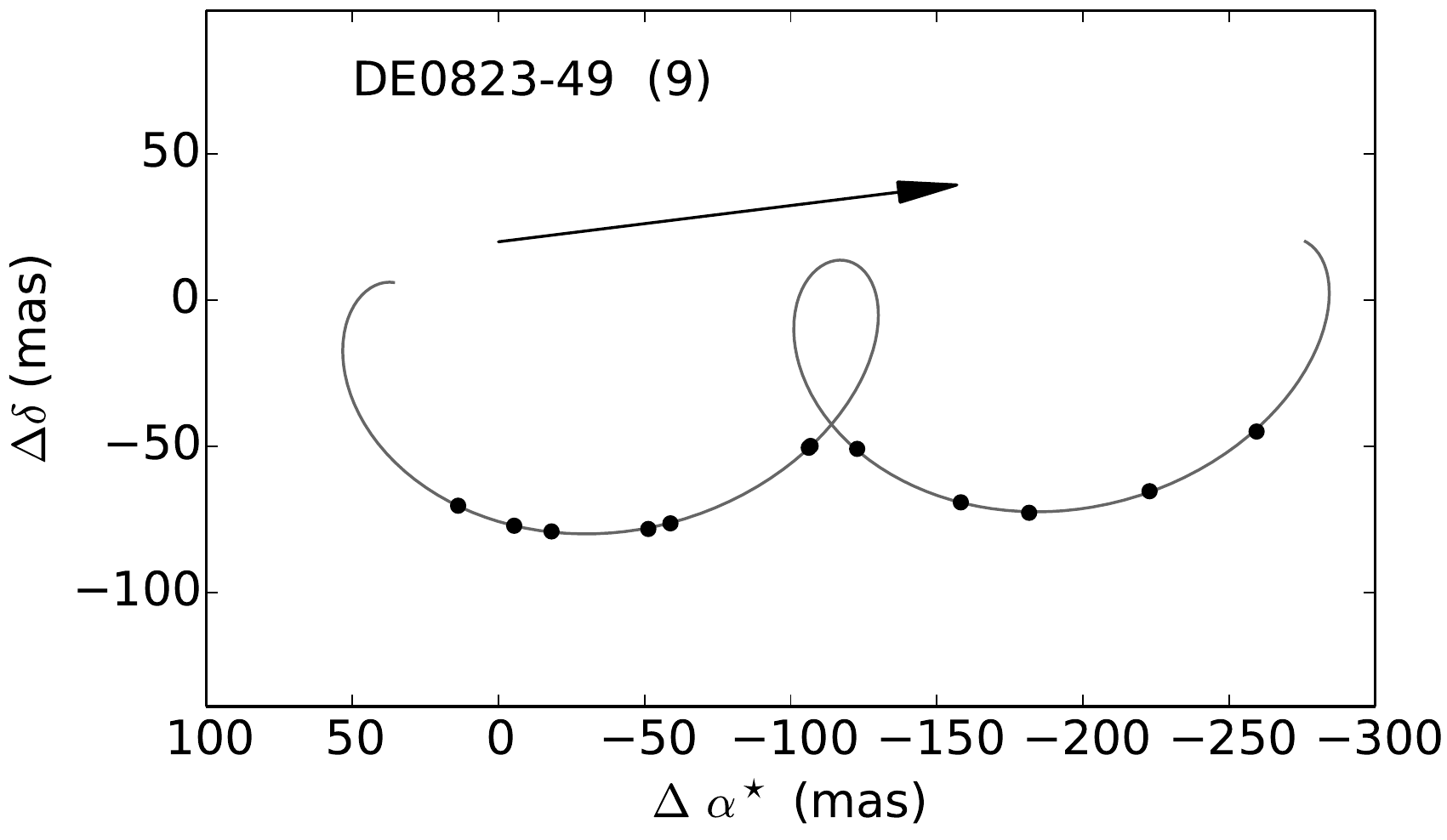}
\caption{Parallactic and proper motion of the two tight binaries \dwtwo\ (\emph{left}, \citealt{Sahlmann:2014}) and \dwnine\
 (\emph{right}, \citetalias{Sahlmann:2013kk}) during the first two survey years. The orbital motion was subtracted and the displays are equivalent to Fig. \ref{fig:dw01_2D}.}
\label{fig:dwarfs1_2D_3}
\end{figure*}

Figure \ref{fig:dw01_2D} shows the sky-projected motion of \dwfour\ over the course of our measurements and the equivalent graphs for all targets are displayed in Figs. \ref{fig:dwarfs1_2D}--\ref{fig:dwarfs1_2D_3}. 

\begin{table} 
\caption{Parameter correlations for \dwfour}
\label{tab:dw4corr} 
\centering 
\begin{tabular}{l |r@{\;\;} r@{\;\;} r@{\;\;} r@{\;\;} r@{\;\;} r@{\;\;} r@{\;\;}} 
\hline\hline 
& $\Delta \alpha^\star_0$ &$\Delta \delta_0$ &$\varpi$ &$\mu_{\alpha^\star}$ &$\mu_{\delta}$ &$\rho$ &$d$ \\ \hline
$\Delta \alpha^\star_0$ &$+1.00$ & \\ 
$\Delta \delta_0$ &$+0.49$ &$+1.00$ & \\ 
$\varpi$ &$+0.27$ &$+0.52$ &$+1.00$ & \\ 
$\mu_{\alpha^\star}$ &$-0.01$ &$+0.16$ &$+0.14$ &$+1.00$ & \\ 
$\mu_{\delta}$ &$+0.04$ &$+0.10$ &$-0.15$ &$+0.48$ &$+1.00$ & \\ 
$\rho$ &$+0.46$ &$+0.05$ &$+0.12$ &$-0.40$ &$-0.02$ &$+1.00$ & \\ 
$d$ &$-0.36$ &$-0.06$ &$-0.13$ &$+0.39$ &$+0.03$ &$-0.97$ &$+1.00$   \\ 
\hline 
\end{tabular} 
\end{table}


\subsection{Parallax correction and absolute parallaxes}\label{sec:parcor}
Because the astrometric reference stars are not located at infinity, a correction has to be applied to the relative parallax value of an individual target to convert it into an absolute parallax that determines the distance. As discussed in \citetalias{Sahlmann:2013kk}, this correction can be derived essentially in three ways: by using extragalactic references, by relying on photometric distance estimations of reference stars, or by using a Galaxy model. Because extragalactic sources identified in our fields usually are extended objects, the photocentre computation does nor reach the required $\sim$0.1 mas precision. Photometric distance estimates of reference stars rely on a detailed stellar classification with external data, which would have  exceeded the scope of our project. We therefore adopted the Galaxy-model method (e.g.\ \citealt{Dupuy:2012fk}) and followed the same procedure as described in \citetalias{Sahlmann:2013kk}: the Galaxy model of \cite{Robin:2003fk} was used to obtain a large sample of pseudo-stars in every target region. The comparison between the model parallaxes and the measured relative parallaxes of stars covering the same magnitude range yields an average offset, which is the parallax correction. 

\begin{table} 
\caption{Absolute parallaxes.}       
\label{tab:galcorr}      
\centering         
\begin{tabular}{r@{\;\;} c@{\;\;} r@{\;\;} r@{\;\;} r@{\;\;} r}    
\hline\hline      
Nr & ID & $\Delta \varpi_{galax}$ & $\sigma_{galax}$ & $N_{stars}$ &$ \varpi_{abs}$\\
    & & (mas)                             & (mas)        &                 &(mas) \\
\hline    
1 & \dwone & $-0.445$ & $0.877$& 194 & $45.700 \pm 0.112$ \\
2 & \dwtwo & $-0.428$ & $0.493$& 141 & $51.719 \pm 0.099$\tablefootmark{a} \\
3 & \dwthree & $-0.332$ & $0.714$& 135 & $25.094 \pm 0.094$ \\
4 & \dwfour & $-0.526$ & $0.390$& 106 & $62.023 \pm 0.070$ \\
5 & \dwfive & $-0.389$ & $1.561$& 373 & $40.918 \pm 0.144$ \\
6 & \dwsix & $-0.327$ & $0.429$& 342 & $56.304 \pm 0.085$ \\
7 & \dwseven & $-0.336$ & $0.625$& 376 & $42.428 \pm 0.083$ \\
8 & \dweight & $-0.323$ & $0.919$& 364 & $47.282 \pm 0.094$ \\
9\tablefootmark{b} & \dwnine & $-0.062$ & $0.643$& 283 & $48.16 \pm 0.19$ \\
10 & \dwten & $-0.578$ & $0.855$& 123 & $85.838 \pm 0.148$ \\
11 & \dweleven & $-0.275$ & $0.674$& 565 & $36.212 \pm 0.077$ \\
12 & \dwtwelve & $-0.245$ & $0.679$& 323 & $34.633 \pm 0.082$ \\
13 & \dwthirt & $-0.332$ & $0.495$& 237 & $105.538 \pm 0.120$ \\
14 & \dwfourt & $-0.192$ & $0.425$& 478 & $60.064 \pm 0.054$ \\
15 & \dwfift & $-0.159$ & $0.660$& 414 & $53.995 \pm 0.109$ \\
16 & \dwsixt & $-0.038$ & $1.188$& 1184 & $37.549 \pm 0.087$ \\
17 & \dwsevent & $-0.164$ & $0.791$& 1530 & $55.272 \pm 0.073$ \\
18 & \dweightt & $-0.030$ & $0.833$& 1511 & $50.871 \pm 0.096$ \\
19 & \dwninet & $-0.194$ & $0.411$& 631 & $43.577 \pm 0.064$ \\
20 & \dwtwenty & $-0.057$ & $0.560$& 783 & $47.039 \pm 0.058$ \\
\hline
\end{tabular}
\tablefoot{
\tablefoottext{a}{Preliminary value to be updated by \cite{Sahlmann:2014}.}
\tablefoottext{b}{Values from \citetalias{Sahlmann:2013kk}.}
}
\end{table}

Table \ref{tab:galcorr} shows the results. The parallax correction $\Delta \varpi_{galax}$ is the mean difference value between the model and the measured parallaxes of $N_{stars}$ reference stars in the {\small FORS2} images, and the r.m.s. value $\sigma_{galax}$ of these differences is tabulated as well. The absolute parallax $\varpi_{abs} = \varpi - \Delta \varpi_{galax}$ is larger than the relative parallax because the reference stars absorbed a small portion of the parallactic motion. The uncertainty of $\varpi_{abs}$ is computed by adding $\sigma_{galax}/\sqrt{N_{stars}}$ in quadrature to the relative parallax uncertainty.

On average, the parallax correction amounts to $-0.27$\,mas with an uncertainty of 0.040\,mas. The correction amplitude tends to be smaller towards the Galactic centre, which roughly corresponds to targets with large RA, because the density of faint, distant stars is higher in these regions and the relative parallax is therefore closer to the absolute parallax.

\subsection{Photometric variability}\label{sect:variability}
We examined photometric variability by measuring the targets' brightness variations relative to field stars. As part of the data reduction process, we estimated the flux of an object by computing the sum of counts $N_{\rm ADU}$ within the central $11$$\times$$11$ pixel area around the object's photocentre. Using 20--100 field stars in each exposure $m$, we derived the target's brightness $\mu_{i,m}$  relative to field stars denoted by the index $i$. We normalised this value by the average over all exposures for the star $i$. The subsequent averaging over field stars yielded the target's differential magnitude $\mu_m$ for each exposure. Finally, we averaged $\mu_m$ within an epoch to obtain differential magnitudes $\mu_e$. 

\begin{figure}
\centering
\includegraphics[width=\linewidth]{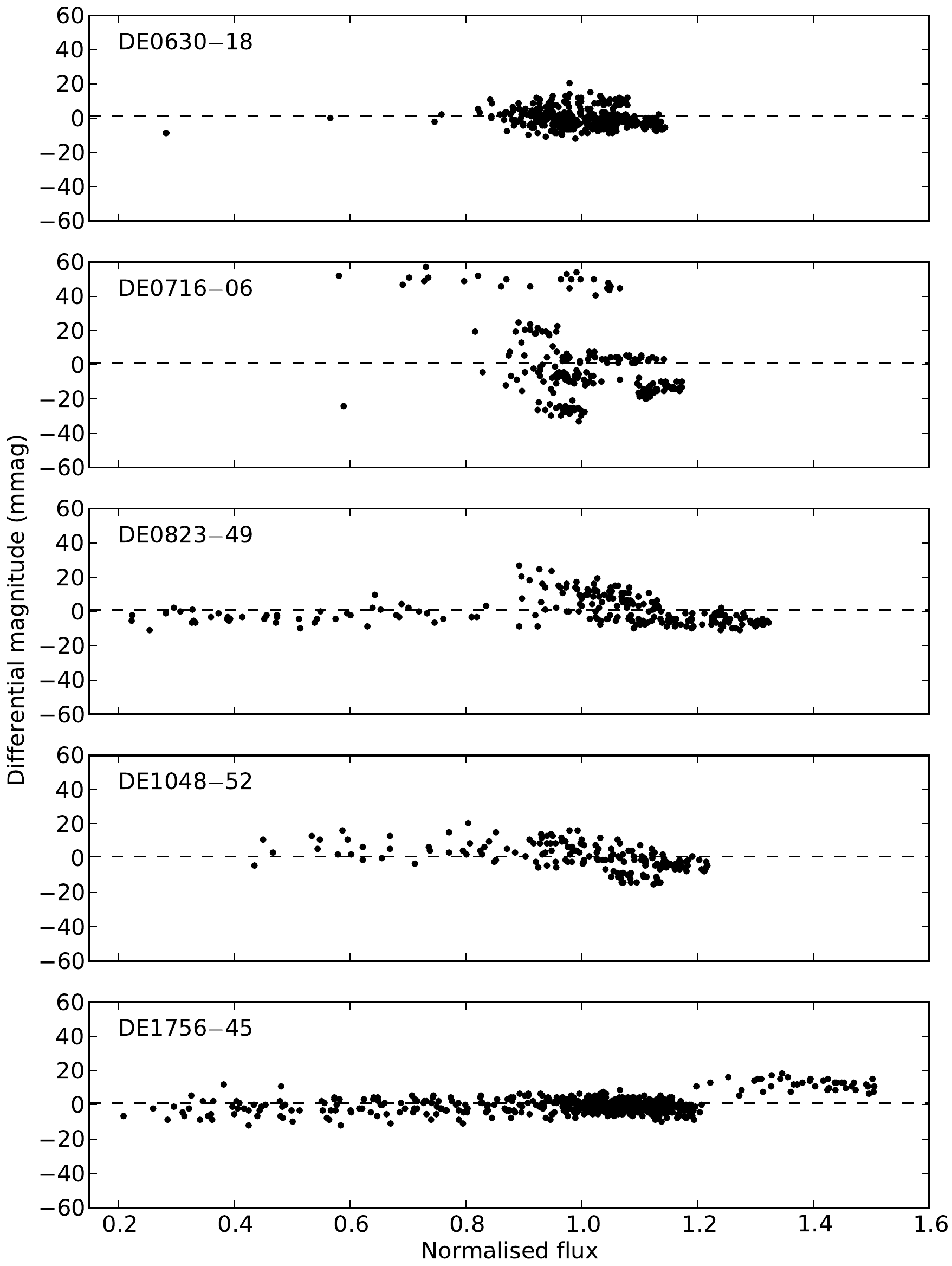}
\caption{Differential magnitude $\mu_m$ for every exposure $m$ as a function of the measured star counts normalised to unity for five targets, showing cases without noticeable variability (\dwtwo), with a broad range of extinction variation (\dwninet), and with clearly detected variability (\dwfive).}
\label{fig:ph_1_10}
\end{figure}

Because the observations were not necessarily obtained in photometric conditions and sometimes through thin cirrus clouds, we considered the effect of extinction on the differential photometry. 
The extinction variation within one epoch can be related to the flux $\bar N_m$, normalised by its mean value. The differential magnitude $\mu_m$ as a function of $\bar N_{m}$ for a few targets is shown in Fig. \ref{fig:ph_1_10}, where the vertical scatter indicates photometric variations and the horizontal spread is related to extinction. These two effects are largely uncorrelated, even in the rare cases of cloudy conditions with $\bar N_{m}$ values as low as 0.2. 

The typical accuracy of $\mu_e$ is $\sim$3 mmag, which corresponds to the r.m.s. of this quantity computed for bright reference stars. The r.m.s. of $\mu_e$ for the survey targets is shown in Table \ref{tab:pht} and typically amounts to 3--5 mmag. For \dweleven\ (spectral type L1.5), the detected r.m.s. of 6 mmag is slightly higher. Significant photometric variation is detected only for \dwfive\ with an r.m.s. of 20 mmag. In Fig. \ref{fig:ph_1_10}, the data of \dwfive\ are grouped in horizontal strips, each corresponding to a different epoch with a mean magnitude $\mu_e$. $I$-band photometric variability of UCDs with similar amplitude has previously been reported \citep{Martin:2001ly, Bailer-Jones:2001qy, Martin:2013aa, Gizis:2013aa} and was attributed to inhomogeneous cloud coverage, dark surface spots, or binarity. 

\begin{figure}
\centering
\includegraphics[width=\linewidth]{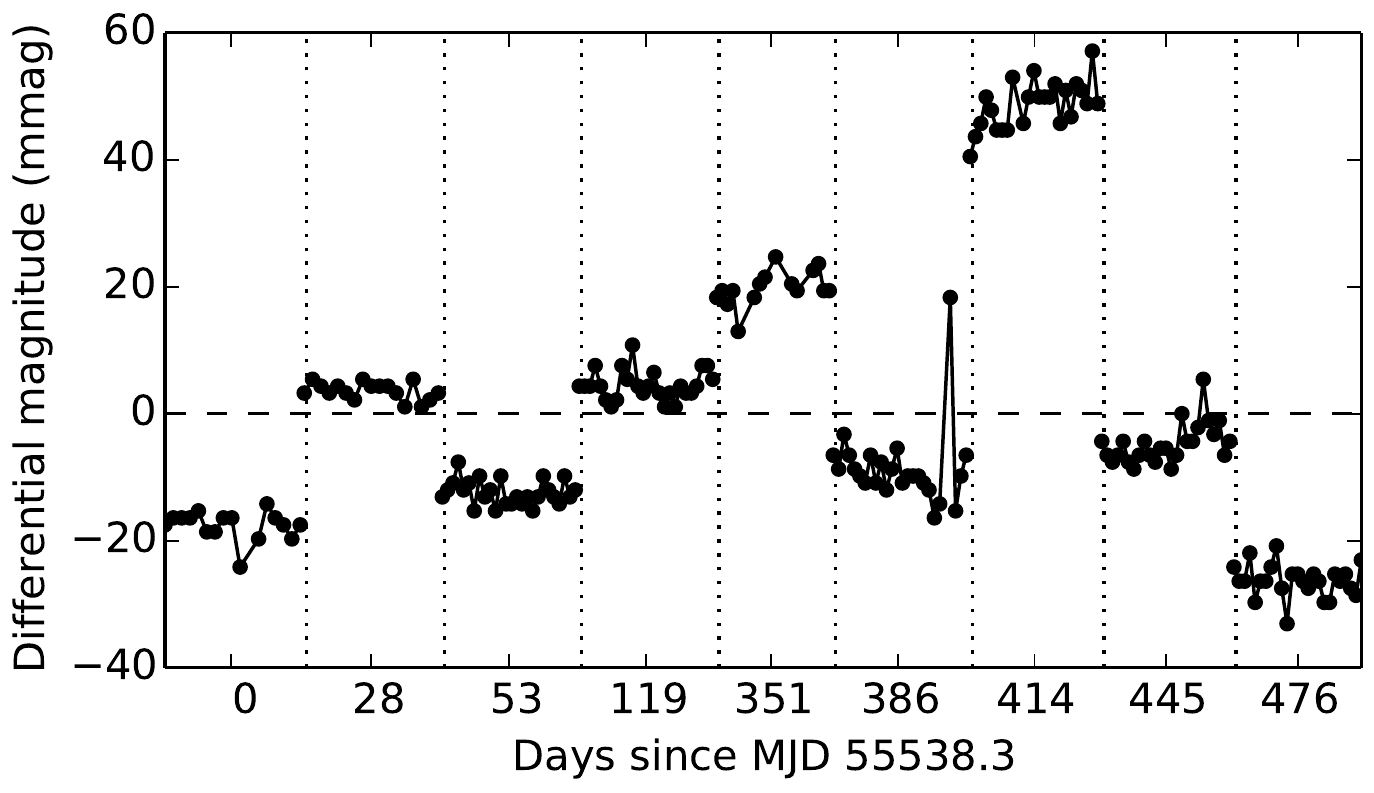}
\caption{Differential magnitude of \dwfive\ as a function of time. For better readability, the time axis is not continuous. Dotted vertical lines separate the different epochs, each consisting of a $\sim$30 min time-series.}
\label{fig:dw5photTime}
\end{figure}

Fourteen out of our twenty targets were searched for optical variability on time-scales of hours by \cite{Koen:2013uq}, who found four common objects to be variable. In contrast to \cite{Koen:2013uq}, we do not observe significant $I$-band variability for \dwsix, \dwthirt, and \dwten. We confirm the variability of \dwfive, but on much longer time-scales. We searched for periodic variations in the light curve (Fig. \ref{fig:dw5photTime}) by computing periodograms of the complete dataset and of individual epochs. In both cases, no significant periodicity was detected.

The photometric variability of \dwfive\ has no noticeable effect on the astrometric measurement or its precision and the astrometric data quality for this target has no distinguishing feature compared with the remaining sample. This is expected, because the astrometry relies on the photocentre determination, which is invariant for a centro-symmetric brightness change of the target. Assuming a radius of 0.1\,$R_\sun$ $\approx 1 R_J$ for \dwfive\ \citep{Demory:2009sf, Triaud:2013ab}, the apparent angular diameter is $\sim$38 $\mu$as, which is several time smaller than the measurement precision. Even if the brightness changes were asymmetric across the stellar disk, for example during the $\sim$30 mmag flare recorded on MJD 55924 (see Fig. \ref{fig:dw5photTime}), we did not expect to detect the corresponding photocentre shift. Future astrometric surveys with even higher precision may be able to detect these effects of activity in UCDs. 

\begin{table}
\caption{$I$-band photometric variability, measured as the r.m.s. of epoch-averaged differential magnitudes $\mu_e$. The observation time-span $\Delta T$ is the same as in Table \ref{tab:paltaobs} with an average value of 477 days.}
\begin{tabular}{rcc|rcc}
\hline
\hline
Nr  & ID  &  $\sigma_{\mu_e}$  &  Nr   &ID   & $\sigma_{\mu_e}$  \\ 
      &       &  (mmag) & & & (mmag)\\
\hline                                                                
1    &  \dwone   &    3 &       11   &  \dweleven     &    6   \\
2    &  \dwtwo     &    5 &  12   &  \dwtwelve     &    4   \\
3     &  \dwthree    &    3 &  13   &  \dwthirt     &    4   \\
4    &  \dwfour     &    3 &   14   &  \dwfourt     &    5   \\
5    &  \dwfive     &   20 &   15 &  \dwfift       &    3   \\
 6    &  \dwsix     &    4 & 16  &  \dwsixt      &    4\\
 7    &  \dwseven     &    5&17   &  \dwsevent     &    4\\
 8    &  \dweight     &    4&18   &  \dweightt     &    5\\
 9     &  \dwnine    &    5&19   &  \dwninet     &    4\\
 10   &  \dwten     &    5&20   &  \dwtwenty     &    4\\
\hline 
\end{tabular}
\label{tab:pht}
\end{table}

\subsection{Primary-mass estimates}
To infer the masses of potential companions, we first have to estimate the masses of the primaries. Accurate mass determinations of UCDs are usually reserved for visual or eclipsing binary systems (e.g. \citealt{Close:2005fk, Zapatero-Osorio:2004fu,Stassun:2006pt}), so that for objects in the field we have to use evolutionary models that describe the relationships between ages, luminosities, and masses of UCDs.

We inspected the low-resolution optical spectra presented in \cite{Phan-Bao:2008fr} for signs of youth, in particular \ion{Li}-absorption features. We also initiated a spectroscopic characterisation campaign, whose results will be reported in a forthcoming paper. With the exception of \dwnine\ \citepalias{Sahlmann:2013kk}, all targets appear to be strongly lithium-depleted and we assumed an possible age range of 1--10 Gyr.  

\begin{figure}
\begin{center}
\includegraphics[width=\linewidth]{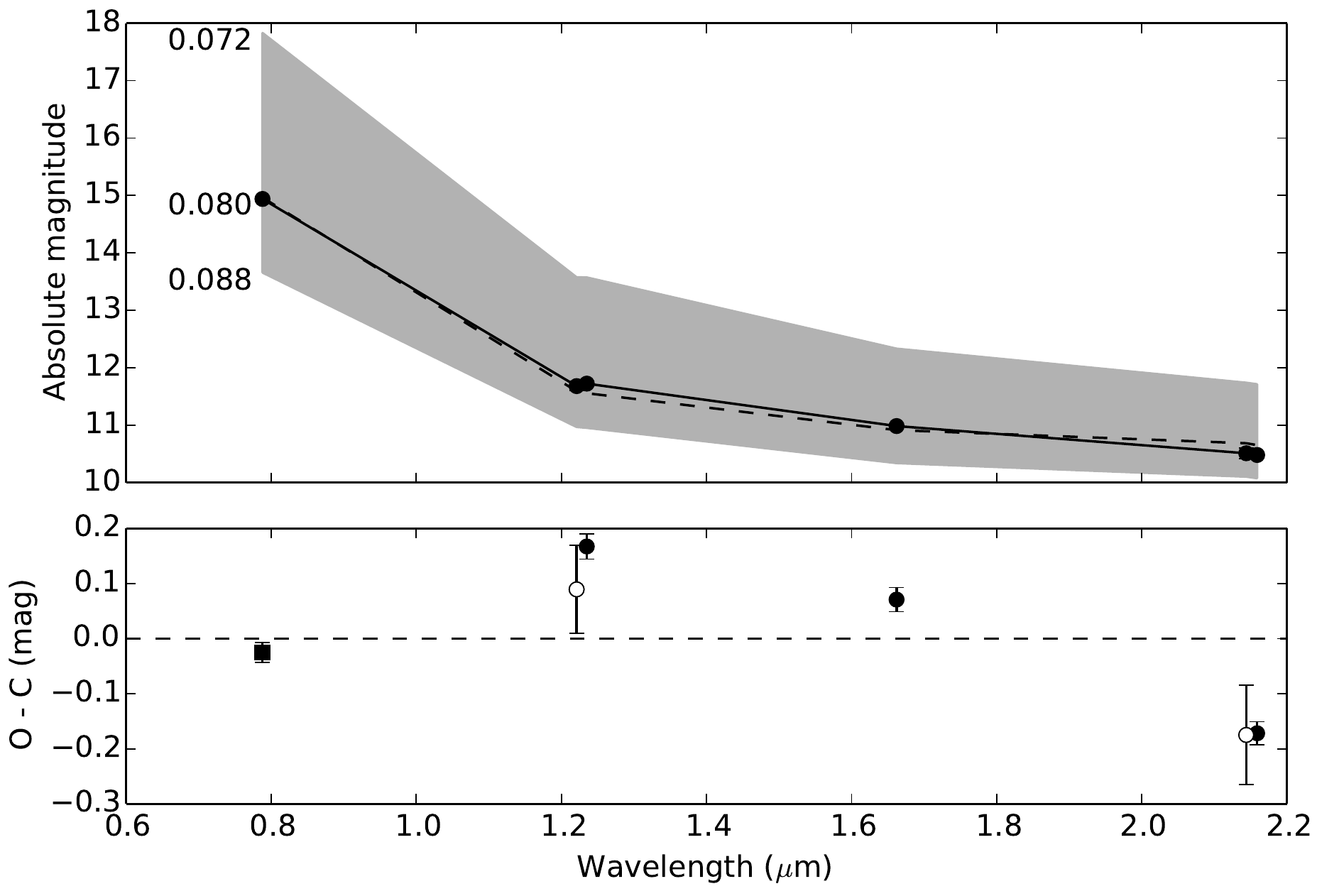}
\caption{Absolute magnitudes of \dwfour\ as a function of wavelength {assuming an age of 3 Gyr}. Six photometric measurements are used for primary-mass estimation. \emph{Top}: magnitudes (black circles), the best-fit model (dashed line), and the magnitude range spanned by the best-fit model with a 10 \% mass uncertainty (shaded area) are shown. {The corresponding masses in $M_\sun$ are indicated to the left of the curves}. \emph{Bottom}: residuals of the best fit. Filled and open circles mark data from 2MASS and \cite{Phan-Bao:2008fr}, respectively, and the square corresponds to the FORS2 measurements.}
\label{fig:massestimation1}
\end{center}
\end{figure}

We retrieved apparent $J$,$H$,$K$-magnitudes from 2MASS \citep{Cutri:2003nx}, $H$- and $K$-magnitudes from \cite{Phan-Bao:2008fr}, and the $I$-band magnitudes from our {\small FORS2} observations (see \citetalias{Lazorenko:2013kk} for details) and converted them into absolute magnitudes using the parallax determinations in Table \ref{tab:galcorr}. The updated $I$-band magnitudes are shown in Table \ref{tab:targsobj} and are usually compatible with the DENIS values. Like in \citetalias{Sahlmann:2013kk}, we used the BT-Settl models \citep{Chabrier:2000kx,Allard:2012uq} for a given age to find the UCD mass that yielded the best fit to the optical and infrared magnitudes as illustrated in Fig. \ref{fig:massestimation1}. Although we formally included magnitude and parallax uncertainties, the dominant uncertainty comes from the model itself, and we globally adopted a 10 \% uncertainty on the derived masses. Still, using a different set of evolutionary models may lead to best-fit masses that differ from those given here. Table \ref{tab:massestimates} lists the masses in the 1--10 Gyr range and at 3 Gyr with the 10 \% uncertainty. In the following, we use the 3 Gyr value, whose uncertainty usually encompasses the variation caused by the acceptable age range.

\begin{table}
\caption{Primary-mass estimates.}       
\label{tab:massestimates}      
\centering         
\begin{tabular}{r c c c}    
\hline\hline      
 Nr & ID & $M_1$ (1--10 Gyr) & $M_1$ (3 Gyr) \\
      &     &($M_\sun$)& ($M_\sun$) \\
\hline
1 & \dwone & $0.074-0.078$ &  $0.078\pm 0.008$ \\
2\tablefootmark{a} & \dwtwo & $0.084-0.086$ &  $0.086\pm 0.009$ \\
3 & \dwthree & $0.087-0.087$ &  $0.087\pm 0.009$ \\
4 & \dwfour & $0.077-0.080$ &  $0.080\pm 0.008$ \\
5 & \dwfive & $0.073-0.078$ &  $0.078\pm 0.008$ \\
6 & \dwsix & $0.074-0.079$ &  $0.078\pm 0.008$ \\
7 & \dwseven & $0.083-0.084$ &  $0.084\pm 0.008$ \\
8 & \dweight & $0.072-0.078$ &  $0.077\pm 0.008$ \\
10 & \dwten & $0.070-0.077$ &  $0.076\pm 0.008$ \\
11 & \dweleven & $0.075-0.079$ &  $0.079\pm 0.008$ \\
12 & \dwtwelve & $0.077-0.080$ &  $0.080\pm 0.008$ \\
13 & \dwthirt & $0.079-0.081$ &  $0.081\pm 0.008$ \\
14 & \dwfourt & $0.072-0.077$ &  $0.077\pm 0.008$ \\
15 & \dwfift & $0.074-0.079$ &  $0.078\pm 0.008$ \\
16 & \dwsixt & $0.081-0.083$ &  $0.083\pm 0.008$ \\
17 & \dwsevent & $0.073-0.078$ &  $0.077\pm 0.008$ \\
18 & \dweightt & $0.073-0.078$ &  $0.078\pm 0.008$ \\
19 & \dwninet & $0.087-0.088$ &  $0.088\pm 0.009$ \\
20 & \dwtwenty & $0.076-0.079$ &  $0.079\pm 0.008$ \\

\hline
9\tablefootmark{b} & \dwnine & $0.067-0.079$ &  $0.075\pm 0.007$ \\

\hline   \end{tabular}
\tablefoot{
\tablefoottext{a}{Preliminary values to be updated by \cite{Sahlmann:2014}.}
\tablefoottext{b}{Values from \citetalias{Sahlmann:2013kk} corresponding to an age range of 0.6--3 Gyr and an adopted age of 1 Gyr.}
}
\end{table}

\subsection{Planet detection and exclusion limits}
Even when no orbital motion is detected, the astrometry allows us to set constraints on the presence of companions by determining the range of companion parameters that are incompatible with the data. The computation of detection limits is common practice in radial-velocity surveys (e.g. \citealt{Murdoch:1993uq, Howard:2010lr, Mayor:2011fj}) and the principle applies equally to the astrometry case, but some variations are necessary. For instance, the use of periodograms is impracticable because of the relatively small number of epochs. We implemented a method on the basis of the observed residual r.m.s. amplitude similar to \cite{Lagrange:2009lr}. 

We started from the null hypothesis that the observed residuals are caused by intrinsic noise sources such as the measurement uncertainty and small systematic errors, and not by an orbital signal. For each target, we generated a grid of period and companion-mass (1\,$M_J$ resolution) values and simulated a set of 5000 single-companion systems for each grid point. These pseudo-orbits have randomly assigned parameters $\Omega$ and orbital phase ($T_0$), an inclination $i$ according to a $\sin{i}$ probability density function, and we assumed circular orbits, that is $e=0$ and $\omega = 0$. The latter does not lead to a loss of generality, which we verified by dedicated simulations with $e\neq0$, but to a considerable gain in computation time. The corresponding astrometric signal at the actual observation dates was computed and added to the best-fit astrometric motion of the original data set, that is the intrinsic noise of the original observation is not present in the simulated data. Then the linear fit with the standard astrometric model was performed for all simulations separately. To improve efficiency, we performed these simulations only for epoch-averaged observations.

A given system of period and companion mass is considered detected if 99.7 \% of the corresponding pseudo-orbits have a residual r.m.s. larger than the observed r.m.s. for the target. This procedure yields a companion-mass limit as a function of period, above which we can exclude the presence of companions at the 3-$\sigma$ level. We ran these simulations for every target excluding the two tight binaries.

\begin{figure}
\center
\includegraphics[width= \linewidth]{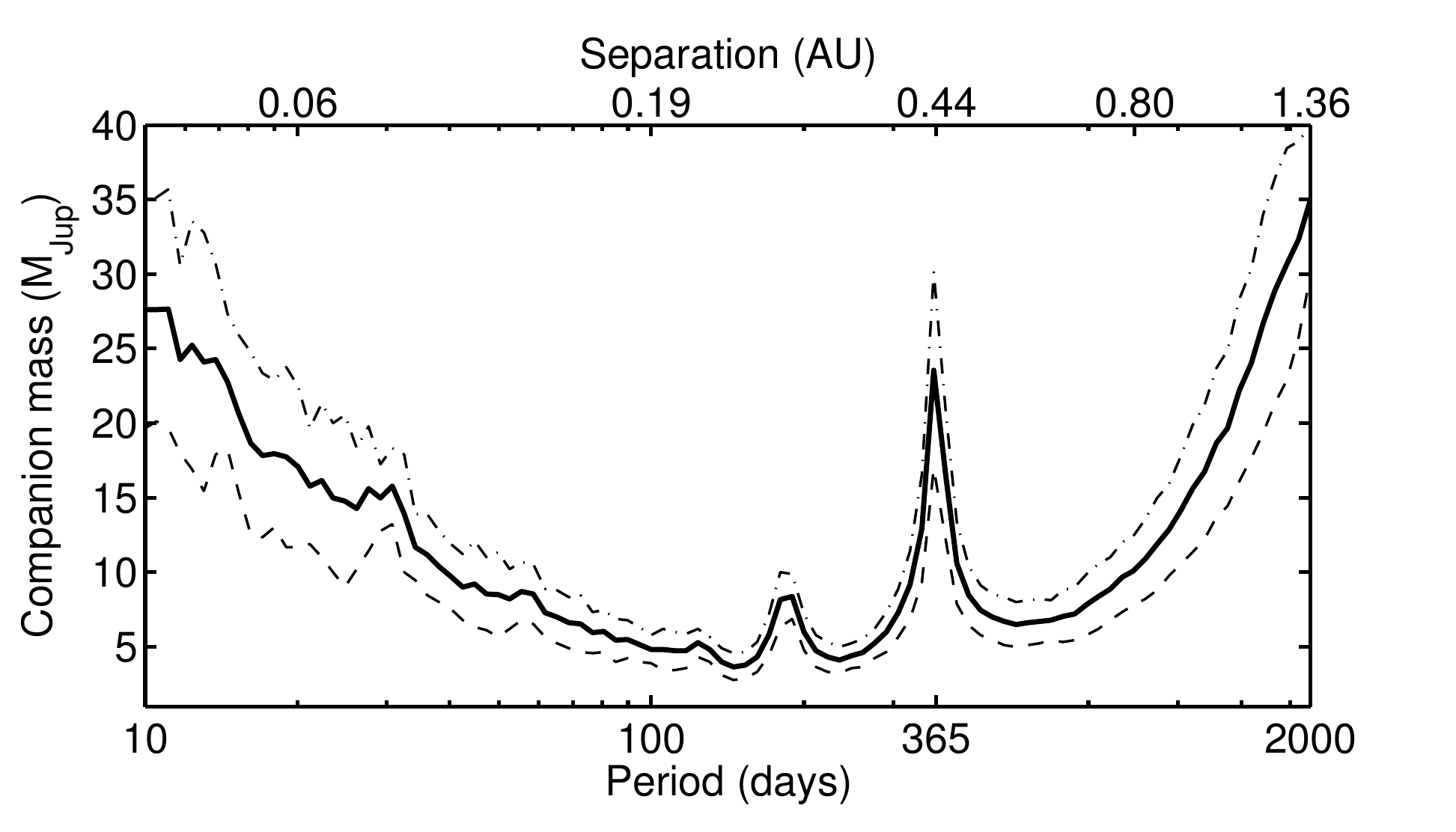}
\caption{Sample-averaged companion exclusion limits (solid line) as a function of orbital period (bottom label) and relative primary-secondary separation (top label, computed for a 5 $M_J$ companion around a 0.08 $M_\sun$ primary). The minimum companion mass incompatible with the measurements is shown, i.e.\ companions on and above the curve are excluded by the data. To illustrate the variation across the sample, we divided it into two groups of targets with lower (9 objets, dashed line) and higher (9 objects, dash-dotted line) exclusion limits. The two tight binaries were excluded.}
\label{fig:detLimits}
\end{figure}

In Fig. \ref{fig:detLimits}, the sample-averaged exclusion limits are shown. We used the best-fit primary masses from Table \ref{tab:massestimates} without accounting for their uncertainty and we excluded the two tight binaries for this analysis. Since all primary masses are approximately compatible with each other, we can globally evaluate the exclusion limits in terms of companions mass instead of mass ratio. At short periods, the sensitivity is reduced because the astrometric signal amplitude decreases. In the period range between 50 and 400 days, the sensitivity is approximately constant except for resonances at periods of 1 year and 1/2 year, which are caused by correlations with the parallax. At periods longer than the observation time-span, that is $\gtrsim480$ days, the sensitivity decreases although the signal becomes stronger.

\subsection{Visual binary \dwfift}
\begin{figure}
\centering
\hspace{8mm}\includegraphics[width=0.85\linewidth]{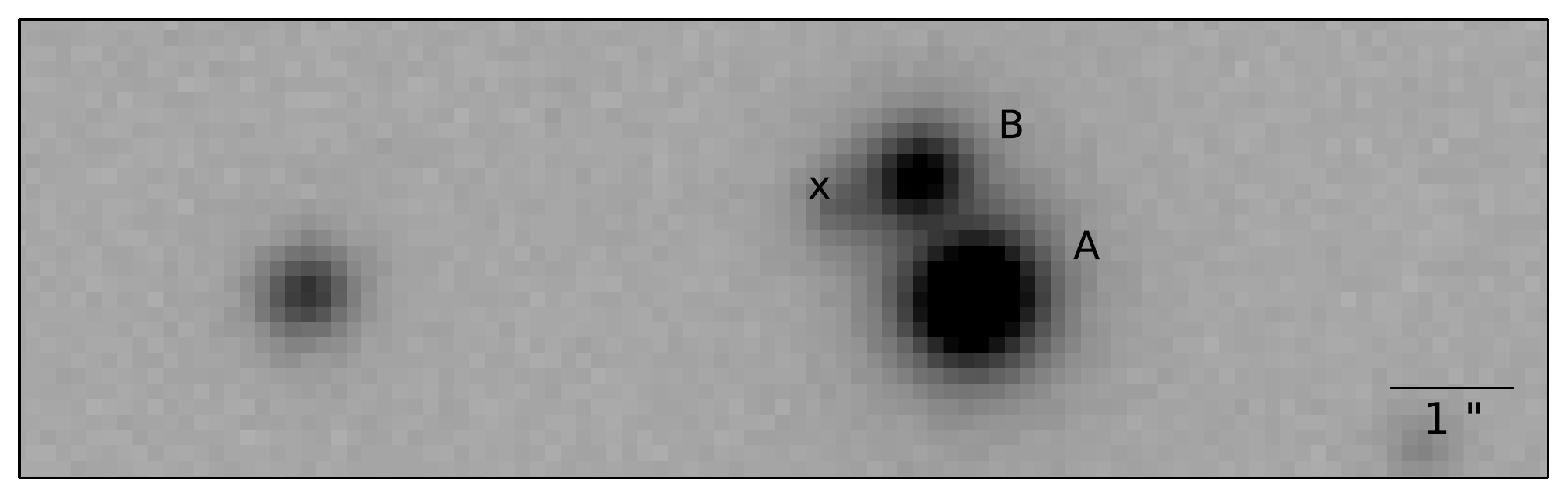}\vspace{1mm}
\includegraphics[width=\linewidth]{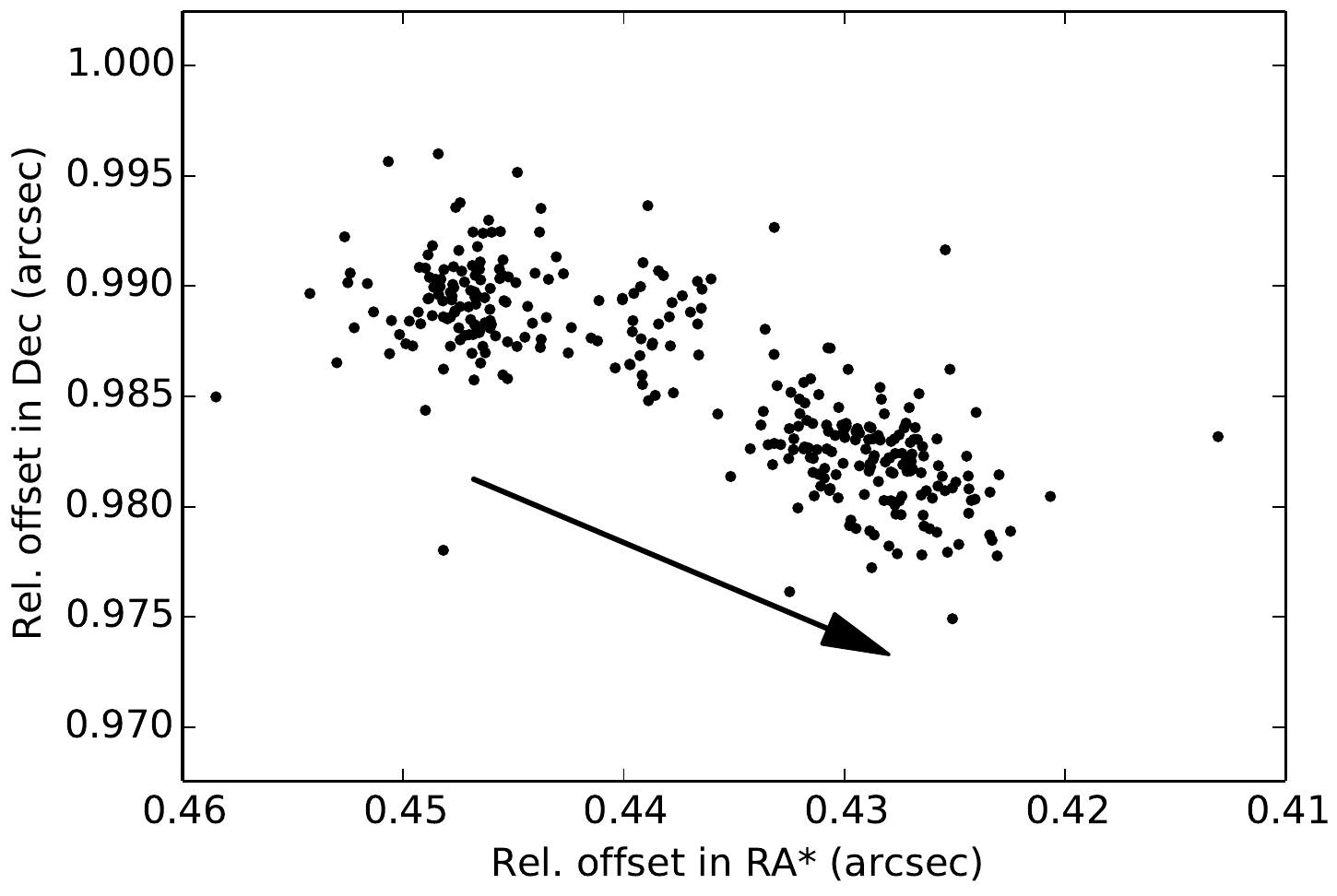}
\caption{\emph{Top}: FORS2 image of \dwfift\ in 0.47\arcsec\ seeing showing the primary (A), its companion (B), and the faint background object (x). \emph{Bottom}: Motion of B relative to A over 460 days, where every dot corresponds to one FORS2 frame and we assumed linear motion with a yearly amplitude and direction indicated by the arrow length and orientation, respectively. In both panels North is up, east is left.}
\label{fig:dw15B_image}
\end{figure}

{\label{sec:dw15}}
\dwfift\ was discovered by \cite{Burgasser:2007gd} and is the only known visually resolved binary in our survey. The companion (denoted B) is 1.6 $I$-band magnitudes fainter and is located 0.4\arcsec\,east and 1\arcsec\,north of the primary (A), see Fig. \ref{fig:dw15B_image}, which corresponds to a minimum relative separation of $\sim$20 AU. The {\small FORS2} images also contain a background star, which is 4 mag fainter than the B-component and located east of B at a distance increasing from 0.7\arcsec\ to 1.1\arcsec\ in time, due to the proper motion of \dwfift. 

Using the {\small FORS2} observations, we measured the relative motion of the \dwfift\,A/B system. Due to light contamination by the background object, the astrometry of the B component is slightly biased with a seeing-dependent amplitude. 
We modelled this systematic photocentre shift as a linear dependence on seeing within every epoch separately and consequently removed its contribution from the astrometry of \dwfift B. We then applied the standard astrometric reduction to both components separately using identical sets of reference stars, hence mutually excluding the gravitationally bound object, but fixing the parallax value of the B component to the one found for the A component. The relative position and proper motions are given in Table \ref{tab:dw15} and the data are shown in Fig. \ref{fig:dw15B_image}. The corresponding DCR parameters are $\rho_A=22.2 \pm 1.8$ mas for \dwfift A and $\rho_B=27.2 \pm 4.9$ mas for \dwfift B, which are compatible within their uncertainties, but because $\rho_B$ is larger, it indicates that the faint component is slightly redder, as expected (\citealt{Burgasser:2007gd} classified the system as L1.5 + L4.5). 

We attribute the measured relative motion to the orbital motion of \dwfift\,A/B. The small arc covered by our measurements is well described with a linear model and we can exclude the presence of a tight companion around \dwfift B that would introduce an astrometric signature larger than a few mas with a period $\lesssim$500 days.

\begin{table}
\centering
\caption{Offset and proper motion of \dwfift B relative to \dwfift A with reference epoch MJD = 55700.}
\begin{tabular}{cccc}
\hline
\hline
$\Delta \alpha^\star_{\mathrm{rel}}$       &$\Delta \delta_{\mathrm{rel}}$    & $\mu_{\alpha^\star,{\mathrm{rel}}}$  & $\mu_{\delta,{\mathrm{rel}}}$\\ 
(mas) & (mas) & (mas yr$^{-1}$) & (mas yr$^{-1}$) \\
\hline                                                    
$444.0 \pm0.2$  & $988.5 \pm0.2$  & $-16.1 \pm0.3 $&  $-6.8 \pm0.2$  \\
\hline                
\end{tabular}
\label{tab:dw15}
\end{table}

\subsection{Notes on individual objects}\label{sec:targets}
We present information in particular related to binarity:
\begin{description}
  \item[\dwtwo] was discovered to be a tight binary and will be the subject of a forthcoming paper \citep{Sahlmann:2014}.    
  \item[\dwnine] has a 28.5 $M_J$ companion in a 246 day orbit, which we discovered and characterised in \citetalias{Sahlmann:2013kk}.
  \item[\dwten] was discovered by \cite{Scholz:2002kl} and studied extensively with spectroscopy \citep{Reiners:2008cr, Seifahrt:2010fj, Faherty:2010fk}. \cite{Blake:2010lr} collected five NIR radial velocity measurements spanning 360 days that do not show the signature of a companion.
  \item[\dwthirt] was reported to be an X-ray source and both flares \citep{Hambaryan:2004lr} and quasi-quiescent emission \citep{Robrade:2009aa} have been observed. It is the brightest target in our sample and the only one that is closer than 10~pc.
\end{description}

\section{Results and discussion}\label{sec:results} 
The astrometry data collected over two years allowed us to study a variety of UCD characteristics in detail, among them are the distances, proper motions, photometric variability, and the occurrence of planetary and binary companions.

\subsection{Precision parallaxes}
\begin{figure}
\center
\includegraphics[width= \linewidth]{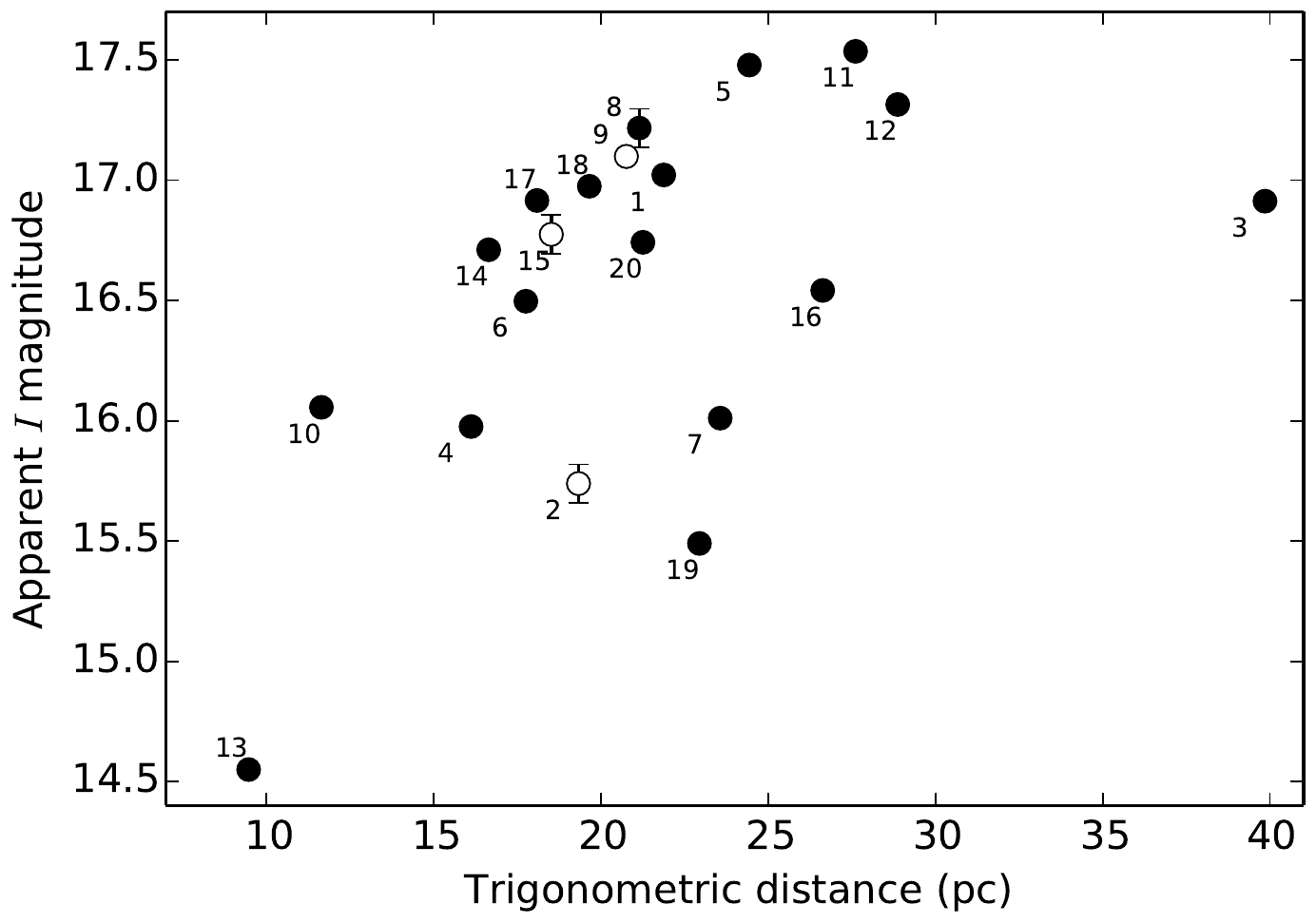}
\caption{Apparent $I$-band magnitudes and trigonometric distances determined in this work. The target number is shown next to each symbol. Open circles mark binaries. The uncertainties are usually smaller than the symbol size.}
\label{fig:appmag}
\end{figure}

\begin{figure}[h!]
\centering
\includegraphics[width=0.8\linewidth]{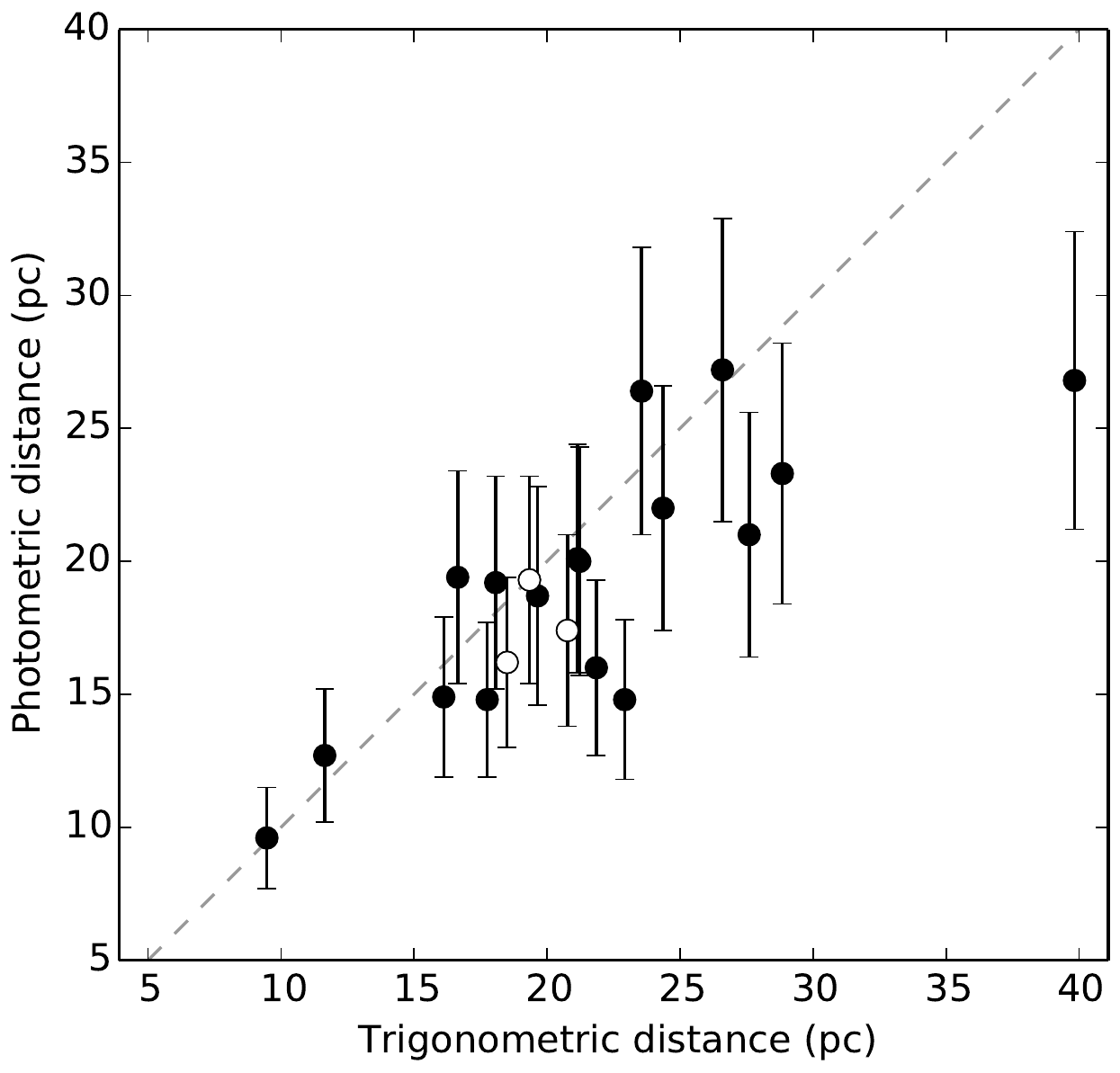}
\caption{Photometric distance estimates from \cite{Phan-Bao:2008fr} as a function of the distance measurements on the basis of absolute trigonometric parallaxes. For the latter, the uncertainties are smaller than the symbol size. Binaries are shown as open circles and the dashed line indicates consensus.}
\label{fig:photplx}
\end{figure}

We obtained trigonometric parallaxes for 20 UCDs that previously only had photometric distance estimates\footnote{Parallaxes of four targets were independently determined by \cite{Dieterich:2013ab}.}; they are displayed in Fig. \ref{fig:appmag}. Most are located in the 15--25 pc range, two are located closer and \dwthirt\ lies within 10 pc, and the most remote target is \dwthree\ at $\sim$40 pc. Figure \ref{fig:photplx} shows that there is general agreement between trigonometric and photometric distances, but also that the photometric method of \cite{Phan-Bao:2008fr} tends to underestimate the distance for targets beyond 20 pc. {For the eleven targets with trigonometric distance $>$20 pc, the photometric distance is too small by 1.0 $\sigma_{d,phot}$ on average, where $\sigma_{d,phot}$ is the uncertainty of the photometric distance estimate.} The average trigonometric parallax uncertainty amounts to 0.09 mas, which corresponds to an average fractional uncertainty of 0.19 \%, where we excluded the two tight binaries.

\begin{figure}[h!]
\begin{center}
\includegraphics[width=\linewidth]{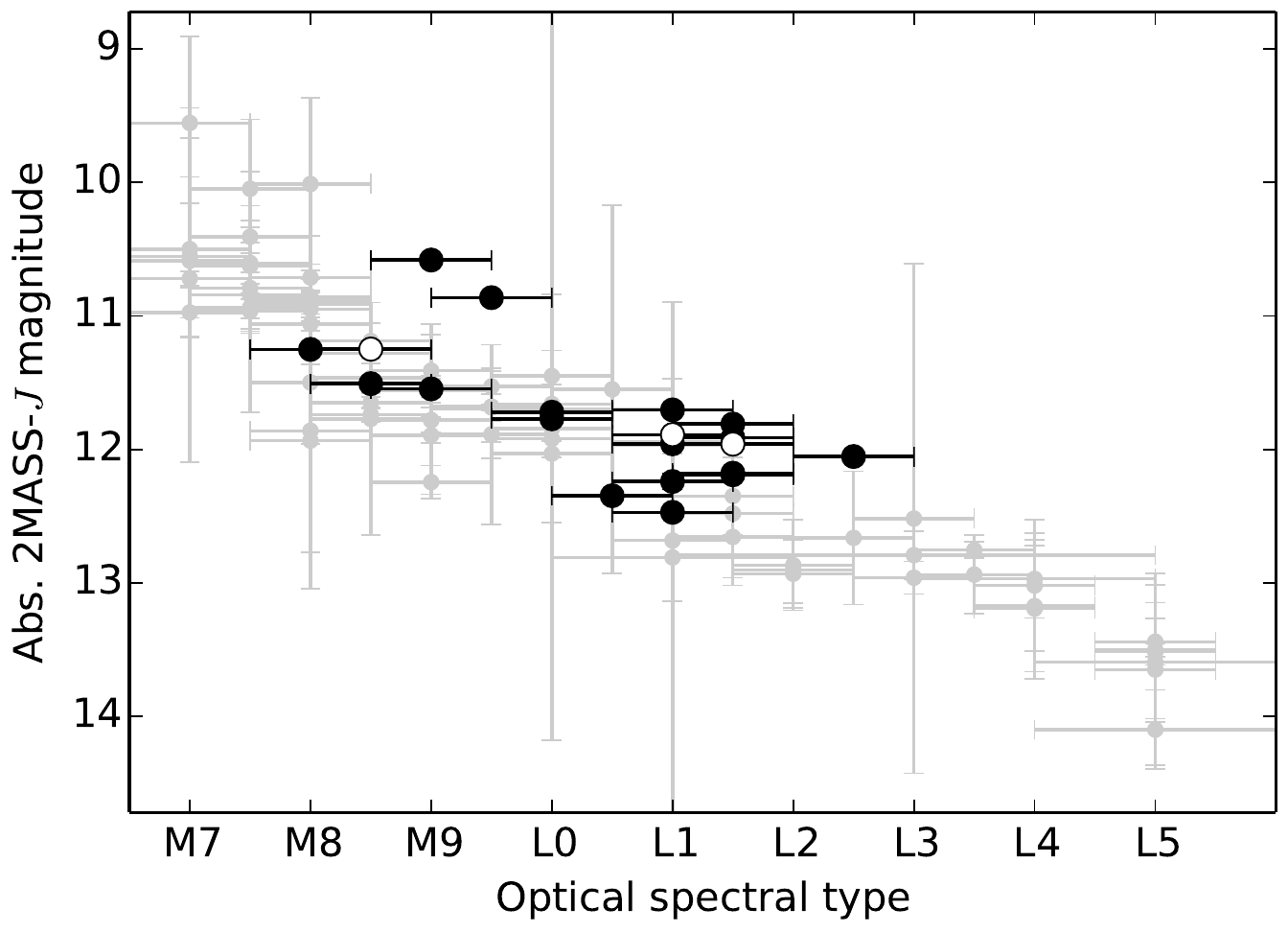}
\caption{Absolute 2MASS $J$-band magnitude as a function of spectral type for M7--L5 dwarfs in the database of ultracool parallaxes \citep{Dupuy:2012fk} (grey symbols) and for our sample (filled black symbols). Magnitude uncertainties of the latter are smaller than the symbol size. Binaries in our sample are shown with open circles.}\label{fig:HRD}\end{center}
\end{figure}

Using these parallaxes, we can set the target sample into context with the known population of UCDs as shown in Fig. \ref{fig:HRD}, where we assumed 0.5 subclasses of spectral type uncertainty. In particular, the dominant uncertainty on the absolute magnitude now stems from the magnitude measurement itself and not from the distance determination. For the 2MASS $J$-band, for instance, the average photometric uncertainty is 28 mmag, while the magnitude uncertainty caused by the parallax uncertainty is 10 mmag.

In the infrared $J$-band (Fig. \ref{fig:HRD}), the two tight binaries with primary spectral types M8.5 and L1.5 do not appear to be over-luminous, that is the flux contribution of the companion should be small at this wavelength. The two slightly over-luminous objects are \dwthree\ (M9.5) and \dwninet\ (M9.0), which otherwise are unremarkable objects.

\begin{figure}[h!]
\begin{center}
\includegraphics[width=\linewidth]{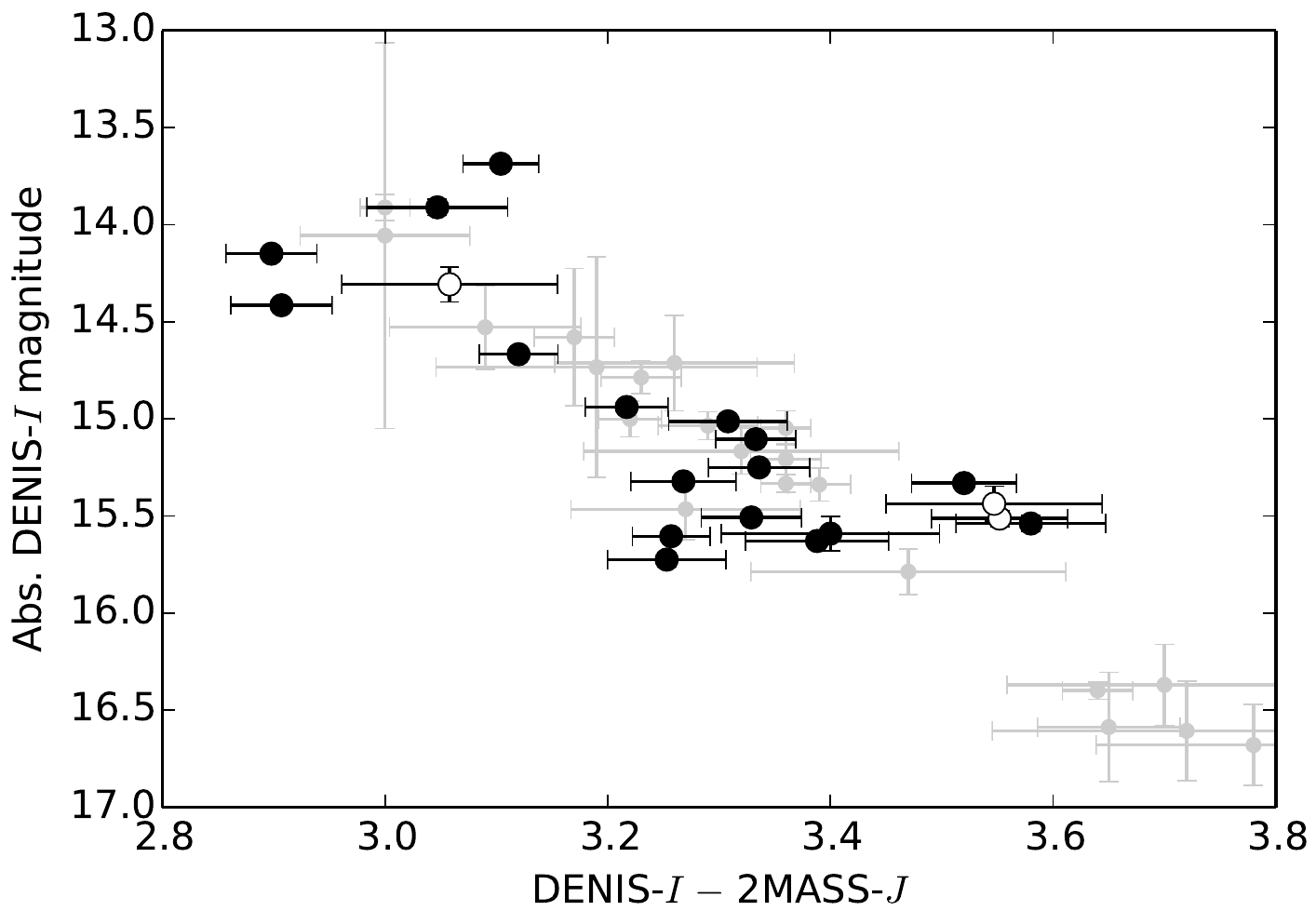}
\caption{Absolute $I$-band magnitudes as determined with FORS2 (given in the DENIS-$I$ magnitude system) as a function of $I$--$J$-colour for our sample (filled symbols). Binaries are shown with open circles. The comparison sample (grey symbols) was taken from \citet[Table 1]{Phan-Bao:2008fr}}\label{fig:CCD}\end{center}
\end{figure}

In Fig. \ref{fig:CCD}, we show the optical $I$-band absolute magnitude as a function of $I$--$J$-colour. As expected from Fig. \ref{fig:HRD}, the binaries do not appear over-luminous at shorter wavelength either.

\subsection{Proper motions and tangential velocities}
\begin{figure}[h!]
\begin{center}
\includegraphics[width=\linewidth]{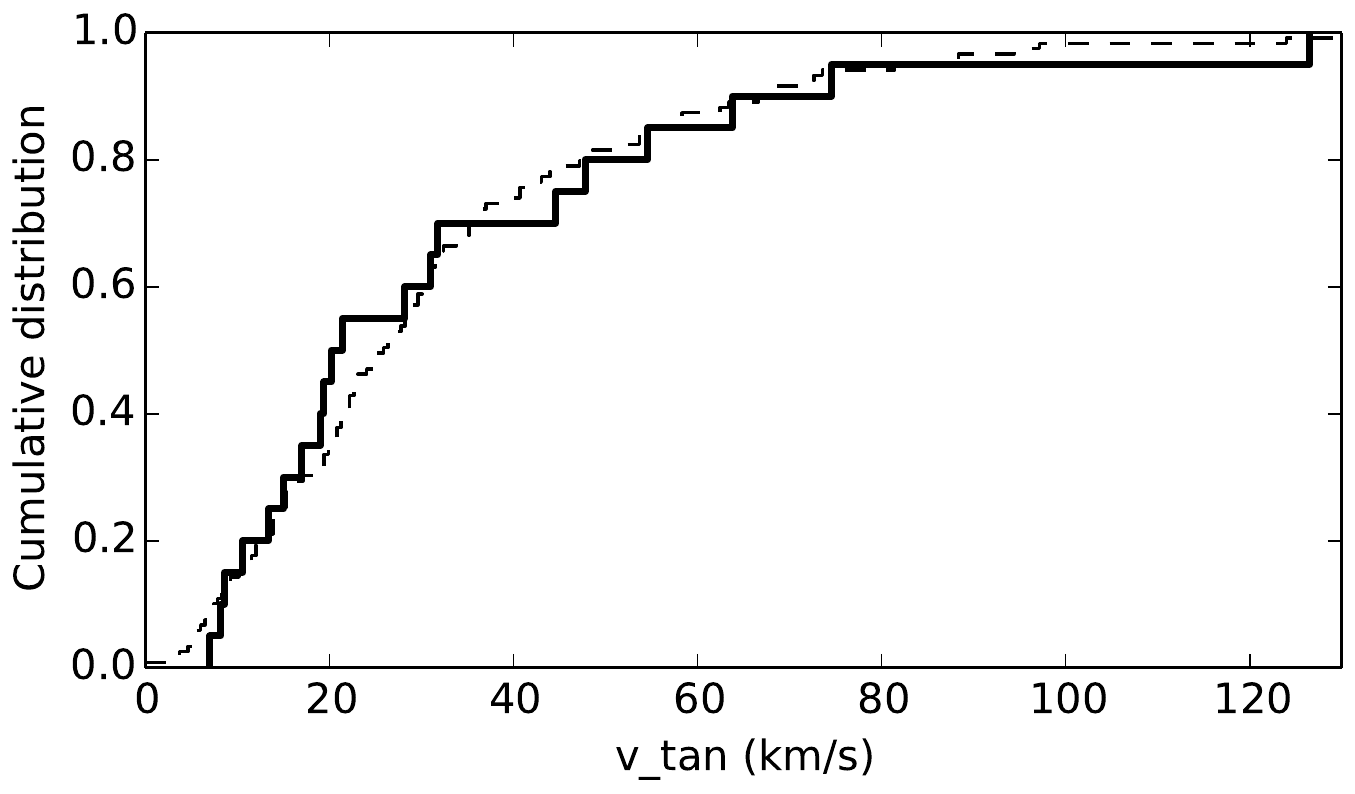}
\caption{Distribution of tangential velocities of our sample (solid line). The dashed line shows 119 M7--L3 dwarfs from \cite{Schmidt:2007ve}.}
\label{fig:vtrans}\end{center}
\end{figure}

The proper motions reported in Table \ref{tab:galcorr} are relative to the local reference field. We decided not to perform the correction to absolute proper motions because this would inevitable be dominated by the uncertainty of the correction. An accurate correction will be made possible by the second intermediate data release of the \emph{Gaia} mission \citep{T.Prusti:2012vn}.\\
Our proper motions agree in general with the values given in \cite{Phan-Bao:2008fr} and, when applicable, in \cite{Faherty:2009fk}, but they have significantly higher precision. Given the large distance of our reference stars, a conservative estimate is that our proper motions can be considered accurate at the 1\% level. Because we usually do not have radial velocity estimates, we cannot determine three-dimensional space velocities, but we examined the tangential velocities of the target sample shown in Fig. \ref{fig:vtrans} and Table \ref{tab:targspar}. The distribution is compatible with other studies of M/L-dwarf transition objects in the field \citep{Schmidt:2007ve}.

\subsection{$I$-band photometric variability}
Using the {\small FORS2} observations collected for astrometry, we were able to obtain $I$-band photometric measurements at the milli-magnitude level, covering time-scales of minutes to several hundred days. We found that the 20 UCDs in our survey are stable at the 3--5 mmag level over $\sim$480 days, with the exception of \dwfive, which exhibits variations of $\pm$40 mmag. {A better characterisation of the photometric variability of \dwfive\ requires additional data and is necessary to distinguish between the different possible causes mentioned in Sect. \ref{sect:variability}.}\\
We therefore found that $5^{+10}_{-2}$\,\% of M8--L2 dwarfs in the field show $I$-band variability higher than 5 mmag r.m.s. over time-scales of minutes to $\sim$500 days, where we quoted uncertainties computed from binomial statistics as in \cite{Burgasser:2003fr}. This is smaller than the variability occurrence of $\sim$ 50 \% found for a sample of comparable size \citep{Bailer-Jones:2001qy} and also smaller than the value of $\sim$16--25 \% obtained for a larger sample \citep{Koen:2013uq}.

\subsection{Occurrence of binary and planetary companions}
The sample of 26 UCDs presented by \cite{Phan-Bao:2008fr} is in principle unbiased with respect to binarity, and our selection of 20 objects by coordinates and magnitude is not expected to alter this property. A potential minor bias towards a higher binary fraction might be introduced by the selection in brightness, because unresolved binaries with similar spectral types are more likely to be included. For the following statistical discussion, we assumed that our sample is unbiased and applied binomial statistics. 

\subsubsection{Tight binary fraction}
We discovered two new binary systems with separations of 0.36 AU \citepalias{Sahlmann:2013kk} and $\sim$1 AU \citep{Sahlmann:2014} and can exclude the presence of additional binaries within $\sim$1.5 AU (Fig. \ref{fig:detLimits}). We thus found that the fraction of M8-L2 dwarfs that form tight ($\lesssim$1\,AU) binary systems is $10^{+11}_{-3}$\,\%. This results obtained from our astrometric observations agrees with the outcome of the radial velocity survey of \cite{Blake:2010lr}.

\subsubsection{Overall binary fraction}
Including the $\gtrsim20$ AU binary \dwfift, there are a total of three binaries in our sample, resulting in an overall binary fraction of  $15^{+11}_{-5}$\,\%. This agrees well with estimates from high-resolution imaging \citep{Bouy:2003kx} and spectroscopic surveys of M/L dwarfs \citep{Reid:2008vn, Burgasser:2010kx}.

\subsubsection{Occurrence of giant planets}
On the basis of Fig. \ref{fig:detLimits}, which excludes tight binaries, we found that none of the 18 targets hosts a giant planet more massive than $\sim$$5\, M_J$ within 0.1--0.8 AU, where strictly speaking we cannot exclude their presence at orbital periods of one year (0.44 AU). Consequently, the upper limit on the occurrence of these giant planets within 0.1--0.8 AU around M/L-transition dwarfs is 9 \%.

We herewith established a low occurrence rate of giant planets around M/L dwarfs in the 0.1--0.8 AU separation range and closed the gap between the previously found low occurrence at smaller $<$ 0.05 AU \citep{Blake:2010lr} and larger $>$1--2 AU \citep{Stumpf:2010lr} separations.

\subsection{Astrometric accuracy of FORS2}
\begin{figure}[h!]
\begin{center}
\includegraphics[width=\linewidth]{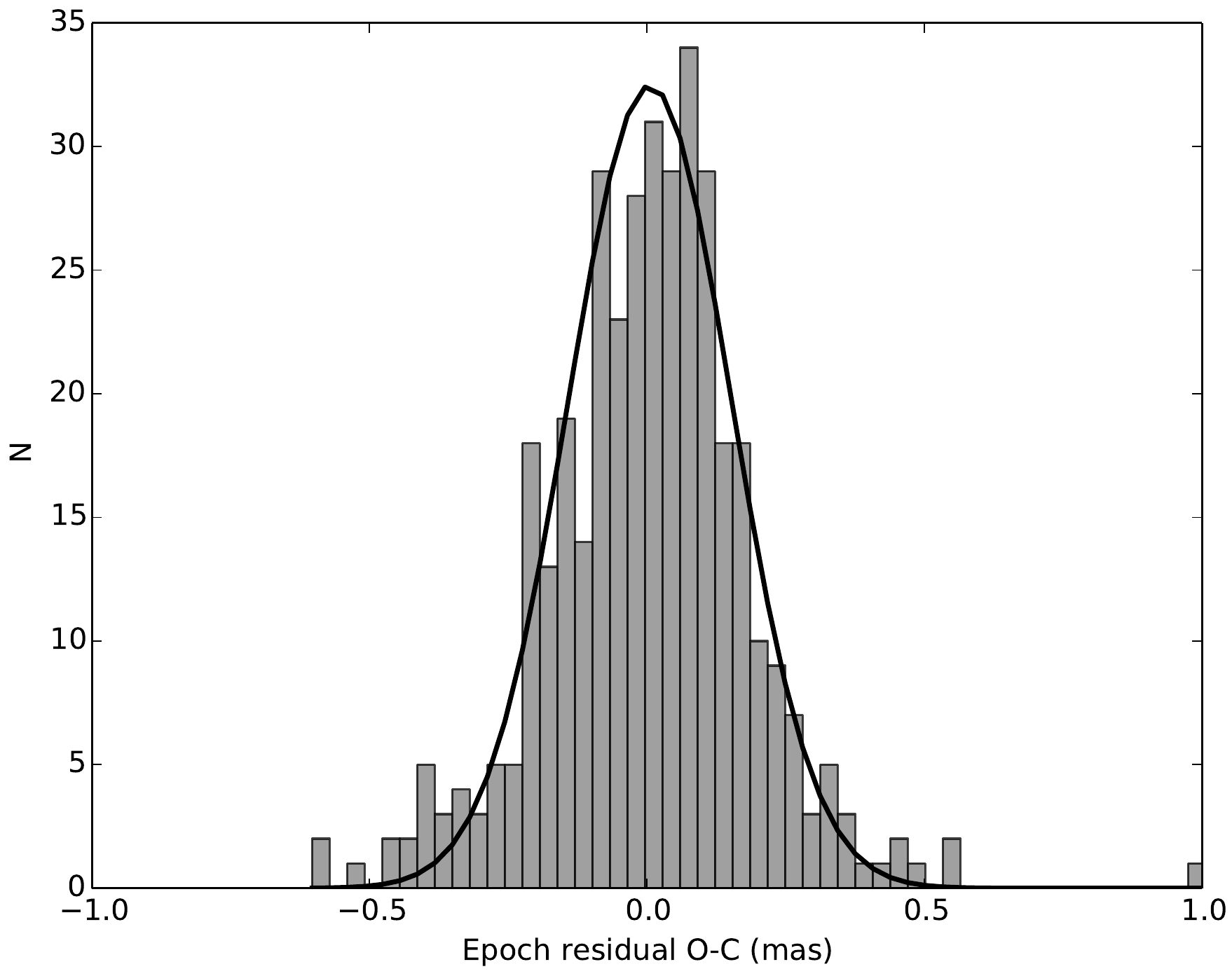}
\caption{Histogram of 2$\times$190 epoch-average residuals after adjustment of parallax and proper motions. The Gaussian curve was adjusted to the 95\% quantile and has a width parameter of $\sigma = 146$\,$\mu$as.}
\label{fig:palta2}
\end{center}
\end{figure}

To assess the global astrometric performance in our survey, we examined the epoch-averaged residuals of the standard fit (Eq.~(\ref{eq:axmodel})) for targets that are not tight binaries. These are 190 epochs of eighteen objects, and the corresponding histogram of O--C residuals in both RA and Dec is shown in Fig. \ref{fig:palta2}. The r.m.s. of this data is 181 $\mu$as. Because we cannot exclude the presence of orbital signals in the data, we rejected 5 \% of the largest residuals and adjusted the data with a normal distribution. The r.m.s. dispersion of the remaining 180 epoch residuals is 146 $\mu$as, which can be seen as an estimate of the global astrometric performance of our programme over a time-base of two years. In some cases, residual r.m.s. of 120 $\mu$as were achieved, see Table~\ref{tab:ppmres}.

A more detailed discussion of the astrometric performance of {\small FORS2} within this survey is given in \citetalias{Lazorenko:2013kk}, where we relate the obtained accuracy with the estimated measurement precision.

\subsection{Our project in the context of {Gaia}}
ESA's \emph{Gaia} space astrometry mission \citep{Perryman:2001vn, de-Bruijne:2012kx} will perform astrometry of stellar objects brighter than \emph{Gaia} magnitude $G\simeq20$ and will therefore deliver data for several hundred UCDs \citep{Smart:2008fk, Sarro:2013uq}. The M/L transition objects studied here have an average magnitude of $I\sim17$ and $G-I\sim1.6$, using $V-I\sim4.5$ \citep{Dahn:2008fk} and the colour-colour-relations of \cite{Jordi:2010kx}, which means that they lie at the faint end of the \emph{Gaia} magnitude range. Therefore, the precision of a single \emph{Gaia} astrometric measurement is expected to be $\sim$\,300--500\,$\mu$as \citep{2011AdSpR..47..356M}, a value that favours the demonstrated {\small FORS2} performance. Our survey will thus not be superseded by \emph{Gaia}. 
On the contrary, it will be possible to tie the local {\small FORS2} astrometric reference frame to the global \emph{Gaia} solution and thereby obtain model-independent absolute parallaxes and proper motions of our targets. 

\section{Conclusions}\label{sec:concl}
We presented the first results of an astrometric survey targeting 20 ultracool dwarfs at the M/L transition obtained after two years. The project's primary goal is to detect planetary companions, but the {\small FORS2} observations provide us with a rich dataset that covers a variety of science cases. 

We determined trigonometric parallaxes of 20 nearby ultracool dwarfs at the M/L transition with unprecedented accuracy of 0.09 mas ($\sim$0.2 \%) on average. Most targets are located at distances of 15--25 pc, and the closest member is at 9.5 pc. In the future, this sample can serve as a reference for the study of ultracool dwarfs at the M/L transition, in particular for the refinement of theoretical models and the search for small transiting planets. 

Applying the planet-search strategy and dedicated tools for the detection and adjustment of astrometric orbits, we discovered two new tight ultracool binary systems and fully characterised their orbital motions. In particular, the low-mass companion of \dwnine\ indicated that tight binary systems with low mass-ratios may not be as rare as previously thought \citepalias{Sahlmann:2013kk}. The overall binary fraction of $15^{+11}_{-5}$\,\% that we found in our sample is compatible with previous surveys using different observing techniques.

The astrometry data collected during the two-year initial phase of the project yielded limits on the occurrence of giant planets around M/L dwarfs in a previously unexplored separation range of $\sim$0.1--0.8 AU and thus closed a gap in detection space left by radial-velocity and direct-imaging planet searches. For the first time, we showed that the upper limit for the occurrence of giant planets $\gtrsim$$5\, M_J$ in this separation range is 9 \%. {This is consistent with the theoretical expectations of planet formation through core accretion that predicts a low occurrence rate of giant planets around M/L-transition dwarfs. If giant planets form via gravitational instability, our results indicate that the occurrence rate of UCD disks that are massive enough to become unstable is low.}

Constraining the planet population around UCDs and obtaining their high-precision distances is relevant for future searches for small, close-in planets that transit their ultracool hosts (e.g.\ \citealt{Triaud:2013aa}). In this context, we also found that optical variability at the M/L transition may not be as widespread as previous studies have indicated: only $5^{+10}_{-2}$\,\% of the M8--L2 dwarfs in our sample of field objects show an $I$-band variability higher than 5 mmag r.m.s. over time-scales of minutes to $\sim$500 days. 

Finally, we demonstrated that astrometric trajectories of faint optical sources can be determined with an accuracy of 120--150~$\mu$as using ground-based observations with an 8 m telescope. The photocentre measurement precision corresponds to 1/1000 of the {\small FORS2} CCD pixel size and is similar to the precision of the spectrum position determination with radial-velocity spectrographs \citep{Pepe:2008kx}.
In \citetalias{Lazorenko:2013kk}, we show that the discrepancy between the above value and the 50 $\mu$as demonstrated by \cite{Lazorenko:2009ph} is due to compromises we had to make to implement the survey. Our observations are executed in queue-scheduling service mode to guarantee good seeing conditions. The exposure times are set to avoid saturation even in the best seeing conditions, consequently, the S/N during an epoch of normal seeing is sub-optimal. Therefore, the performance demonstrated here is not the limit for this type of ground-based astrometry work.

In the future, we will expand this planet-search survey towards lower detectable planet masses and longer periods by continuing the astrometric monitoring and increasing the number of measurements and their time-span. The advent of the \emph{Gaia} mission will not supersede our project. On the contrary, the \emph{Gaia} survey will be complementary in the astrometric search for exoplanets around ultracool dwarfs.

\begin{acknowledgements}
J.S. is supported by an ESA research fellowship. J.S., D.S., M.M., D.Q., and S.U. thank the Swiss National Science Foundation for supporting this research. E.M. was supported by the Spanish Ministerio de Economia y Competitividad through grant AyA2011-30147-C03-03 and thanks the Geosciences Department at the University of Florida for a visiting appointment. J.S. kindly acknowledges support as a visitor at the Centro de Astrobiolog\'ia in Villanueva de la Ca\~nada (Madrid). We thank N. Phan-Bao for making the low-resolution optical spectra available to us. We thank the ESO staff for efficiently scheduling and executing our observations. This publication makes use of data products from the Two Micron All Sky Survey, which is a joint project of the University of Massachusetts and the Infrared Processing and Analysis Center/California Institute of Technology, funded by the National Aeronautics and Space Administration and the National Science Foundation and of the Database of Ultracool Parallaxes maintained by Trent Dupuy. This research made use of Astropy, a community-developed core Python package for Astronomy \citep{Astropy-Collaboration:2013aa}, of alipy (\url{http://obswww.unige.ch/~tewes/alipy/}), of APLpy, an open-source plotting package for Python hosted at \url{http://aplpy.github.com}, of the databases at the Centre de Donn\'ees astronomiques de Strasbourg (\url{http: //cds.u-strasbg.fr/}), and of NASA's Astrophysics Data System Service (\url{http://adsabs.harvard.edu/abstract_service.html}).
\end{acknowledgements}
\bibliographystyle{aa} 
\bibliography{/Users/sahlmann/astro/papers} 

\begin{appendix}

\section{Finding charts}
\begin{figure*}[h!]
\begin{center}
\includegraphics[width= 0.32\linewidth]{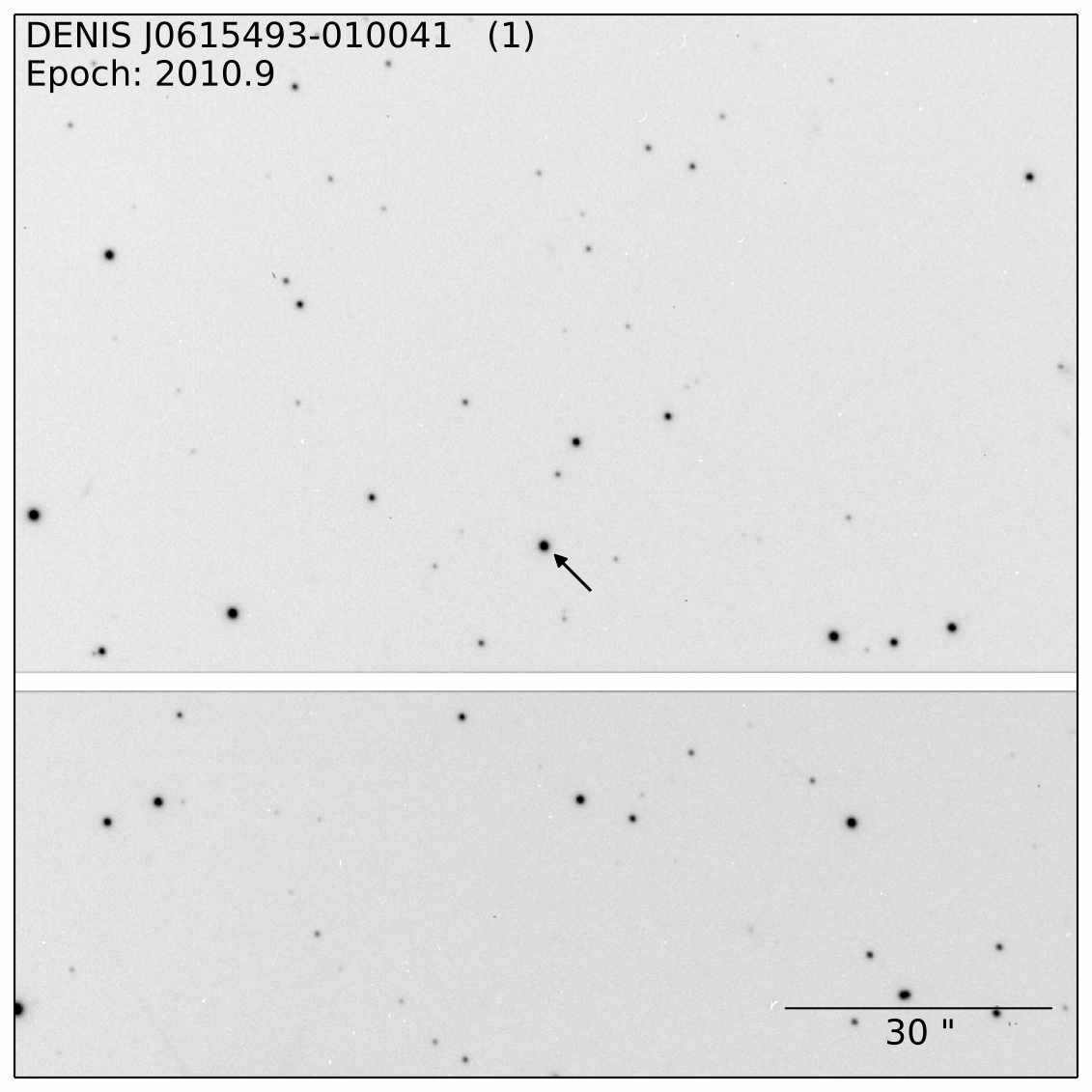}\hspace{0.5mm}
\includegraphics[width= 0.32\linewidth]{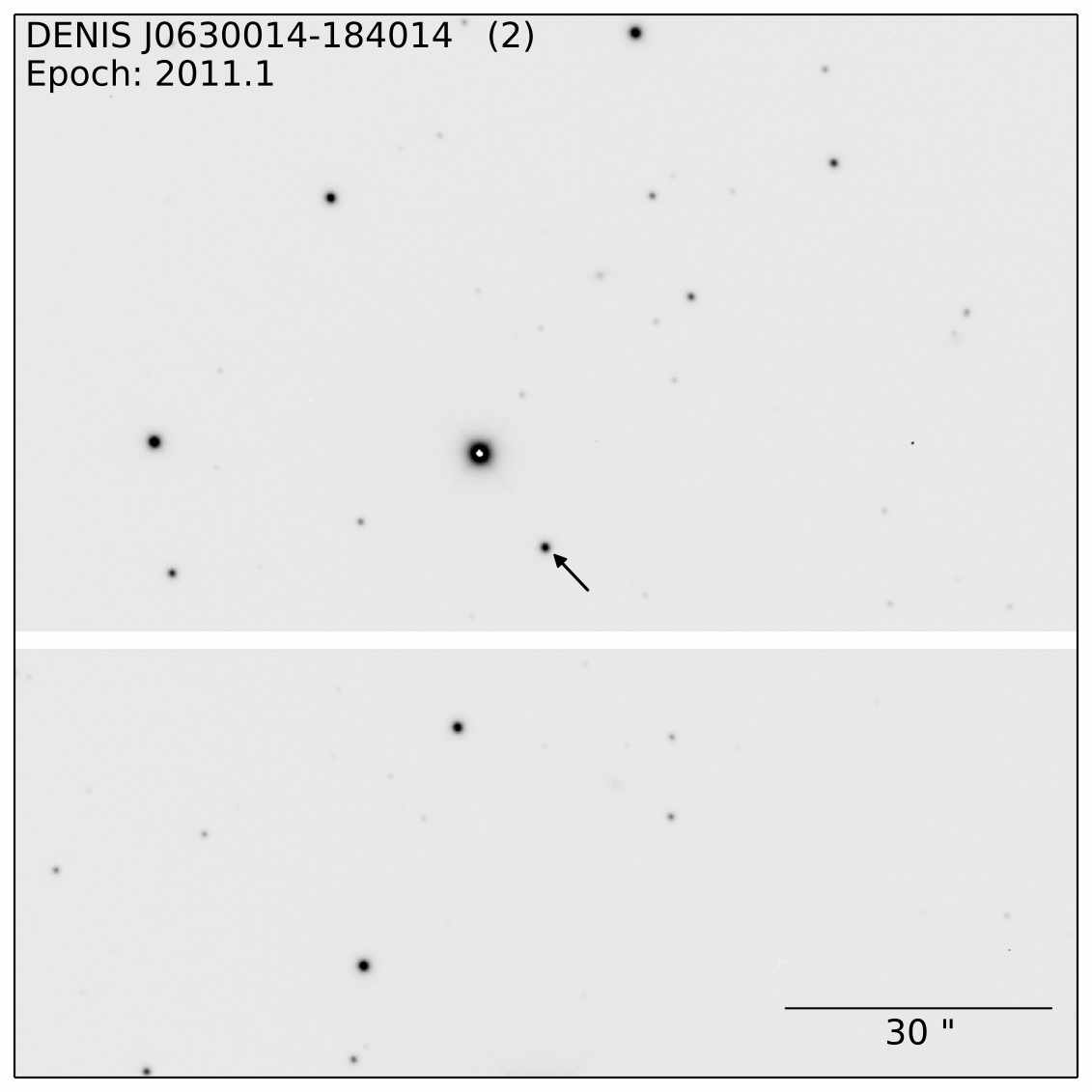}\hspace{0.5mm}
\includegraphics[width= 0.32\linewidth]{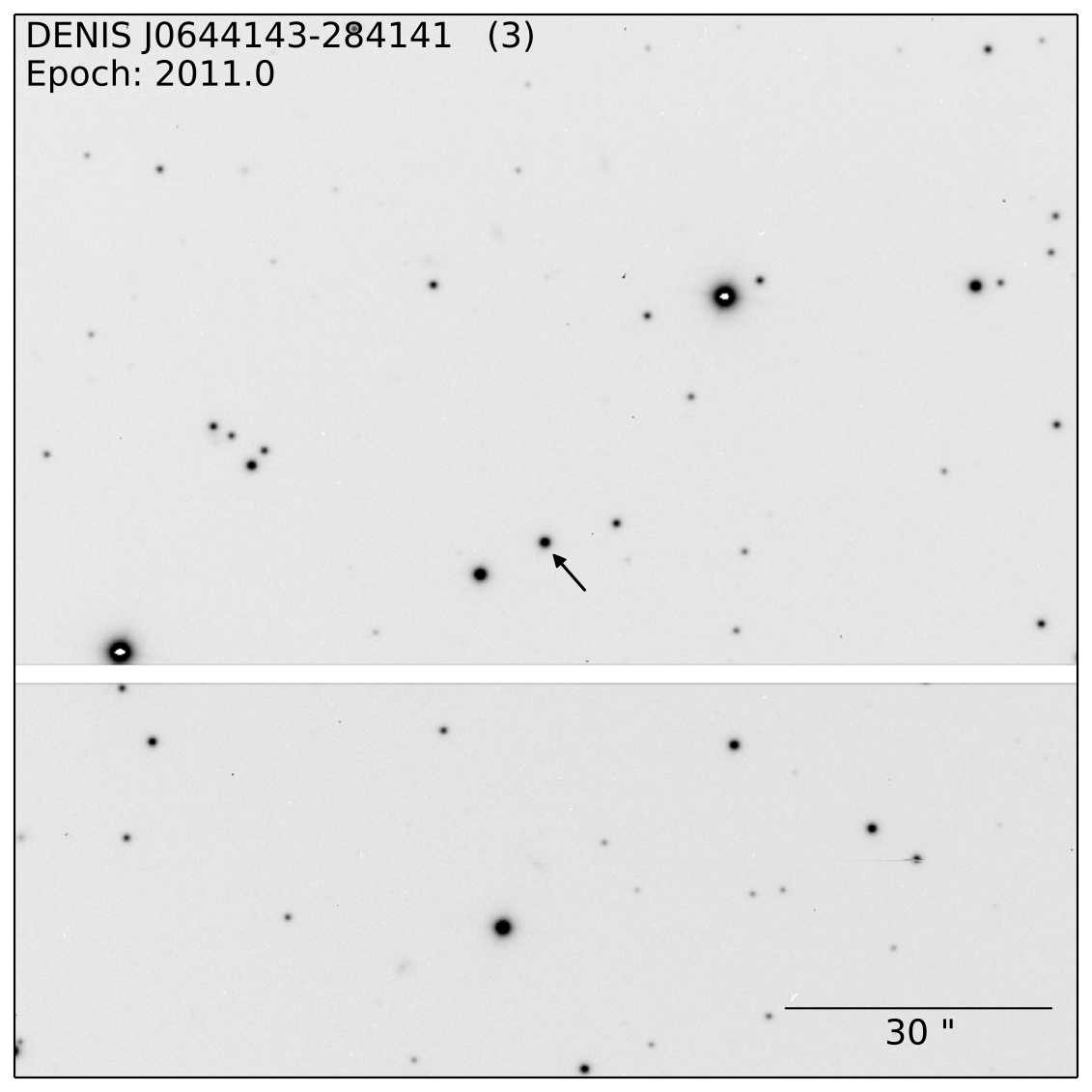}\vspace{1mm}
\includegraphics[width= 0.32\linewidth]{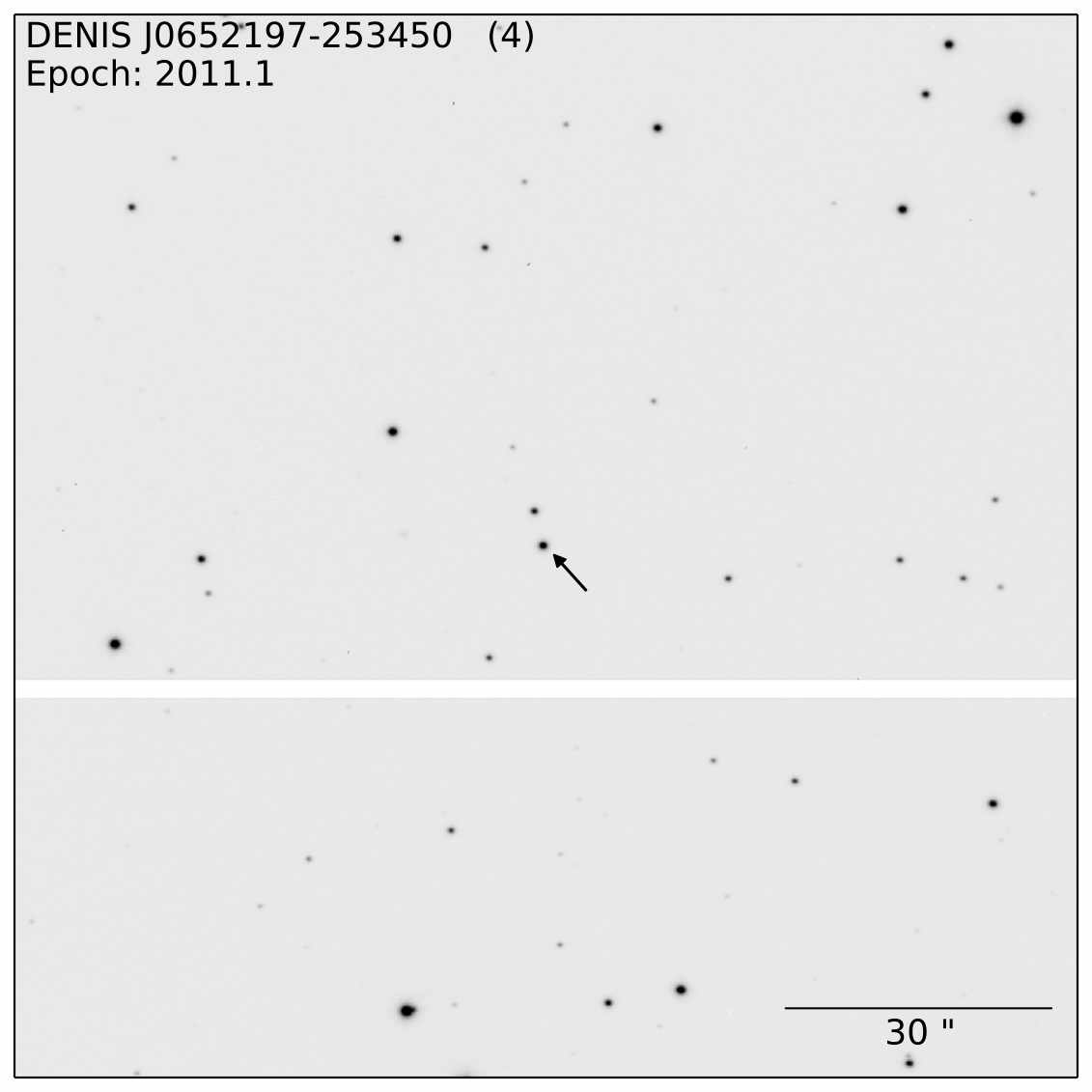}\hspace{0.5mm}
\includegraphics[width= 0.32\linewidth]{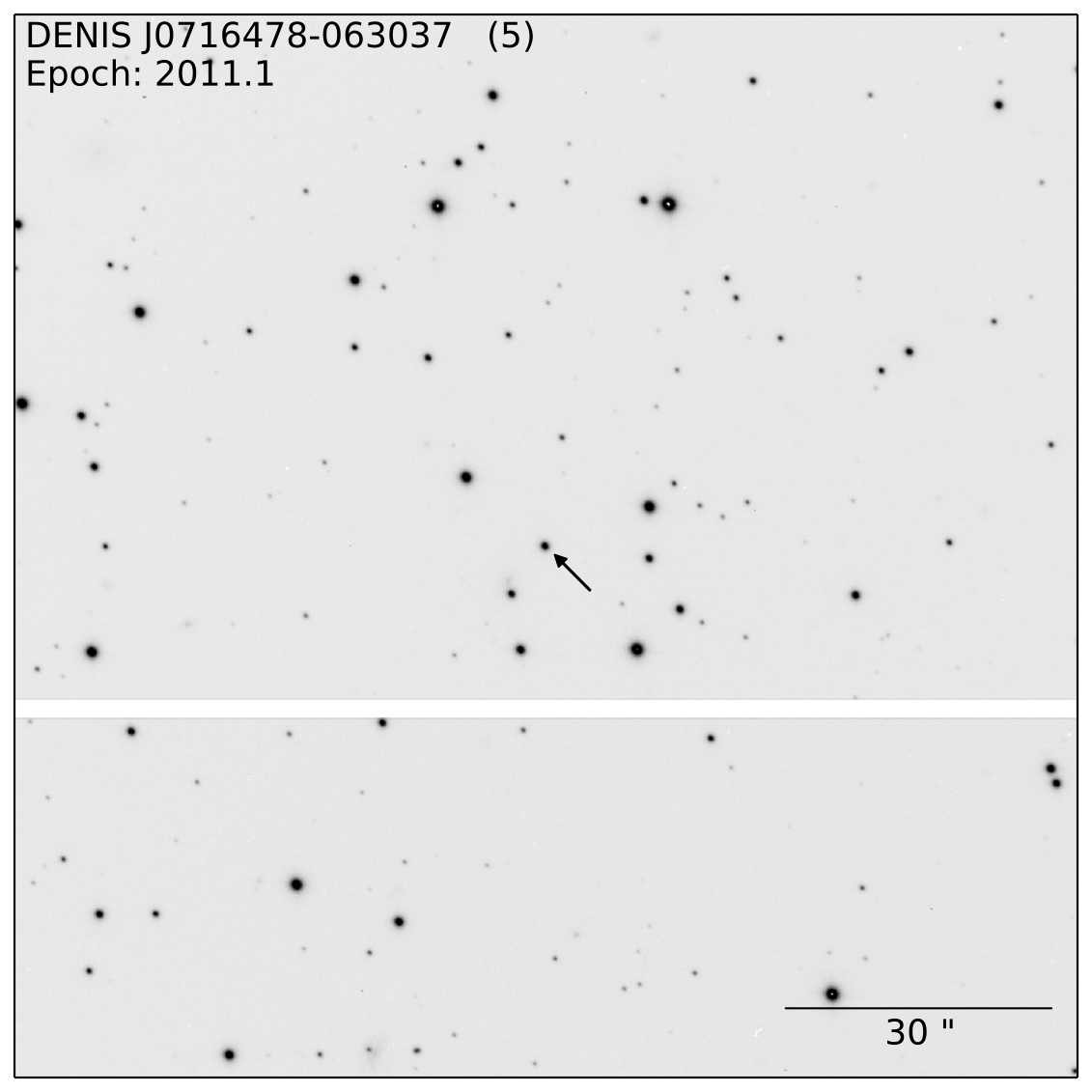}\hspace{0.5mm}
\includegraphics[width= 0.32\linewidth]{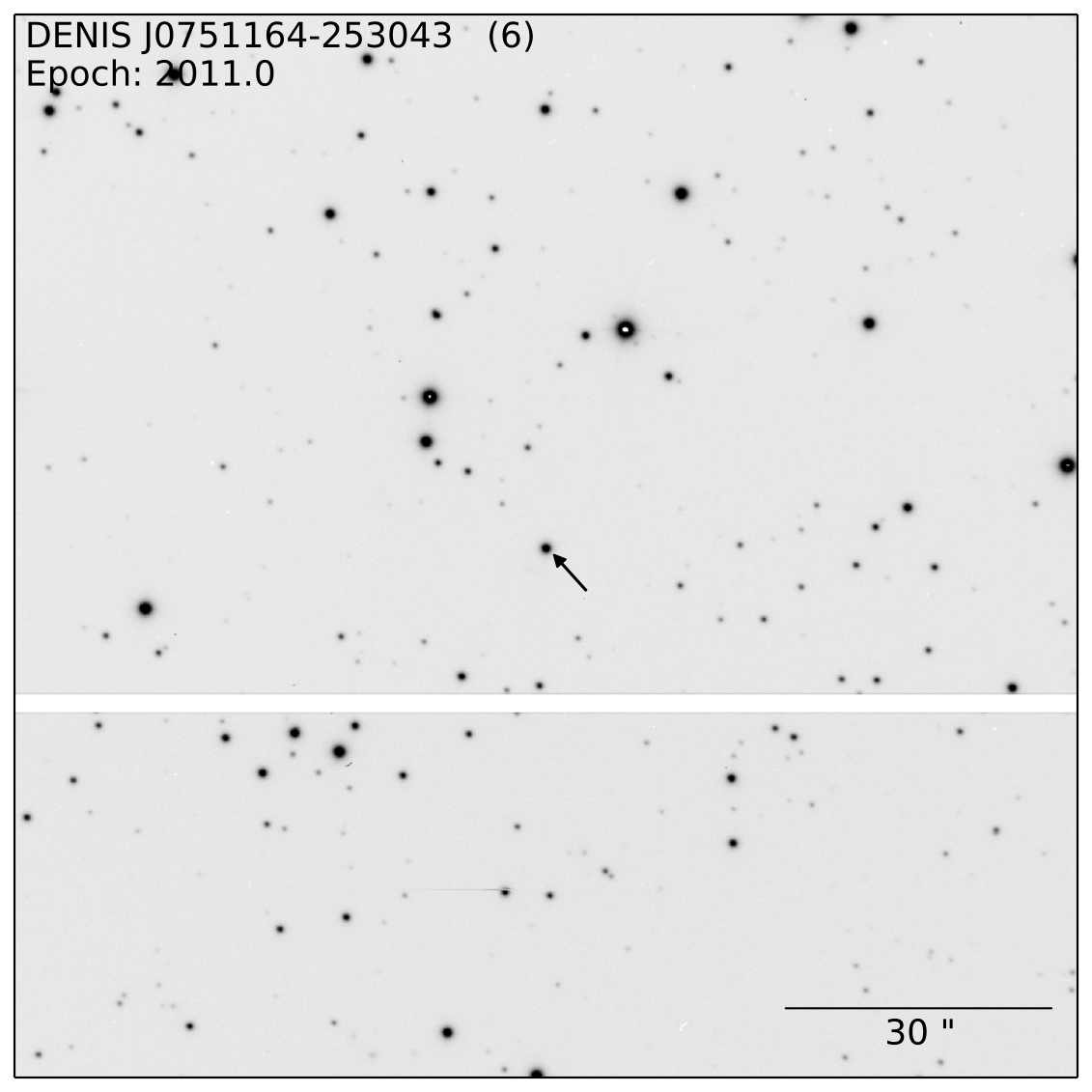}\vspace{1mm}
\includegraphics[width= 0.32\linewidth]{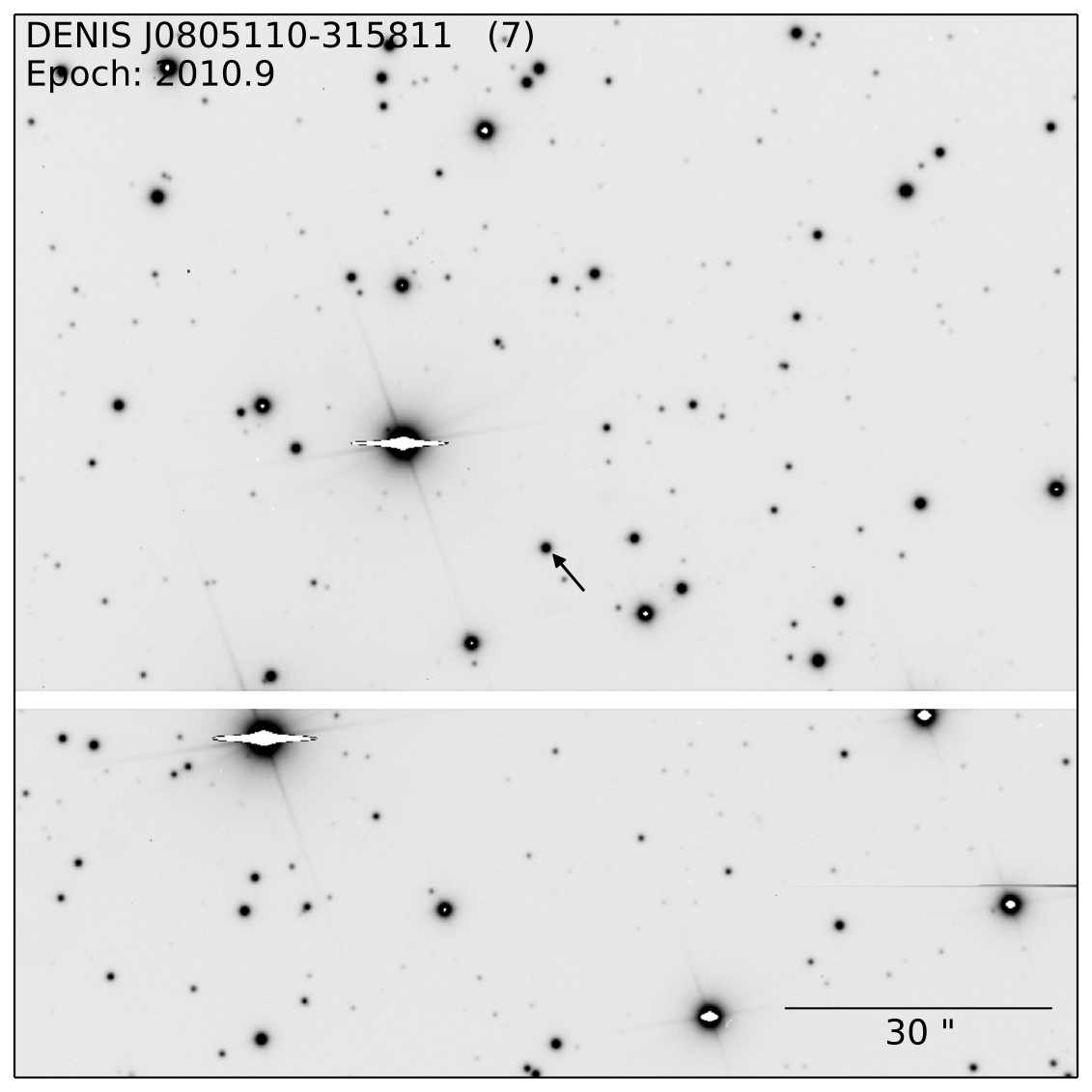}\hspace{0.5mm}
\includegraphics[width= 0.32\linewidth]{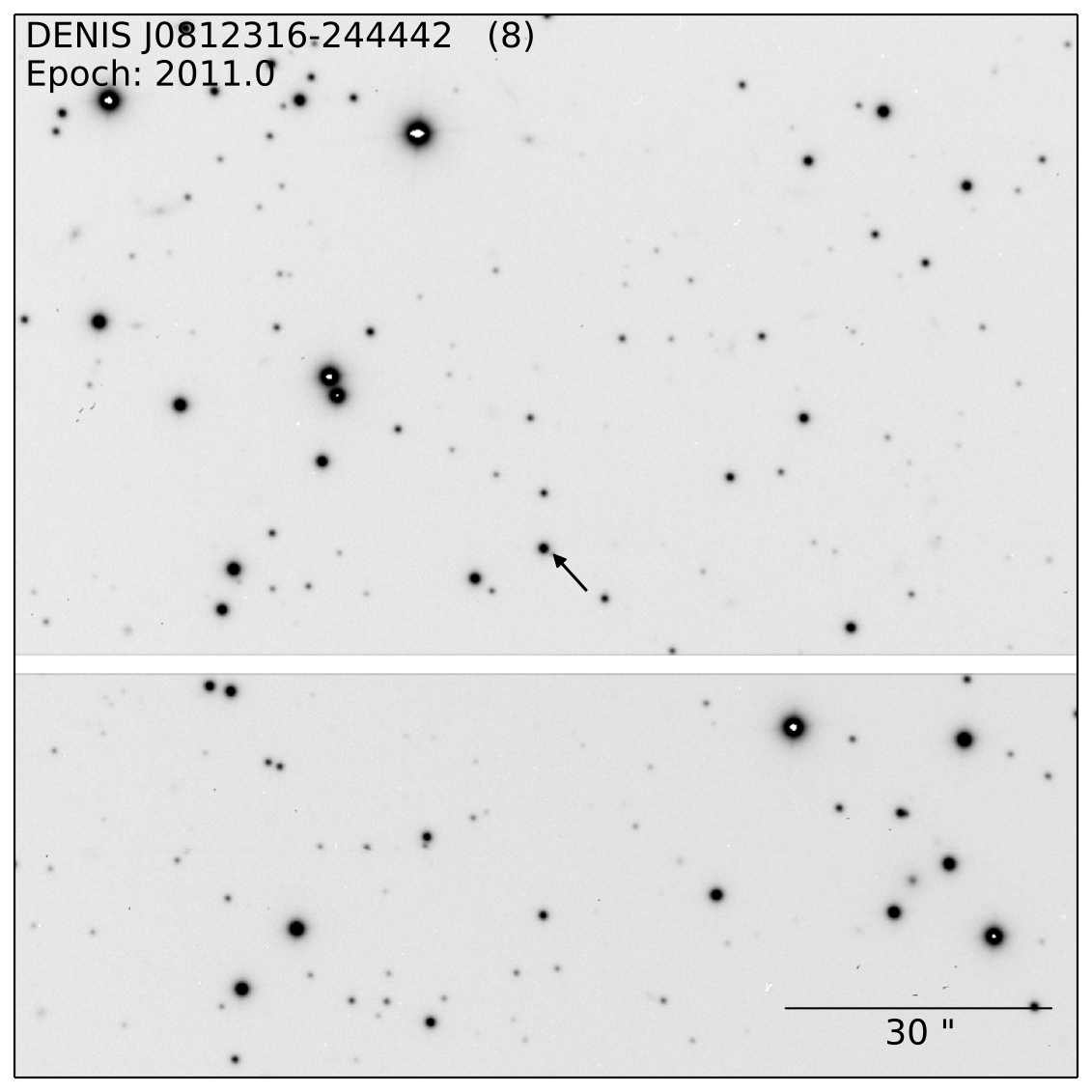}\hspace{0.5mm}
\includegraphics[width= 0.32\linewidth]{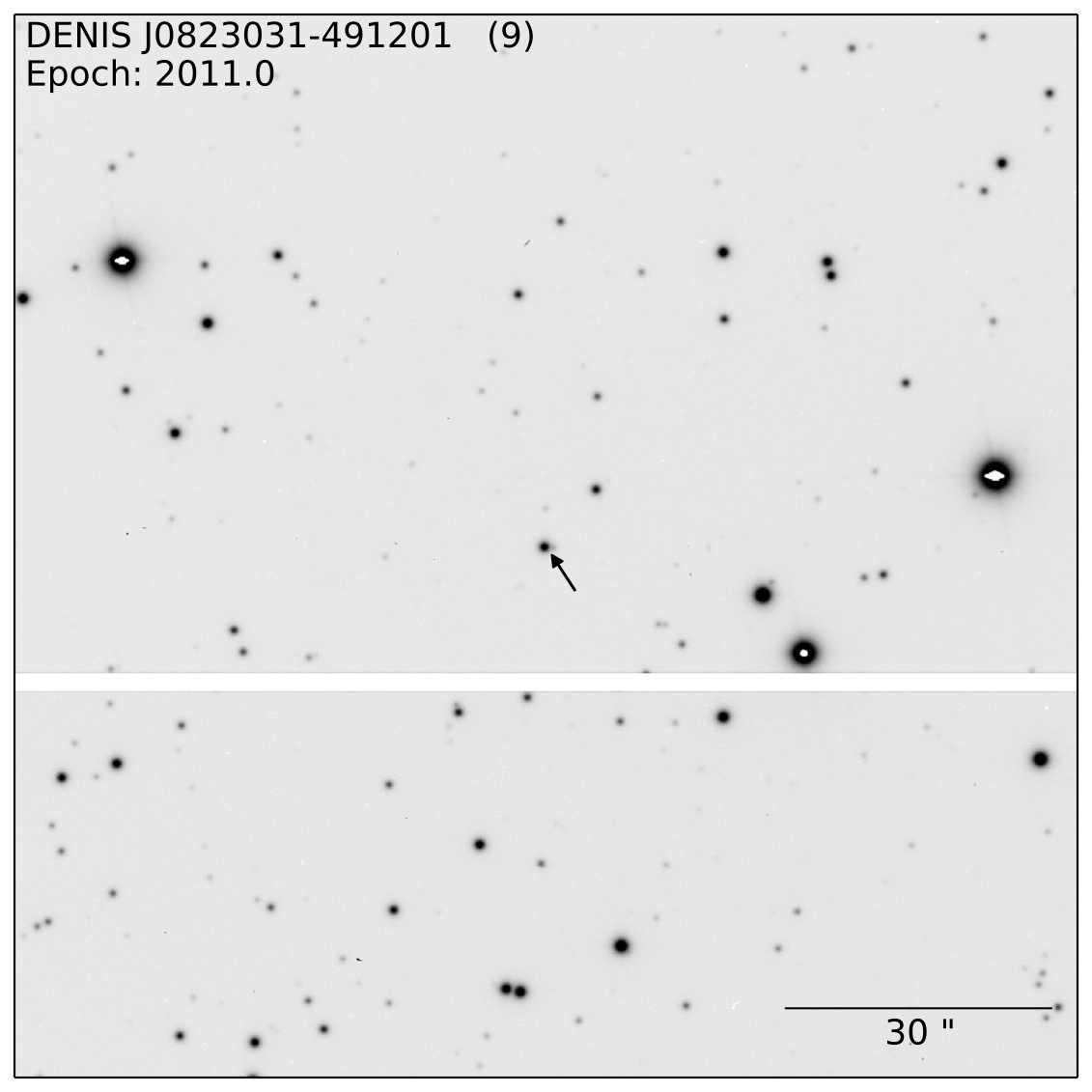}\vspace{1mm}
\includegraphics[width= 0.32\linewidth]{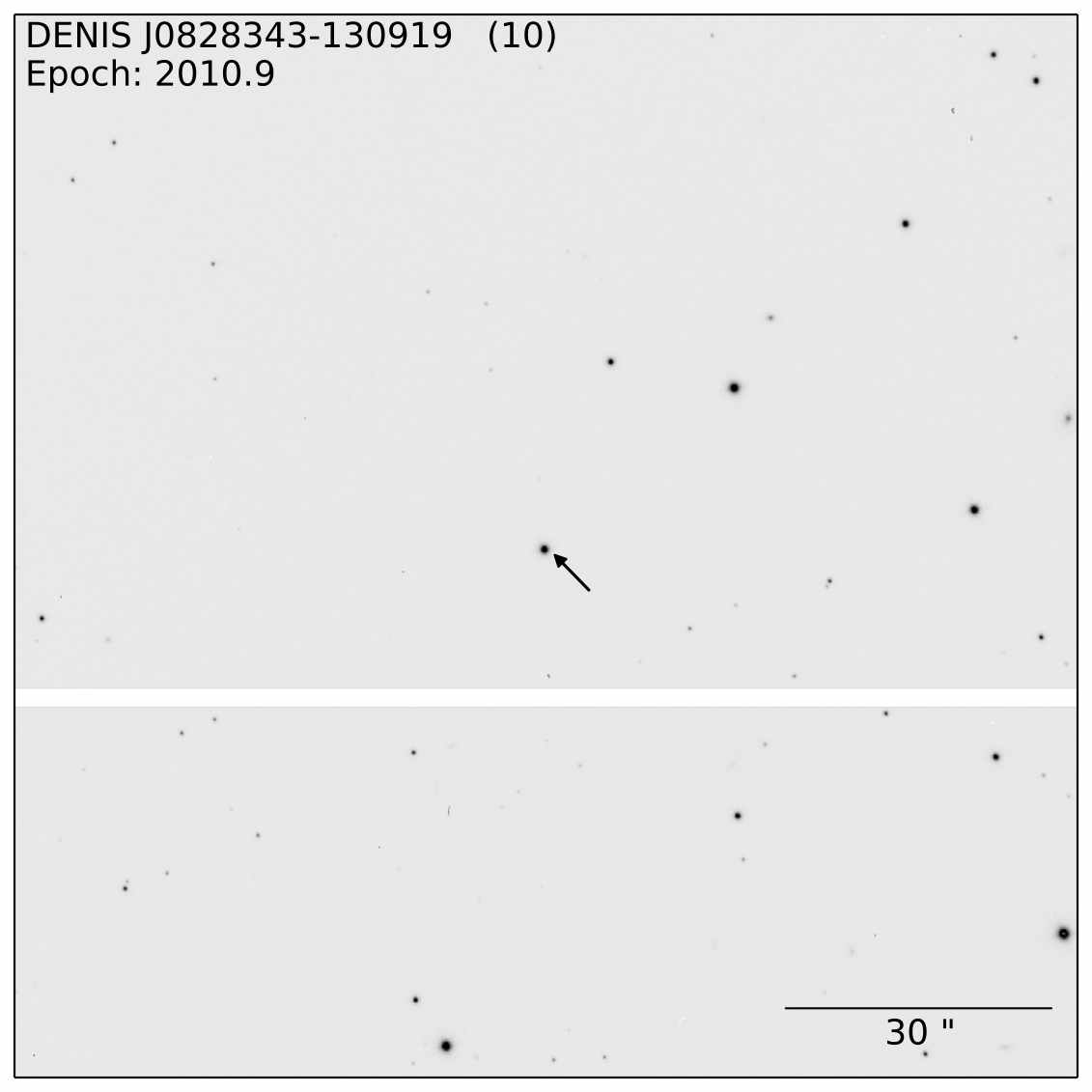}\hspace{0.5mm}
\includegraphics[width= 0.32\linewidth]{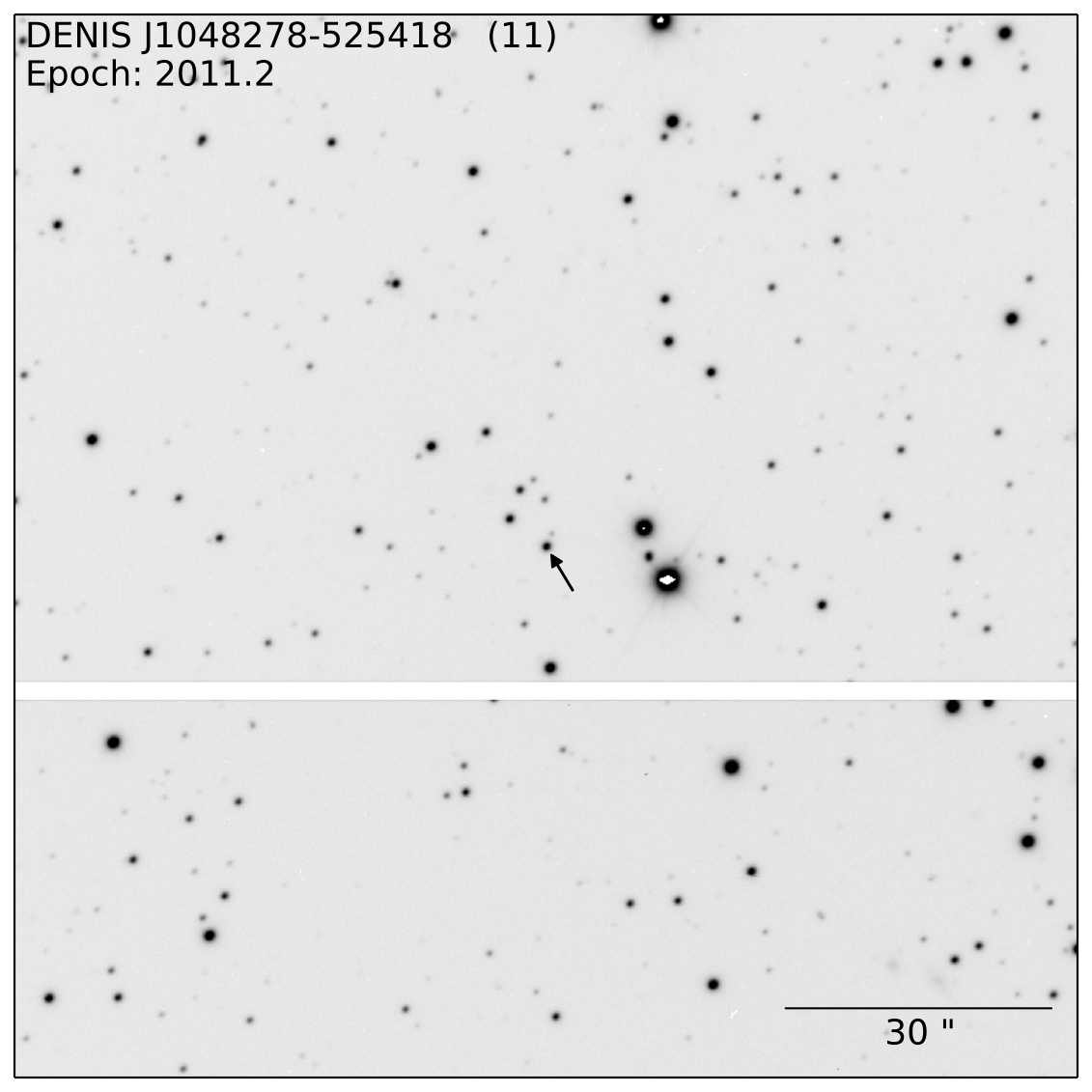}\hspace{0.5mm}
\includegraphics[width= 0.32\linewidth]{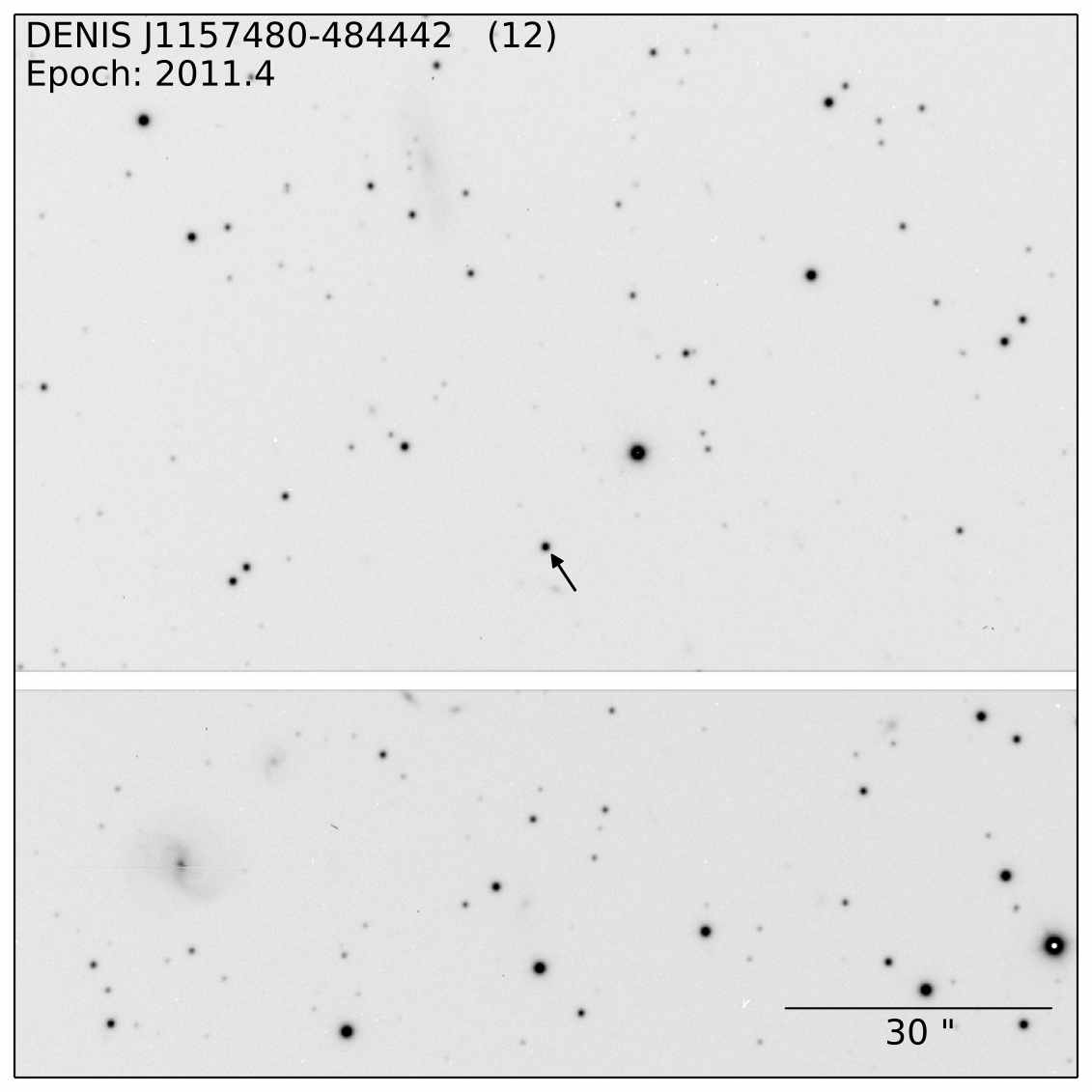}
\caption{Finding charts of target numbers 1--12 generated from FORS2 images in $I$-band. The target name and epoch are indicated in the top-left corner of every panel and an arrow identifies the target. In all panels North is up, east is left. The white strip corresponds to the inter-chip gap.}
\label{fig:fcs1}
\end{center}
\end{figure*}

\begin{figure*}[h]
\begin{center}
\includegraphics[width= 0.32\linewidth]{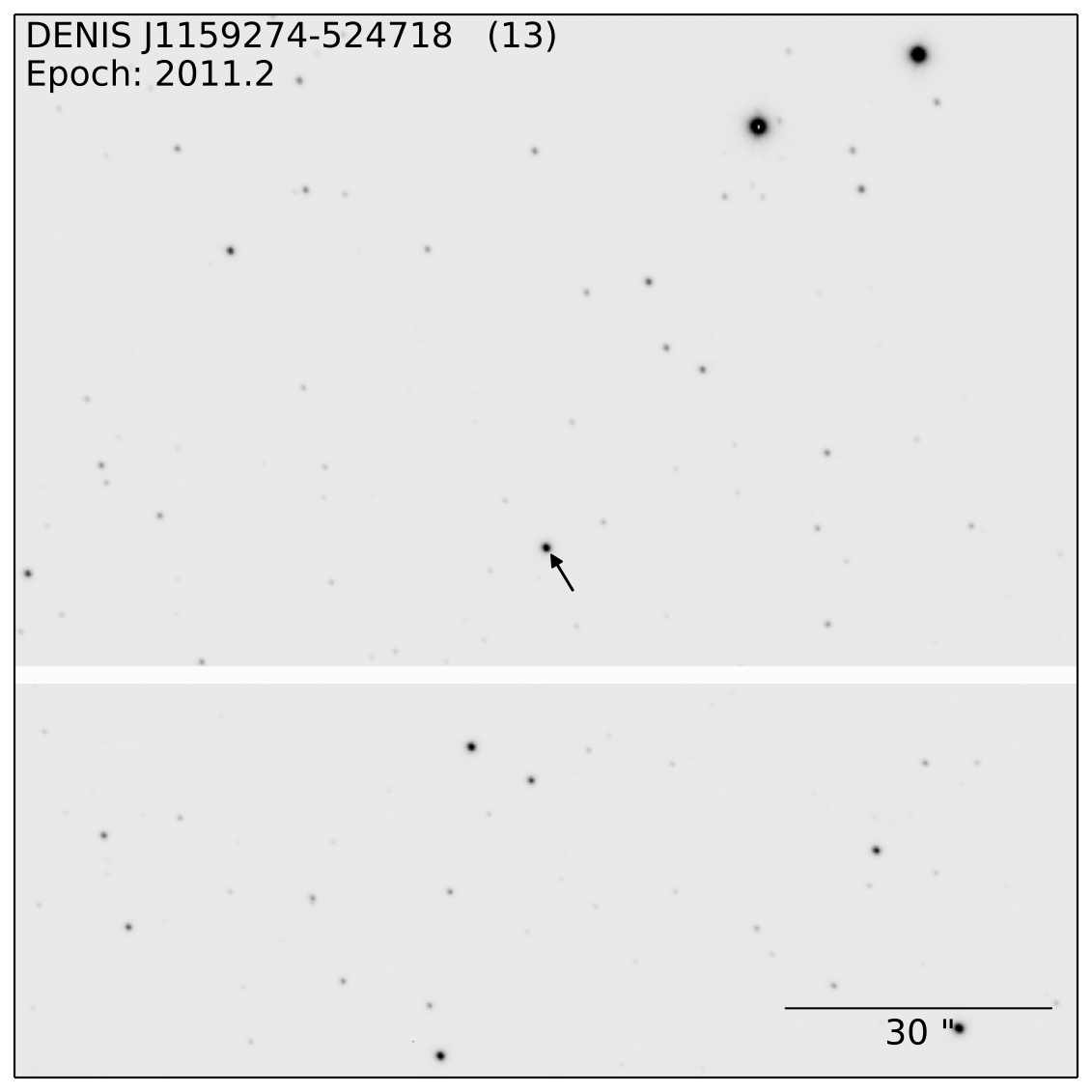}\hspace{0.5mm}
\includegraphics[width= 0.32\linewidth]{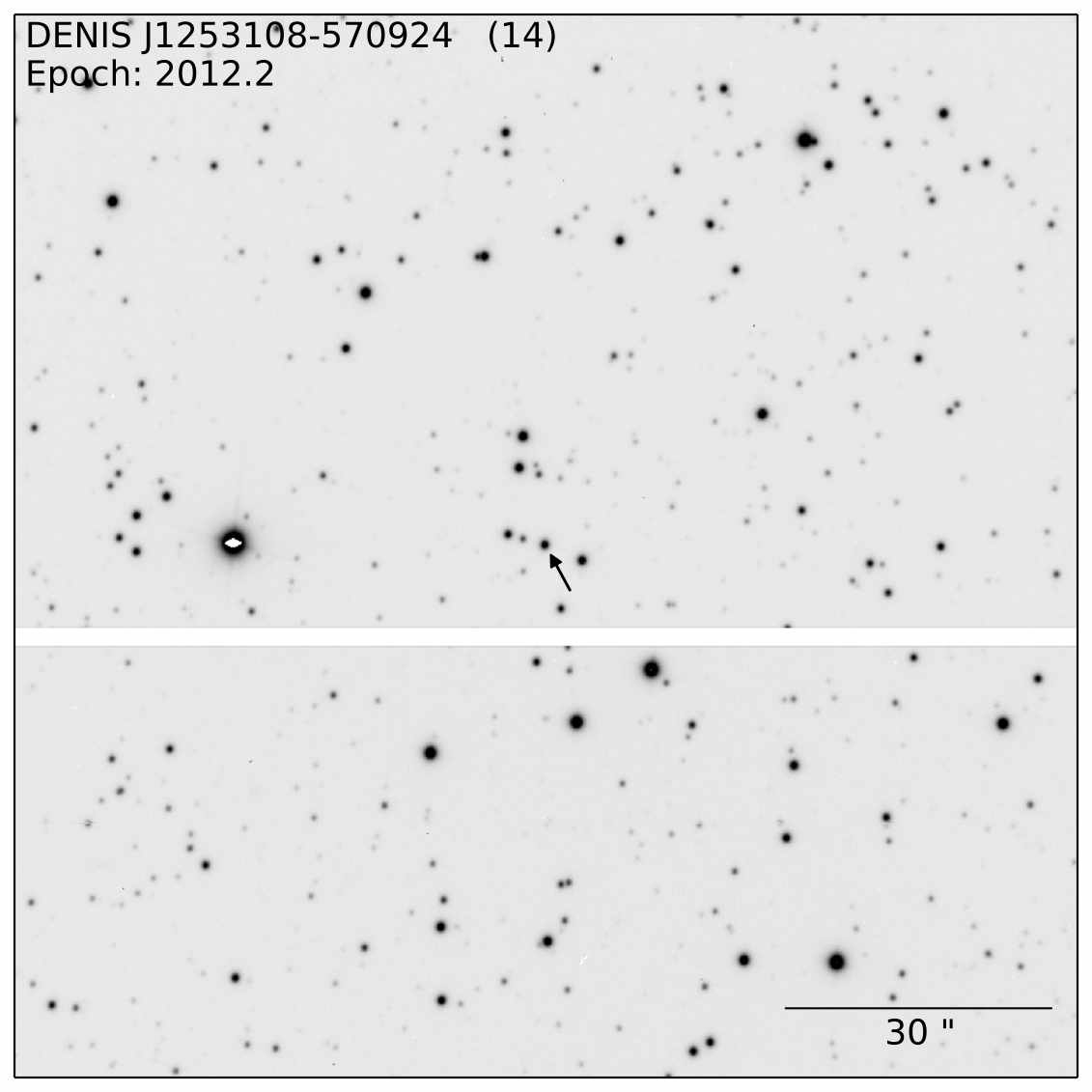}\hspace{0.5mm}
\includegraphics[width= 0.32\linewidth]{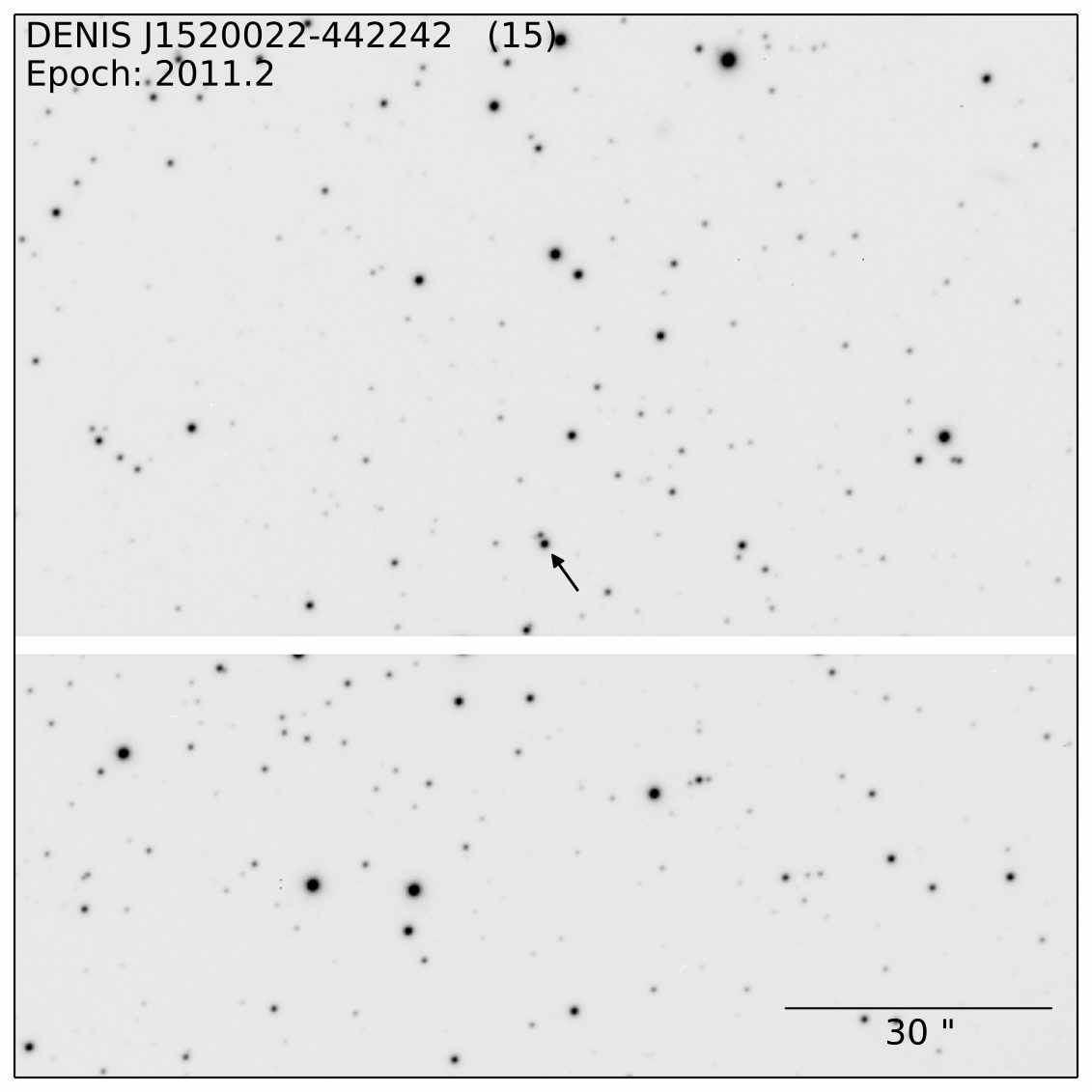}\vspace{1mm}
\includegraphics[width= 0.32\linewidth]{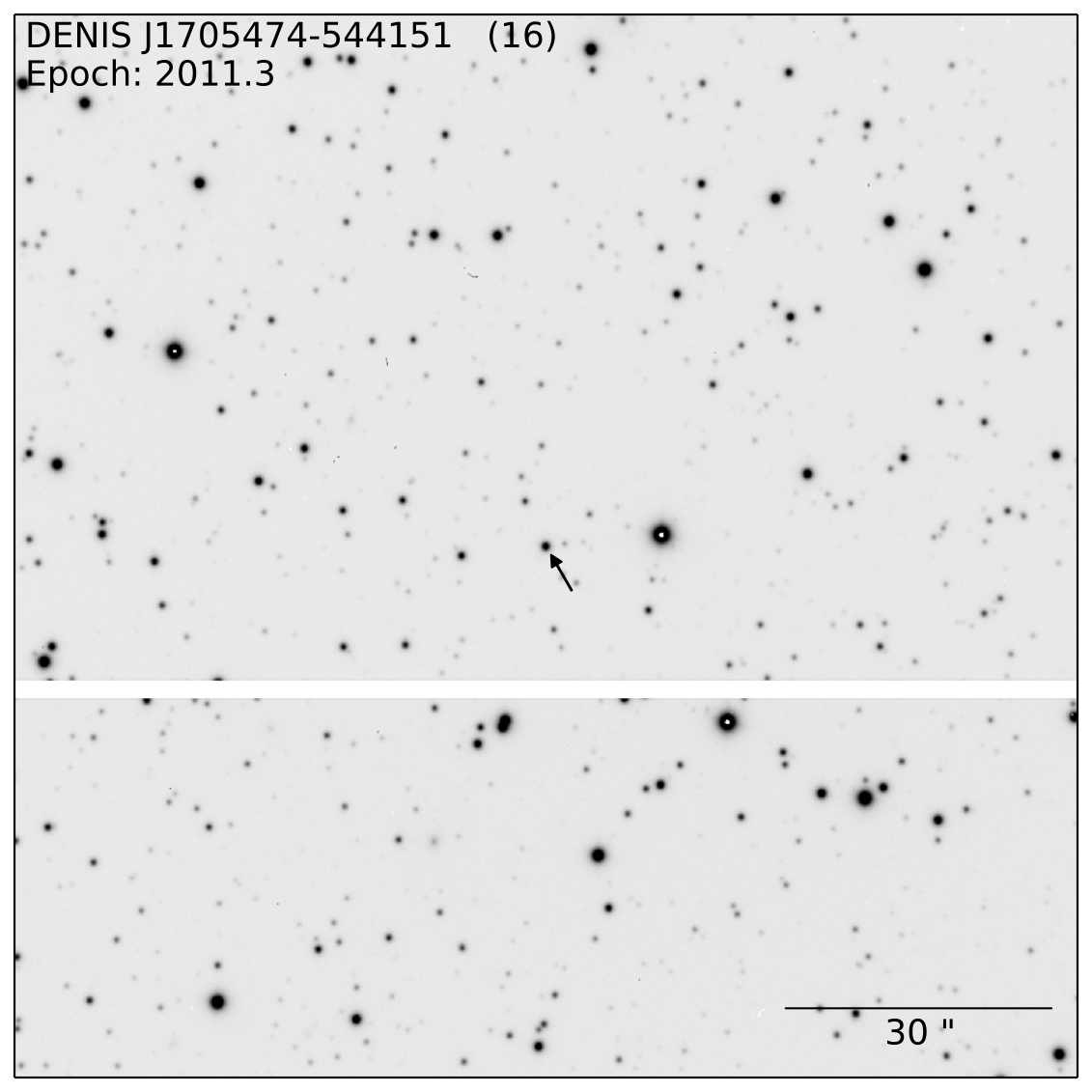}\hspace{0.5mm}
\includegraphics[width= 0.32\linewidth]{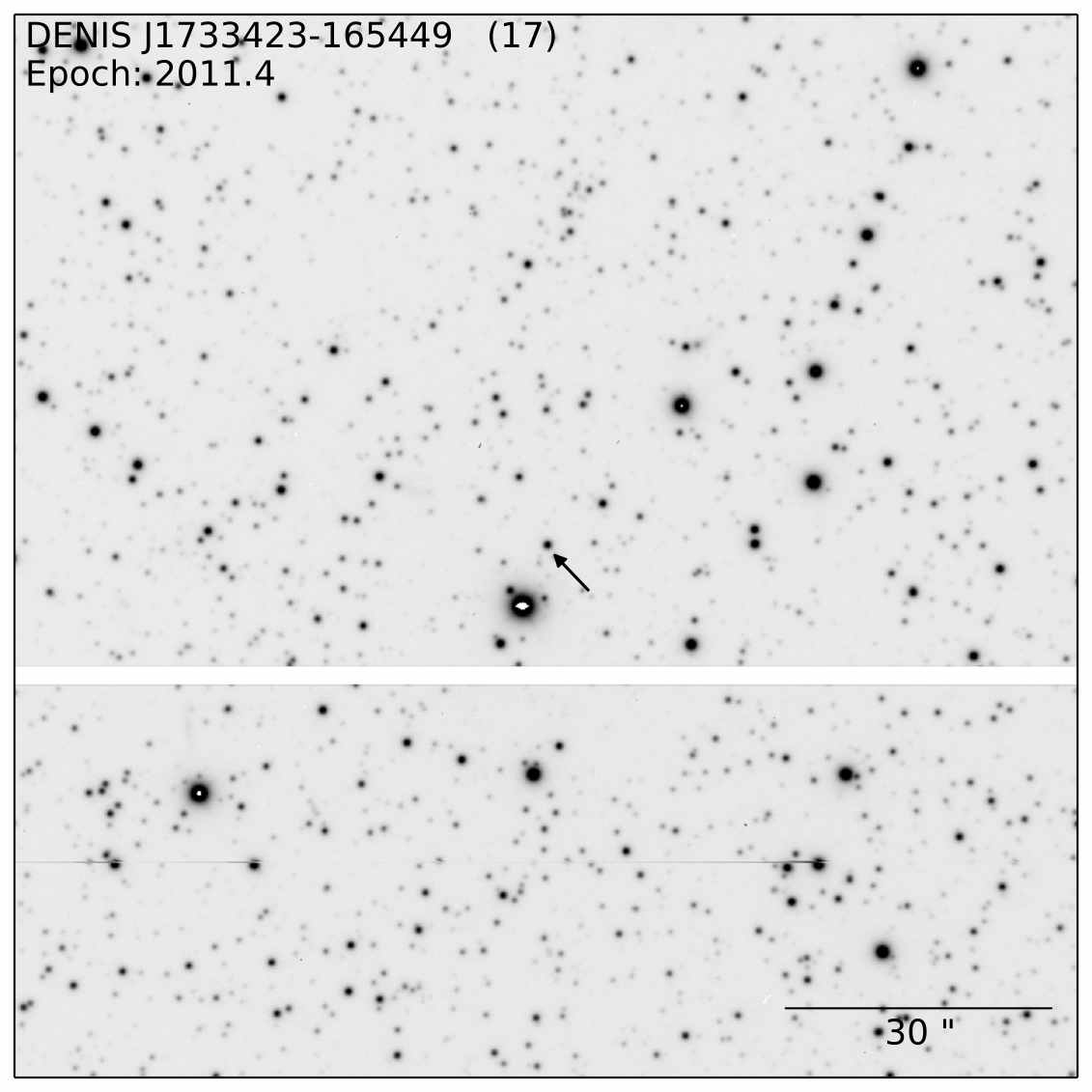}\hspace{0.5mm}
\includegraphics[width= 0.32\linewidth]{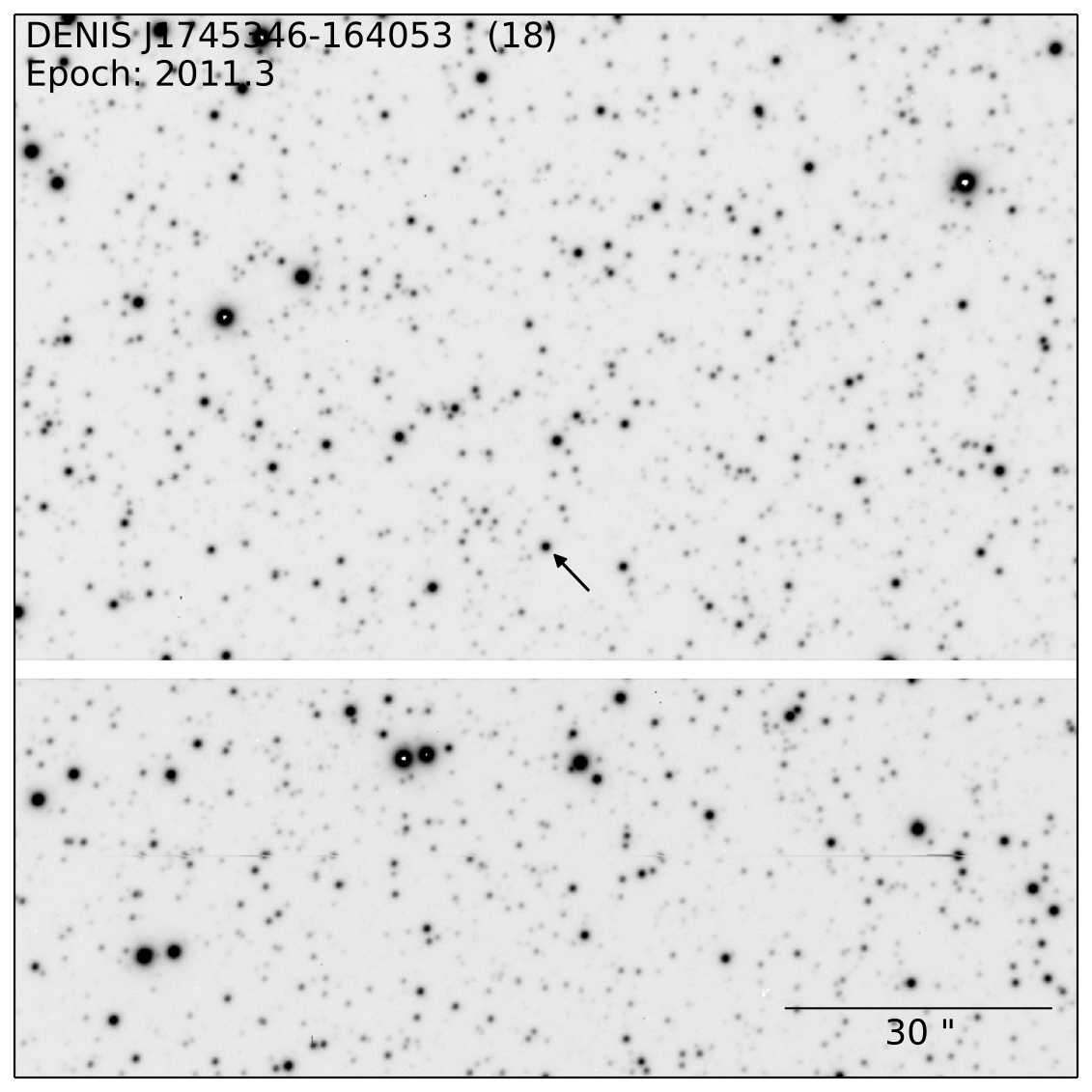}\vspace{1mm}
\includegraphics[width= 0.32\linewidth]{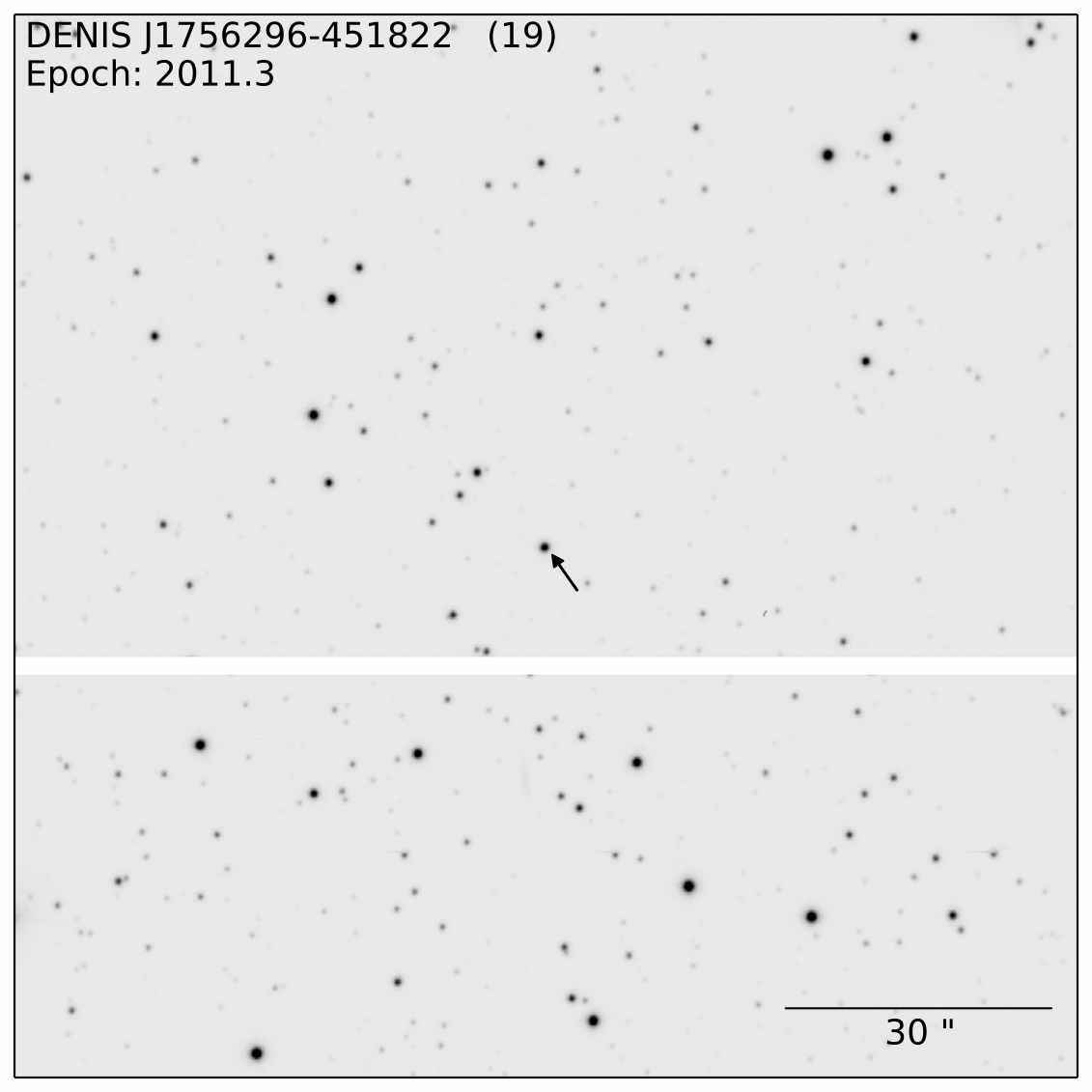}\hspace{0.5mm}
\includegraphics[width= 0.32\linewidth]{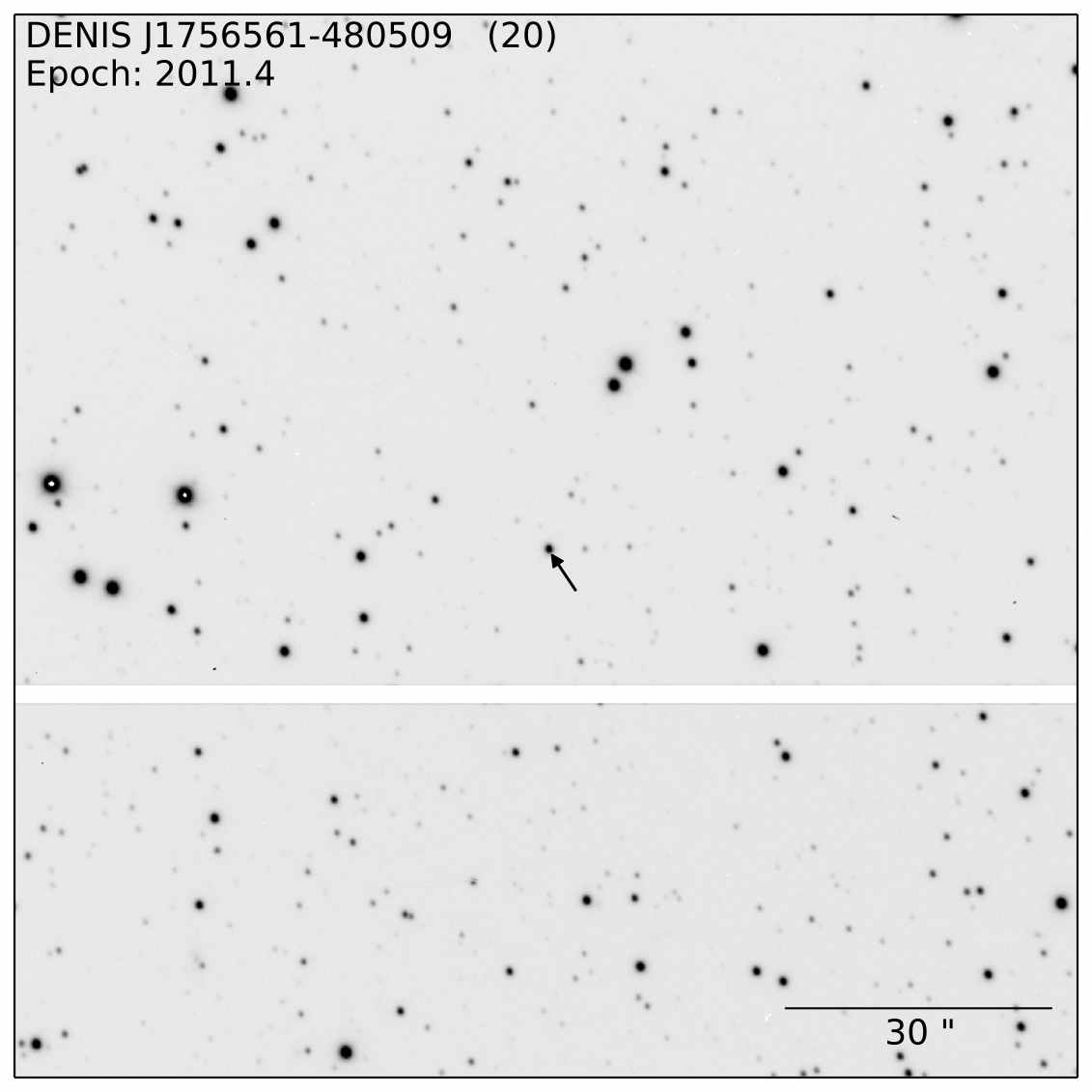}\hspace{0.5mm}
\caption{Finding charts of target numbers 13--20, analogously to Fig.~\ref{fig:fcs1}.}
\label{fig:fcs2}
\end{center}
\end{figure*}

\section{Tables with auxiliary data}
\begin{table*}[h!]
\caption{Approximate Galactic coordinates, space motions assuming zero radial velocity, and tangential velocities.}       
\label{tab:targspar}      
\centering         
\begin{tabular}{r c r r r r r r}    
\hline\hline      
Nr & ID &$l$ & $b$ & $U$& $V$& $W$ & $v_{tan}$\\
 &         & (\degr)& (\degr) & (km/s)& (km/s)& (km/s)& (km/s)\\
\hline   
1 & \dwone & 209.7534 & -8.3824 & -5.3 &  -13.8 & 15.5 & 21.4 \\
2 & \dwtwo & 227.4700 & -12.9818 & -38.7 &  -37.9 & 7.6 & 54.7 \\
3 & \dwthree & 238.2492 & -14.0907 & -22.0 &  -22.9 & 31.6 & 44.7 \\
4 & \dwfour & 236.0941 & -11.1726 & 9.9 &  9.8 & -13.2 & 19.2 \\
5 & \dwfive & 221.6226 & 2.6334 & 10.3 &  12.0 & 6.9 & 17.3 \\
6 & \dwsix & 242.2445 & 0.6267 & 42.8 &  21.8 & -57.1 & 74.6 \\
7 & \dwseven & 249.3422 & -0.1525 & 21.0 &  8.0 & -17.1 & 28.2 \\
8 & \dweight & 244.1357 & 5.1168 & -17.7 &  -8.2 & 3.7 & 19.8 \\
9 & \dwnine & 265.6169 & -6.8262 & 9.0 &  2.1 & -12.0 & 15.2 \\
10 & \dwten & 236.3897 & 14.6329 & 19.1 &  4.8 & -25.0 & 31.9 \\
11 & \dweleven & 284.8847 & 5.6030 & 28.6 &  -8.5 & -8.6 & 31.1 \\
12 & \dwtwelve & 293.8958 & 13.1832 & 6.4 &  -3.9 & -4.0 & 8.5 \\
13 & \dwthirt & 295.0116 & 9.2819 & 40.0 &  -21.4 & -15.1 & 47.8 \\
14 & \dwfourt & 303.1695 & 5.7139 & 100.8 &  -69.8 & -32.9 & 126.9 \\
15 & \dwfift & 329.0024 & 10.8420 & 33.2 &  -54.5 & 1.9 & 63.8 \\
16 & \dwsixt & 334.4639 & -8.2421 & 0.0 &  -3.6 & 10.4 & 11.0 \\
17 & \dwsevent & 8.8064 & 8.6638 & -0.9 &  0.9 & -6.9 & 7.0 \\
18 & \dweightt & 10.4979 & 6.3581 & -1.9 &  -2.2 & -13.4 & 13.7 \\
19 & \dwninet & 346.8777 & -10.1089 & 5.7 &  -13.9 & -13.6 & 20.2 \\
20 & \dwtwenty & 344.4229 & -11.4941 & -0.8 &  6.9 & -5.3 & 8.7 \\

\hline                
\end{tabular}
\end{table*}

\begin{table*}[h!]
\caption{Simbad object identifiers, $I$-band magnitudes, and mean epoch of FORS2 observations.}      
\label{tab:targsobj}      
\centering         
\begin{tabular}{r c r r r r r r r}    
\hline\hline      
Nr & Simbad Object &$m_{I,\mathrm{P}}$\tablefootmark{a}  & $m_{I,\mathrm{D}}$\tablefootmark{b}  & $m_{I}$\tablefootmark{c}  & $<\!t_m\!>$\\
     &                          & (mag)& (mag) &(mag)& (MJD)\\
     \hline
1 & \object{DENIS-P J061549.3-010041} & 17.0& 17.152& $17.022 \pm 0.014$  & 55765.455944\\
2 & \object{DENIS-P J063001.4-184014} & 15.9& 15.866& $15.739 \pm 0.080$  & 55981.600782\\
3 & \object{DENIS-P J064414.3-284141} & 17.0& 16.895& $16.913 \pm 0.023$  & 55773.238708\\
4 & \object{DENIS-P J065219.7-253450} & 15.9& 16.155& $15.976 \pm 0.018$  & 55769.666540\\
5 & \object{DENIS-P J071647.8-063037} & 17.5& 17.398& $17.479 \pm 0.023$  & 55793.491767\\
6 & \object{DENIS-P J075116.4-253043} & 16.5& 16.625& $16.497 \pm 0.025$  & 55780.454597\\
7 & \object{DENIS-P J080511.0-315811} & 16.5& 16.195& $16.011 \pm 0.016$  & 55758.247346\\
8 & \object{DENIS-P J081231.6-244442} & 17.3& 17.344& $17.217 \pm 0.080$  & 55790.231405\\
9 & \object{DENIS-P J082303.1-491201} & 17.1& 17.325& $17.099 \pm 0.021$  & 55821.933543\\
10 & \object{DENIS-P J082834.3-130919} & 16.1& 16.182& $16.056 \pm 0.028$  & 55784.811740\\
11 & \object{DENIS-P J104827.8-525418} & 17.2& 17.624& $17.536 \pm 0.013$  & 55774.709383\\
12 & \object{DENIS-P J115748.0-484442} & 17.4& 17.565& $17.315 \pm 0.015$  & 55781.585985\\
13 & \object{DENIS-P J115927.4-524718} & 14.5& 14.703& $14.550 \pm 0.010$  & 55820.313659\\
14 & \object{DENIS-P J125310.8-570924} & 16.7& 16.914& $16.711 \pm 0.017$  & 55884.313171\\
15 & \object{DENIS-P J152002.2-442242} & 16.7& 16.902& $16.775 \pm 0.080$  & 55859.681472\\
16 & \object{DENIS-P J170547.4-544151} & 16.7& 16.689& $16.542 \pm 0.009$  & 55900.133788\\
17 & \object{DENIS-P J173342.3-165449} & 16.8& 17.035& $16.916 \pm 0.015$  & 55915.072446\\
18 & \object{DENIS-P J174534.6-164053} & 17.1& 17.016& $16.975 \pm 0.014$  & 55909.746593\\
19 & \object{DENIS-P J175629.6-451822} & 15.5& 15.659& $15.490 \pm 0.008$  & 55872.130183\\
20 & \object{DENIS-P J175656.1-480509} & 16.7& 16.858& $16.742 \pm 0.016$  & 55892.867372\\
\hline                
\end{tabular}
\tablefoot{
\tablefoottext{a}{From \cite{Phan-Bao:2008fr}.}
\tablefoottext{b}{From DENIS catalogue \citep{Epchtein:1999aa}.}
\tablefoottext{c}{This work (see \citetalias{Lazorenko:2013kk} for details).}
}
\end{table*}

\end{appendix}

\end{document}